\documentclass[preprint,3p,12pt]{elsarticle}
\usepackage{amsmath}
\usepackage{stmaryrd}
\usepackage{bbding}
\usepackage{dcolumn}
\usepackage{graphicx}
\usepackage{amsfonts}
\usepackage{amssymb}
\usepackage{psfrag}
\usepackage{wrapfig}
\usepackage{subfigure}
\usepackage{makeidx}
\usepackage{bm}
\usepackage{epsf}
\usepackage{epsfig}
\usepackage{setspace}
\usepackage{color}
\usepackage{algorithm}
\usepackage{algorithmic}
\usepackage{setspace}
\usepackage[figuresright]{rotating}

\begin{document}

\title{High-order gas-kinetic scheme for numerical simulations of wind turbine with nacelle and tower using ALM and IBM}

\author[BNU]{Pengyu Huo}

\author[BNU]{Liang Pan \corref{cor}}
\ead{panliang@bnu.edu.cn}

\author[Oxford,SuStream]{Guiyu Cao \corref{cor}}
\ead{guiyu.cao@eng.ox.ac.uk}

\author[CAS-1,CAS-2]{Baoqing Meng}

\author[BUAA]{Baolin Tian}

\author[BristolU]{Yubo Huang}

\cortext[cor]{Corresponding author}
\address[BNU]{School of Mathematical Science, Beijing Normal University, Beijing, China}
\address[Oxford]{Department of Engineering Science, University of Oxford, Oxford, UK}
\address[SuStream]{Suzhou Stream Energy Technology Co., Ltd., Suzhou, China}
\address[CAS-1]{Institute of Mechanics, Chinese Academic of Sciences, Beijing, China}
\address[CAS-2]{University of Chinese Academy of Sciences, Beijing, China}
\address[BUAA]{School of Aeronautic Science and Engineering, Beihang University, Beijing, China}
\address[BristolU]{School of Computer Science, University of Bristol, Bristol, UK}

\begin{abstract}
For the first time, the actuator line model (ALM) and the immersed boundary method (IBM) are integrated into the high-order gas-kinetic scheme (GKS) to simulate the wind turbine with the nacelle and tower. 
The high-order GKS is extended to the simulation of three-dimensional weakly compressible isothermal flows within a well-developed two-stage fourth-order framework. 
For the wind turbine, the rotor blades are represented by a group of actuator points in ALM, and the nacelle and tower are represented by a group of Lagrangian points in IBM. 
Both ALM and IBM are integrated through an external body force added in the momentum equation within the high-order GKS. The high-order GKS is implemented on Graphics Processing Units (GPU) to achieve the parallel computing capabilities for the large-scale simulation of turbulent wakes. 
Turbulent channel flow and turbulent circular cylinder flow are firstly simulated to validate the numerical accuracy of weakly compressible high-order GKS.
The NREL $5$ MW reference wind turbine is simulated using ALM without the nacelle and tower.
The predicted normal and tangential aerodynamic loadings along the blade, as well as the time-averaged streamwise velocity in wind turbine wakes agree well with reference solutions. 
The numerical results of the high-order GKS are compared with that of the second-order GKS, which demonstrates that the high-order scheme resolves the vortex-dominated wake turbulence more effectively. Furthermore, 
the NTNU "Blind Test 1" wind turbine is simulated with nacelle and tower using IBM. 
The current method yields the periodic power and thrust coefficients of the rotor blade due to the blade-tower interactions, while the steady coefficients are obtained when the tower is omitted. 
Compared to the turbine wakes without the tower, the interaction between  tower vortex and tip vortex causes an earlier transition. 
The high-order GKS with ALM and IBM also well predicts the asymmetric mean flows of turbine wake, including the time-averaged streamwise velocity and turbulent kinetic energy, which are in good agreement with NTNU experimental data. 
The multiple-GPU enabled high-order GKS integrated with ALM and IBM offers an accurate and efficient approach for realistic wind turbine simulations.
\end{abstract}


\maketitle

\section{Introduction}
Increasing demand for wind energy has seen the size of wind turbines grow rapidly in recent years, facilitated in particular by the move to offshore wind farms. 
Recent turbines deployed in the North Sea have exceeded $200$ meters in diameter \cite{moraywest}, and conceptual designs for rotors approaching $300$ meters in diameter have been proposed \cite{zahle2024definition}. The chord-based Reynolds numbers characterizing the rotors and their blade flows for such large turbines are on the order of $10$ million, resulting in a wide range of multi-scale turbulent structures in the wake. 
These features pose significant challenges for numerical simulation at an affordable computational cost.

The actuator line model (ALM) has become widely-used as a mid-fidelity alternative to computationally expensive blade-resolved simulations for the prediction of aerodynamic loadings and wind turbine wakes \cite{ALM-1}.  
Most ALM simulations to date have been implemented in computational fluid dynamics (CFD) codes with second-order numerical schemes. 
However, as the Reynolds number of the problem increases, the mismatch between the spatial and temporal scales of the large-scale flow and turbulence dissipation increases, requiring higher-resolution simulations to resolve the wake turbulence well. 
Therefore, the high-order schemes have been used for ALM simulations of wind turbine wakes, such as the high-order spectral-element code Nek5000 \cite{ALM-high-order-1} and the PyFR solver based on the high-order Flux
Reconstruction (FR) approach \cite{ALM-high-order-2}.  Additionally, 
the lattice Boltzmann method (LBM) provides an alternative solver to the Navier-Stokes equations, which has been used for ALM simulations \cite{ALM-LBM-1,ALM-LBM-2,ALM-LBM-3}.

When simulating wind turbine wakes, the effect of turbine nacelle and tower has often been omitted for simplicity. 
The tower structure decelerates the local incident velocity, and the local blade forces are typically reduced due to a perturbation to the local flow angle \cite{aerodynamic-shkara,tower-effect-Zahle}.
For offshore floating wind turbines, the nacelle and tower are required to simulate accurately, considering the six-degree of freedom motions and the potential interactions between the tower vortex and the tip vortex \cite{NTNU-IBM-Santoni,support-Churchfield}.
The accurate simulation of objects with complicated geometry immersed in a flow remains a challenging problem.

Instead of the body-fitted mesh, the immersed boundary method (IBM) was originally developed  
to simulate cardiac mechanics and associated blood flow with Cartesian meshes \cite{IBM-Peskin}.  
In this approach, an Eulerian mesh for the fluid is used together with a set of Lagrangian points representing the immersed boundary \cite{IBM-Peskin2,IBM-Peskin3,IBM-Guo,IBM-Mittal,IBM-Shu}.
Within the direct-force formulation of IBM, the coupling effects between the particle and the fluid are explicitly imposed at the positions of the Lagrangian points to enforce the boundary condition. 
The force is formulated on the immersed boundary and re-distributed to the fluid elements surrounding the boundary.
The Lagrangian points move with the translational and rotational motion of the particle. 
The direct-force immersed boundary method has been developed within an Eulerian-Lagrangian framework \cite{IBM-Meng}. 
The interaction of particles and fluids, particle translation and rotation, and collisions among complex-shaped particles can be well resolved within a uniform framework. 
The flux conservation over the boundary surface of the particle and conservation of interphase coupling effects can be achieved.

In the last two decades, the gas-kinetic scheme (GKS)
\cite{GKS-Xu1,GKS-Xu2} based on the BGK model \cite{BGK-1,BGK-2} has been developed for
the computations from laminar to turbulent flows. The gas-kinetic
scheme presents a gas evolution process from the kinetic scale to
hydrodynamic scale, and both inviscid and viscous fluxes can be
calculated in one framework. Recently, a reliable framework was
provided to construct high-order GKS with fourth-order and
even higher-order spatial-temporal accuracy \cite{GKS-high-1}. Due
to its high accuracy and efficiency, the high-order GKS has been successfully
applied to the direct numerical simulation of multi-scale turbulent
flows \cite{GKS-high-2}. To further accelerate the large-scale
computation of turbulent flows, the high-order GKS has also been developed for
implementation on graphical processing units (GPU) architectures
\cite{GKS-high-3}.

In this paper, the ALM and IBM are integrated into the high-order GKS framework to simulate wind turbine with  nacelle and tower. 
The second-order GKS has already been developed for the simulation of low-Mach number
weakly compressible isothermal flows \cite{GKS-Xu3}. 
The high-order GKS is first extended to the simulation of
three-dimensional weakly compressible isothermal flows. 
The ALM and IBM are both integrated into the high-order GKS framework as an external body
force term in the momentum equation. To accelerate the computations, the high-order GKS code is implemented
with multiple GPUs to leverage parallel computing capabilities.
Numerical validations are conducted on the low-speed turbulent channel flow and the turbulent circular cylinder flow to confirm the ability of the high-order GKS to accurately resolve weakly compressible
multi-scale turbulent flows.

The NREL $5$~MW reference wind turbine \cite{NREL-turbine} is simulated using the in-house GPU-enabled high-order GKS with ALM. 
Firstly, we study the effect of the smearing kernel width $\varepsilon$ in ALM. We compare the predicted aerodynamic loadings and wake metrics from the high-order GKS with those of the second-order GKS.
The NTNU "Blind Test 1" wind turbine \cite{NTNU-turbine} is also simulated, where the rotor blade is modeled using ALM and the nacelle and tower are represented using IBM. 
The power and thrust coefficients and wake metrics are compared with the NTNU experimental data.

This paper is organized as follows. 
In Section $2$, the high-order GKS for weakly compressible flow is
presented. 
Section $3$ introduces the high-order GKS for ALM and IBM.
Section $4$ presents numerical simulations in wind turbines.
We summarize the findings in the last section.

\section{High-order GKS for weakly compressible flow}
In this section, the high-order gas-kinetic scheme is extended for the simulation of three-dimensional weakly compressible flows.
The high-order GKS is developed based on  the three-dimensional BGK equation \cite{BGK-1,BGK-2}
\begin{equation}\label{bgk-3d}
f_t+uf_x+vf_y+wf_z=\frac{g-f}{\tau},
\end{equation}
where  $f$ is the gas distribution function, $g$ is the three-dimensional Maxwellian
distribution and $\tau$ is the collision time and  $\boldsymbol{u}=(u,v,w)^T$ is the particle velocity.
In this paper, the three-dimensional weakly compressible isothermal
flow is considered \cite{GKS-Xu3}, and the equilibrium state is given by
\begin{equation}
g=\rho\bigg(\frac{\lambda}{\pi}\bigg)^{\frac{3}{2}}e^{-\lambda[(u-U)^2+(v-V)^2+(w-W)^2]},
\end{equation}
where $\rho$ is the density, and $\boldsymbol{U} \equiv (U,V,W)^T$
is the macroscopic velocity. $\lambda$ takes a constant value
determined by $\lambda={Ma}^2/(2{\boldsymbol{U}}_{\infty}^2)$, where
$Ma$ is the Mach number and $\boldsymbol{U}_{\infty}$ is the
free-stream velocity. The right-hand-side collision term in Eq.\eqref{bgk-3d}
satisfies the compatibility condition
\begin{equation}
\int \frac{g-f}{\tau} \boldsymbol{\psi} \text{d}\Xi = \boldsymbol{0},
\end{equation}
where $\boldsymbol{\psi} = (\psi_1,...,\psi_4)^T=(1,u,v,w)^T$ are
four collision invariants and
$\text{d}\Xi=\text{d}u\text{d}v\text{d}w$. Compared with the
BGK model for compressible flows \cite{GKS-Xu2},
the collision invariant for total energy is absent in weakly
compressible flows.

Taking moments of Eq.\eqref{bgk-3d} and integrating with respect to
space, the finite volume scheme can be expressed as
\begin{align}\label{semi}
\frac{\text{d}(\boldsymbol{Q}_{ijk})}{\text{d}t}=\mathcal{L}(\boldsymbol{Q}_{ijk}),
\end{align}
where $\boldsymbol{Q}=(\rho, \rho U, \rho V, \rho W)$ represents the
conservative macroscopic variables. The operator $\mathcal{L}$ is
defined as
\begin{equation}\label{finite}
\mathcal{L}(\boldsymbol{Q}_{ijk})=-\frac{1}{|\Omega_{ijk}|}\sum_{p=1}^6\mathbb{F}_{p}(t),
\end{equation}
where the control volume
$\Omega_{ijk}=\overline{x}_i\times\overline{y}_j\times
\overline{z}_k$ with $\overline{x}_i=[x_i-\Delta x/2,x_i+\Delta
x/2], \overline{y}_j=[y_j-\Delta y/2,y_j+\Delta y/2],
\overline{z}_k=[z_k-\Delta z/2,z_k+\Delta z/2]$, $|\Omega_{ijk}|$ is the volume of $\Omega_{ijk}$ and $\mathbb{F}_{p}(t)$
is the time-dependent numerical flux across the cell interface
$\Sigma_{p}$. The numerical flux in $x$-direction is given as an
example
\begin{equation}\label{x_example}
\mathbb{F}_{p}(t)=\iint_{\Sigma_{p}} F(\boldsymbol{Q})\cdot\boldsymbol{n}\text{d}\sigma=
\sum_{m,n=1}^2\omega_{mn} \int\boldsymbol{\psi} u f(\boldsymbol{x}_{i+1/2,j_m,k_n},t,\boldsymbol{u},\xi)\text{d}\Xi\Delta y\Delta z,
\end{equation}
where $\boldsymbol{n}$ is the outer normal direction, $\omega_{mn}$
the quadrature weight and $(\boldsymbol{x}_{i+1/2,j_m,k_n})$ represents
Gaussian quadrature points at the interface $\overline{y}_j \times
\overline{z}_k$. The numerical flux needs to be constructed by taking moments of gas
distribution function at cell interface. 
For the  weakly compressible flows  \cite{GKS-Xu1}, the gas distribution function $f$ at
a cell interface has been constructed as
\begin{equation}\label{flux}
f(\boldsymbol{x}_{i+1/2,j_m,k_n},t,\boldsymbol{u})=g_0(1-\tau(a_1u+a_2v+a_3w+A)+At),
\end{equation}
where $g_0$ is the collisional equilibrium state. 
In the computation, the collision time $\tau$ in the gas distribution function Eq.\eqref{flux} takes 
\begin{align*}
\tau=\mu/p, 
\end{align*}
where $\mu$ the molecular viscosity and $p$ the pressure. With the reconstruction of macroscopic
variables, the coefficients in Eq.\eqref{flux} can be fully
determined by the reconstructed derivatives and compatibility
condition as
\begin{align}
 \displaystyle
    \langle a_1\rangle=\frac{\partial \boldsymbol{Q}_{0}}{\partial x},
    \langle a_2\rangle=\frac{\partial \boldsymbol{Q}_{0}}{\partial y},
    \langle a_3\rangle=\frac{\partial \boldsymbol{Q}_{0}}{\partial z},
    \langle a_1u+a_2v+a_3w+A\rangle=\boldsymbol{0},
    \end{align}
where $\langle \cdot \rangle$ represents the moments of equilibrium
state, and $\boldsymbol{Q}_0$ is the equilibrium macroscopic
variables. More details about microscopic coefficients are presented in Appendix.

To achieve the high-order spatial accuracy, the fifth-order weighted
essentially non-oscillatory (WENO) reconstruction method \cite{WENO-JS} is adopted. For the three-dimensional computation, the
dimension-by-dimension reconstruction is used to obtain the
macroscopic variables at the Gaussian quadrature points. For the temporal 
discretization, the two-stage fourth-order method is used.
The time step $\Delta t$ is determined by the CFL condition as 
\begin{align*}
\Delta t = CFL \times {\min\{\Delta x, \Delta y, \Delta z\}}/ (|\boldsymbol{U}|_{max}+C), 
\end{align*}
where $|\boldsymbol{U}|_{max}$ is the maximum velocity in the computation
domain, and $C=\sqrt{1/2\lambda}$ denotes the local sound speed, and $CFL$ represents the CFL number. More details about spatial reconstruction can be found in previous work
\cite{GKS-high-1,GKS-high-2}.

\section{Modeling for wind turbine with nacelle and tower}
In this section, the modeling for wind turbine with nacelle and tower is introduced. 
The actuator line model is used for the rotor blades of wind turbine, and the immersed boundary method is used to model the nacelle and tower.

\subsection{Actuator line model for wind turbine}
In the actuator line model, the rotor blades are represented by a spanwise distribution
of actuator points. These lines are rotating to mimic the effect of
the rotor blades without requiring them to be directly meshed. The
ALM consists of three key steps: flow sampling, force calculation
and force imposition on the flow. In the simulation of NREL $5$ MW wind turbine, the point-based velocity sampling \citep{mittal2015improvements} is adopted to determine the flow incident on each blade element. The trilinear interpolation method is used to interpolate the simulated velocity to each actuator point. And the line sampling method \cite{ALM-2} is used for the simulation of the NTNU rotor. 
The blade forces at the actuator points can
be calculated with the sampled velocity and lift-drag polar based on
two-dimensional airfoil theory. Correction factors can be applied to
account for three-dimension flow phenomena; none have been applied
in NREL$5$ MW wind turbine simulation and the tip correction factor in \cite{tip-correction-shen} has been used in NTNU rotor simulation. The obtained blade forces are distributed onto the flow field
as body force with a Gaussian kernel. The two-dimensional blade-local velocity $\boldsymbol{U}^{2D}$ sampled from the flow field is added to the rotational velocity $\omega r$ to produce the blade-relative velocity $\boldsymbol{U_{rel}}$. 
The axial and tangential velocities of the incident flow can be denoted as $\boldsymbol{U}^{2D}=(U_{x},U_{\theta})$, where $x$ and $\theta$ represent the axial and the blade rotational directions,
respectively. The blade-relative velocity is
defined as $\boldsymbol{U}_{rel} = (U_{x},U_{\theta }-\omega r)$.
According to the formula in \citep{Optimal-eps}, the free stream
velocity $\boldsymbol{U_{\infty}}^{2D}$ can be estimated using the
blade-local velocity by
\begin{align*}
\boldsymbol{U_{\infty}}^{2D}=\frac{\boldsymbol{U}^{2D}}{\displaystyle 1-\frac{1}{4\pi^{1/2}}C_{D}\frac{c}{\epsilon}}.
\end{align*}
The blade-relative velocity is updated with $\boldsymbol{U_{\infty}}^{2D}$ by
$\boldsymbol{U}_{rel}=((U_{\infty})_{x},(U_{\infty})_{\theta
}-\omega r)$. The angle of attack $\alpha$ is defined as
\begin{align*}
\alpha=\phi-\gamma,
\end{align*}
where $\gamma$ is the local twist angle and $\phi$ is the flow angle
between $\boldsymbol{U}_{rel}$ and the rotor plane, which can be
calculated by
\begin{align*}
\phi=\arctan\frac{U_{x}}{(\omega r-U_{\theta })}.
\end{align*}

\begin{figure*}[!h]
\centering
\includegraphics[trim=80 100 100 100, clip = true, width=0.99\linewidth]{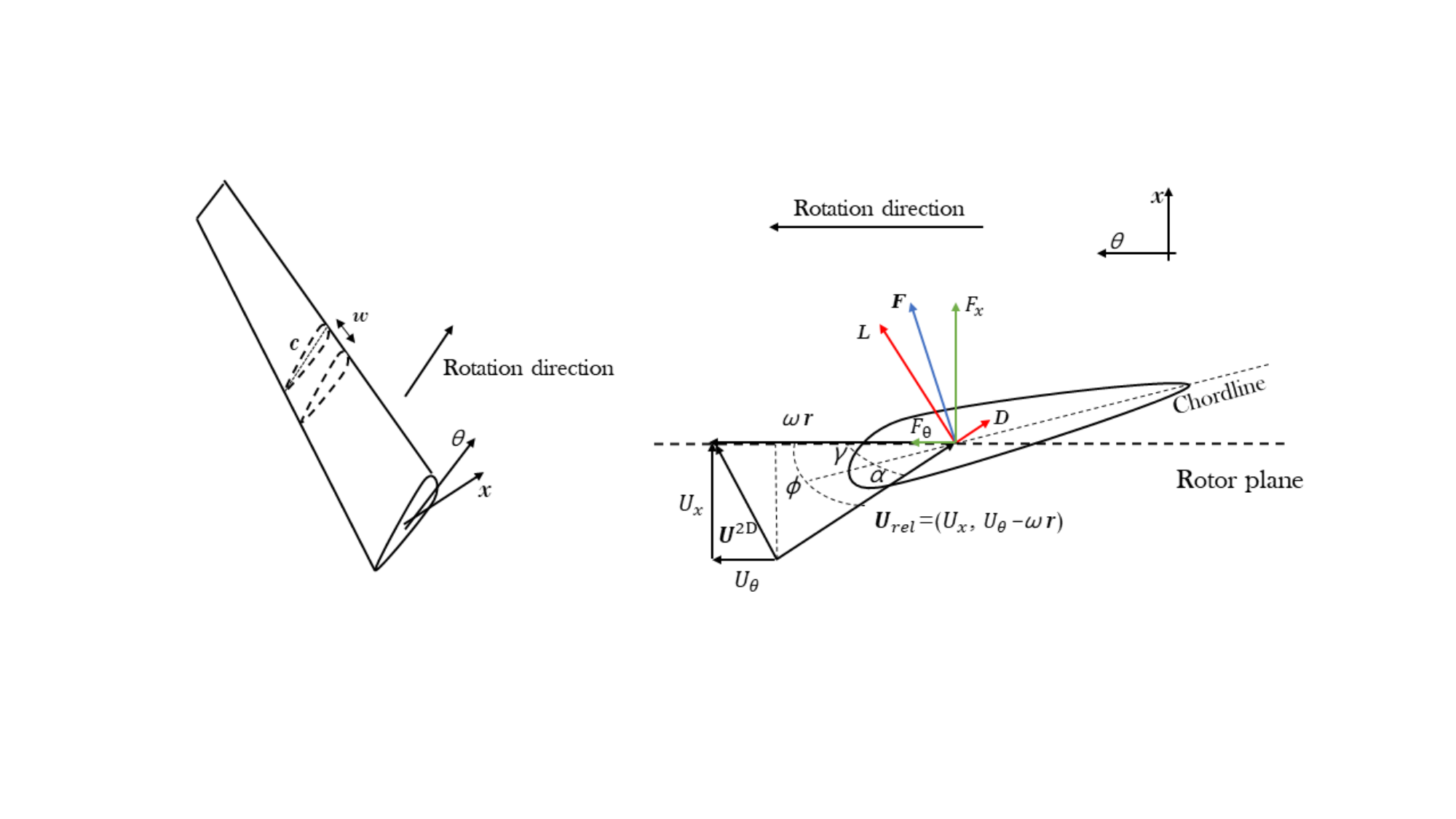}
\caption{\label{Schematic-1} Two-dimensional cross-sectional airfoil
element for the actuator line model.}
\end{figure*}

With the obtained angle of attack $\alpha$, the lift and drag
coefficients $C_L(\alpha)$ and $C_D(\alpha)$ can be evaluated by
using tabulated 2D airfoil data. The lift $L$ and the drag $D$ at each
actuator point can be computed with
\begin{align}\label{axial-tangential}
\begin{split}
L &= \frac12 \rho c w C_L(\alpha)\boldsymbol{U}_{rel}^2,\\
D &=\frac12 \rho c w C_D(\alpha)\boldsymbol{U}_{rel}^2,
\end{split}
\end{align}
where $c$ denotes the airfoil chord length and $w$ is the width of the blade element. The axial and
tangential forces on each blade element are then computed as
follows
\begin{align*}
F_x        &= (L \cos \phi + D \sin \phi),\\
F_\theta&= (L\sin \phi-D\cos \phi).
\end{align*}
To make the force distribution smoothly, the interpolation of the
lift and drag tables between nearby airfoil sections is needed. The
obtained blade forces are projected onto the flow field as body
forces. In order to distribute the point forces onto the surrounding
cells, the forces are projected by a Gaussian function
\begin{align}\label{Gaussian}
    \boldsymbol{f}_{ALM}=\frac{\boldsymbol{F}}{\varepsilon ^{3}\pi ^{3/2}}\exp[-(r/\varepsilon )^{2}],
\end{align}
where $\boldsymbol{f}_{ALM}$ is the distributed body force,
$\boldsymbol{F} \equiv (F_x, F_\theta)$ the point force at each actuator point, $\varepsilon $ the
smearing kernel width and $r$ the distance from the
actuator point where the force is distributed. More details in ALM
can be found in \citep{ALM-2}.

\subsection{Immersed boundary method for nacelle and tower}
To model the effect of nacelle and tower for the wind turbine, the immersed boundary method
is used under an Eulerian-Lagrangian framework \cite{IBM-Meng}. The
fluid equation is solved using an Eulerian frame and an immersed
boundary is represented by a group of Lagrangian points. Each
Lagrangian point provides an external force on the surrounding
Eulerian cells. The fluid field is updated by GKS solver without
external force. The velocities of Lagrangian points are interpolated
to the cell centers
\begin{align*}
\boldsymbol{\bar{v}}_{ijk} = \sum_p W_{p\rightarrow ijk} \boldsymbol{v}_{p},
\end{align*}
where $\boldsymbol{v}_{p}$ is the velocity of Lagrangian point $p$,
$\boldsymbol{\overline{v}}_{ijk}$ represents the interpolated
velocity of Lagrangian points at the Eulerian cell $(i,j,k)$ and $W_{p\to ijk}$ is the
weight from the Lagrangian point to Eulerian cell. Extra
interpolation is performed for the interpolated volume of the Lagrangian points 
at the Eulerian cell $(i,j,k)$ as follows
\begin{align*}
\bar{{V}}_{ijk}= \sum_pW_{p\rightarrow ijk} V_{p},
\end{align*}
where $V_p$ is the volume of Lagrangian point $p$. The volumes of Lagrangian point can be determined as follows
\begin{align*}
V_p=\Delta x\cdot\text{d}S,
\end{align*} 
where $\Delta x$ is the characteristic length of local cell and $\text{d}S$ is the area of the surface element of Lagrangian point $p$. 
The coupling force on each cell is estimated as follows
\begin{align}\label{ibm-1}
\boldsymbol{F}_{ijk}^{s} = \rho_{f,ijk} \bar{{V}}_{ijk}\frac{\boldsymbol{\bar{v}}_{ijk}-\boldsymbol{\bar{U}}_{f,ijk}^{s}}{\Delta t},
\end{align}
where the superscript $s$ represents the iterative step, $\rho_{f,ijk}$ is the gas density of cell $(i,j,k)$ and
$\boldsymbol{\bar{U}}_{f,ijk}^{s}$ represents the fluid
velocity at the Eulerian cell $(i,j,k)$.
With the external force term, the gas velocity at Eulerian cells can
be updated by
\begin{align*}
\rho_{f,ijk}\boldsymbol{\bar{U}}_{f,ijk}^{s+1}=\rho_{f,ijk}\boldsymbol{\bar{U}}_{f,ijk}^{s}+\frac{\Delta t}{\bar{V}_{ijk}}\boldsymbol{F}_{ijk}^{s}.
\end{align*}
And with the weight from Eulerian cell to Lagrangian point $W_{ijk\rightarrow p}$,
the gas velocity at the Lagrangian point can be interpolated by
\begin{align*}
\boldsymbol{\bar{U}}_{f}^{s+1} = \sum W_{ijk\rightarrow p} \boldsymbol{\bar{U}}_{f,ijk}^{s+1},
\end{align*}
If the norm of velocity difference $\left\lvert \boldsymbol{v}_p-\boldsymbol{\bar{U}}_f^{s+1} \right\rvert$
is less than a given small value, the iteration stops.
The total external force on the fluid can be updated as follows
\begin{align}\label{force-ibm}
\boldsymbol{f}_{IBM}=\sum_{s}\boldsymbol{F}_{ijk}^{s}.
\end{align}
In this paper, the nacelle and tower are stationary, and it is not necessary to consider the movement of  Lagrangian point.

 \subsection{Update of momentum equation}
 The immersed boundary method and actuator line model are implemented
 into the GKS framework as an external body force term in the
 momentum equation. With the external body force, the equation for
 momentum can be written as
 \begin{align*}
     \frac{\partial \rho \boldsymbol{U}}{\partial t}&=\mathcal {L}(\rho \boldsymbol{U})+\boldsymbol{f},
 \end{align*}
 where $\mathcal {L}(\rho \boldsymbol{U})$ is the operator for the
 spatial derivative of momentum.  The external body force term reads
\begin{align*}
\boldsymbol{f}=\boldsymbol{f}_{ALM} + \boldsymbol{f}_{IBM},
\end{align*}
where $\boldsymbol{f}_{ALM}$ and $\boldsymbol{f}_{IBM}$ are provided by actuator line model with Eq.\eqref{Gaussian} and
immersed boundary method with Eq.\eqref{force-ibm}. For the
two-stage fourth-order GKS, we update the ${\rho}^{n+1}$ and $({\rho
\boldsymbol{U}})^{n+1}$ without external
force. After obtaining the external body force, the momentum at each
control volume can be updated by
\begin{align}
     (\widetilde{\rho \boldsymbol{U}})_{ijk}^{n+1} = (\rho \boldsymbol{U})_{ijk}^{n+1} + \boldsymbol{f}_{ijk}\frac{\Delta t}{\bar{V}_{ijk}},
\end{align}
where $(\widetilde{\rho \boldsymbol{U}})^{n+1}$ is the final updated
momentum at the $n+1$ step.

\subsection{Multiple-GPU implementation}
\begin{figure}[!h]
\centering
.\includegraphics[width=0.65\linewidth]{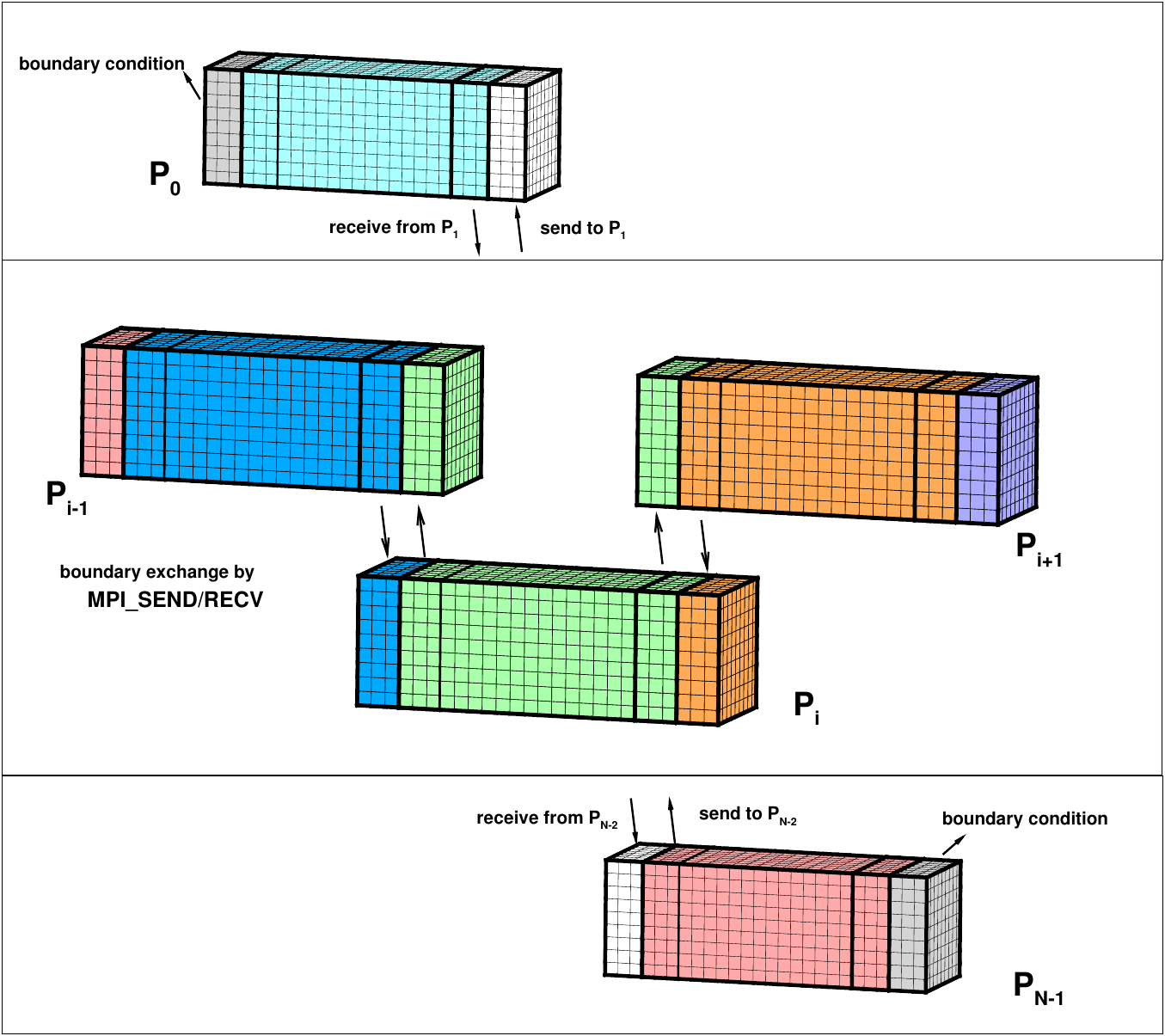}
\caption{\label{schematic-GPU} Domain decomposition and communications for multiple-GPU implementation. The illustration is adapted from \cite{GKS-high-3}.}
\end{figure}

To conduct the large-scale computations efficiently, the high-order GKS code is implemented with multiple GPUs using MPI and CUDA \cite{GKS-high-3}. 
CUDA-aware MPI library is chosen \cite{CUDA-MPI}, where GPU data can be directly passed by MPI function and transferred in the efficient way automatically. 
In terms of GPU computation with MPI, the slab decomposition is used, which reduces the frequency of GPU-GPU communications and improves performances. 
As shown in Fig.\ref{schematic-GPU}, the computational domain is divided into $N$ parts in $x$-direction and $N$ GPUs are used.
The processor $P_i$ exchanges the data with $P_{i-1}$ and $P_{i+1}$, and  the wall boundary condition of $P_0$ and $P_{N-1}$ can be
implemented directly.  For each decomposed domain, the computation is executed with the prescribed GPU. The identical code is running for
each GPU, where kernels and grids are similarly set as single-GPU code. 
In the computations, the Nvidia Tesla V$100$ GPUs  are used and GPU-GPU communication is achieved by Nvidia NVLink.

The detailed parameters of GPU are given
in Tab.\ref{GPU-CPU-B}, respectively.
For RTX A5000  GPU,  the GPU-GPU communication is achieved by
connection traversing PCIe, and there are $2$ RTX A5000 GPUs in one
node. For Tesla V100 GPU, there are $8$ GPUs inside one GPU node,
and more nodes are needed for more than $8$ GPUs. The GPU-GPU
communication in one GPU node is achieved by Nvidia NVLink. The
communication across GPU nodes can be achieved by GPU Direct RDMA
via iWARP, RDMA over Converged Ethernet (RoCE) or InfiniBand. In
this paper, RoCE is used for communication across GPU nodes.

\section{Numerical simulations}
In this section, the numerical examples are presented to validate the current method.
Turbulent channel flow is used to test the performance of classical wall bounded turbulence. 
For the flow past circular cylinder, the performance for the wake turbulence is tested by high-order GKS with IBM. 
The NREL $5$ MW reference wind turbine \cite{NREL-turbine,wind-turbine-modeling} is simulated using high-order GKS with ALM. 
Finally, the NTNU "Blind Test 1" wind turbine  with nacelle and tower  \cite{NTNU-turbine} is tested by high-order GKS with ALM and IBM.
The computational mesh and the settings of GPU for each case are listed in Tab.\ref{computational settings}.


\begin{table}[!h]
\begin{center}
\def\temptablewidth{1.0\textwidth}
{\rule{\temptablewidth}{1.0pt}}
\begin{tabular*}{\temptablewidth}{@{\extracolsep{\fill}}c|c|c}
~                            & Nvidia RTX A5000 & Nvidia Tesla V100 \\
\hline
Clock rate                   & 1.17 GHz         & 1.53 GHz          \\
\hline
Stream multiprocessor        & 64               & 80                \\
\hline
FP$32$ precision performance & 27.77 Tflops     & 15.7 Tflops       \\
\hline
FP$64$ precision performance & 433.9 Gflops  & 7834 Gflops       \\
\hline
GPU memory size              & 24 GB            & 32 GB             \\
\hline
Memory bandwidth             & 768 GB/s         & 897 GB/s          \\
\end{tabular*}
{\rule{\temptablewidth}{1.0pt}}
\caption{\label{GPU-CPU-B} The detailed parameters of GPUs.}
\end{center}
\begin{center}
\def\temptablewidth{1.0\textwidth}
{\rule{\temptablewidth}{1.0pt}}
\begin{tabular*}{\temptablewidth}{@{\extracolsep{\fill}}c|c|c}
Case                          & Mesh  & GPU   \\
\hline
Turbulent channel flow                           & $128\times128\times128$         & 1 GPU,  A5000     \\
\hline
Flow past circular cylinder                     & $460\times160\times360$          & 2 GPUs,  V100     \\
\hline
NREL $5$MW wind turbine  wake         &  $700\times300\times300$         & 2 GPUs,  V100     \\
\hline
NTNU rotor with nacelle and tower       &$1400\times 300\times 220$         & 4 GPUs,  V100    \\
\end{tabular*}
{\rule{\temptablewidth}{1.0pt}}
\caption{\label{computational settings} The setups for computational mesh and settings of GPU in current numerical cases.}
\end{center}
\end{table}

\subsection{Turbulent channel flow}
The turbulent channel flows have been well studied to understand the mechanism of wall-bounded turbulent flows \cite{channel-dns-1,channel-dns-2}. 
In this case, the weakly compressible turbulent channel flow is tested with friction Reynolds number $Re_\tau=180$ and Mach number $Ma=0.1$.  
In the computation, the physical domain is $(x,y,z)\in[0,2 \pi H] \times [-H,H] \times[0,\pi H]$, and the computational mesh is given by the coordinate transformation
\begin{align*}
\begin{cases}
\displaystyle x=\xi,\\
\displaystyle y=\tanh(b_g(\frac{\eta}{1.5\pi}-1))/\tanh(b_g),\\
\displaystyle z=\zeta,
\end{cases}
\end{align*}
where the computational domain takes $(\xi,\eta,\zeta)\in[0, 2\pi H]\times[0, 3\pi H]\times[0, \pi H]$, $H=1$ and $b_g=2$. 
The fluid is initiated with density $\rho = 1$ and the initial streamwise velocity $U(y)$ profile is given by the Poiseuille flow profile. 
The periodic boundary conditions are used in streamwise $x$-direction and spanwise $z$-directions, and
the non-slip and isothermal boundary conditions are used in vertical $y$-direction. The velocity components are denoted as $U$, $V$, and $W$ for $x$-,
$y$-, $z$-directions, respectively.
The constant moment flux in the $x$-direction is used to determine the external force \cite{GKS-high-2, GKS-high-3}.

\begin{figure}[!h]
\centering 
\includegraphics[width=0.55\linewidth]{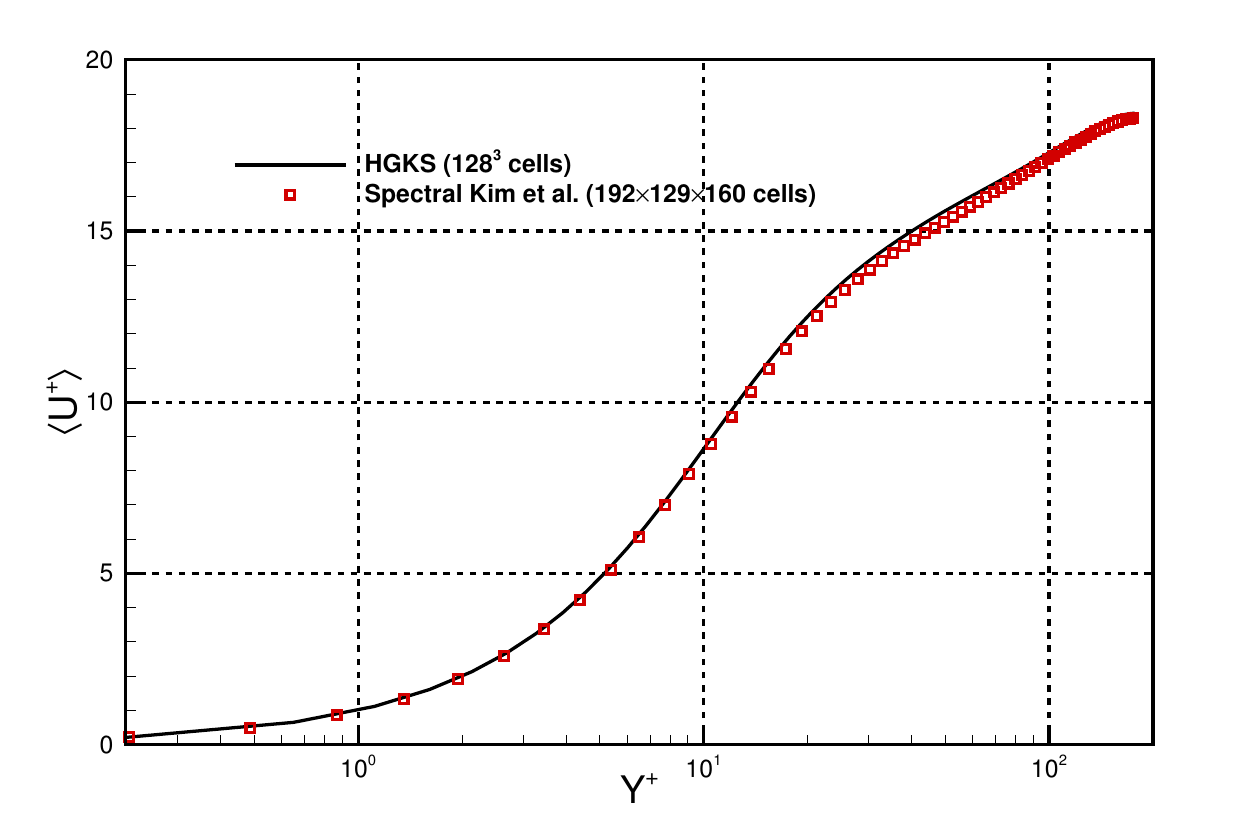}
\caption{\label{channel-flow-1} Turbulent channel flow: the mean streamwise flow velocity profile.}
\end{figure}

\begin{figure}[!h]
\centering
\includegraphics[width=0.475\linewidth]{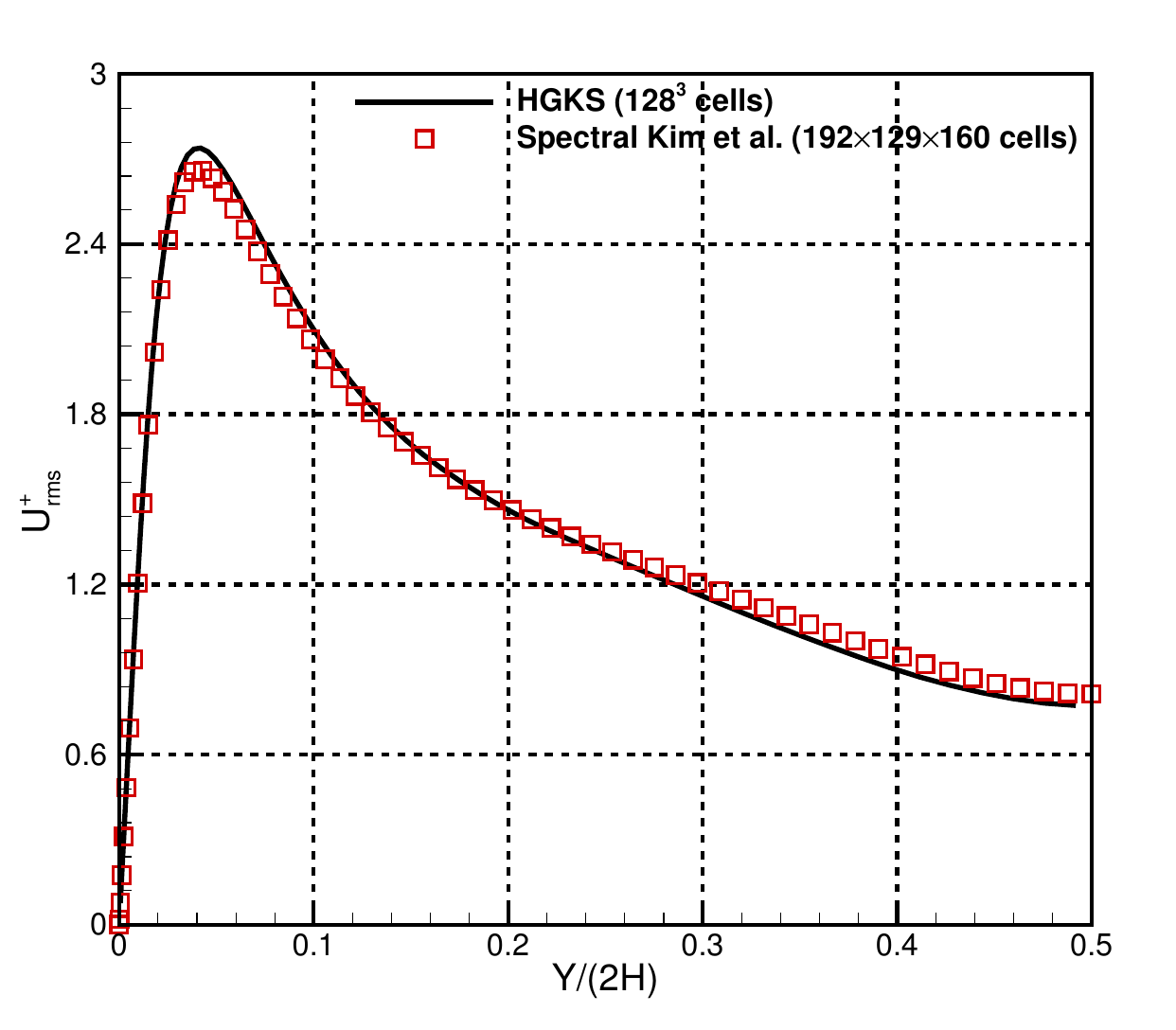}
\includegraphics[width=0.475\linewidth]{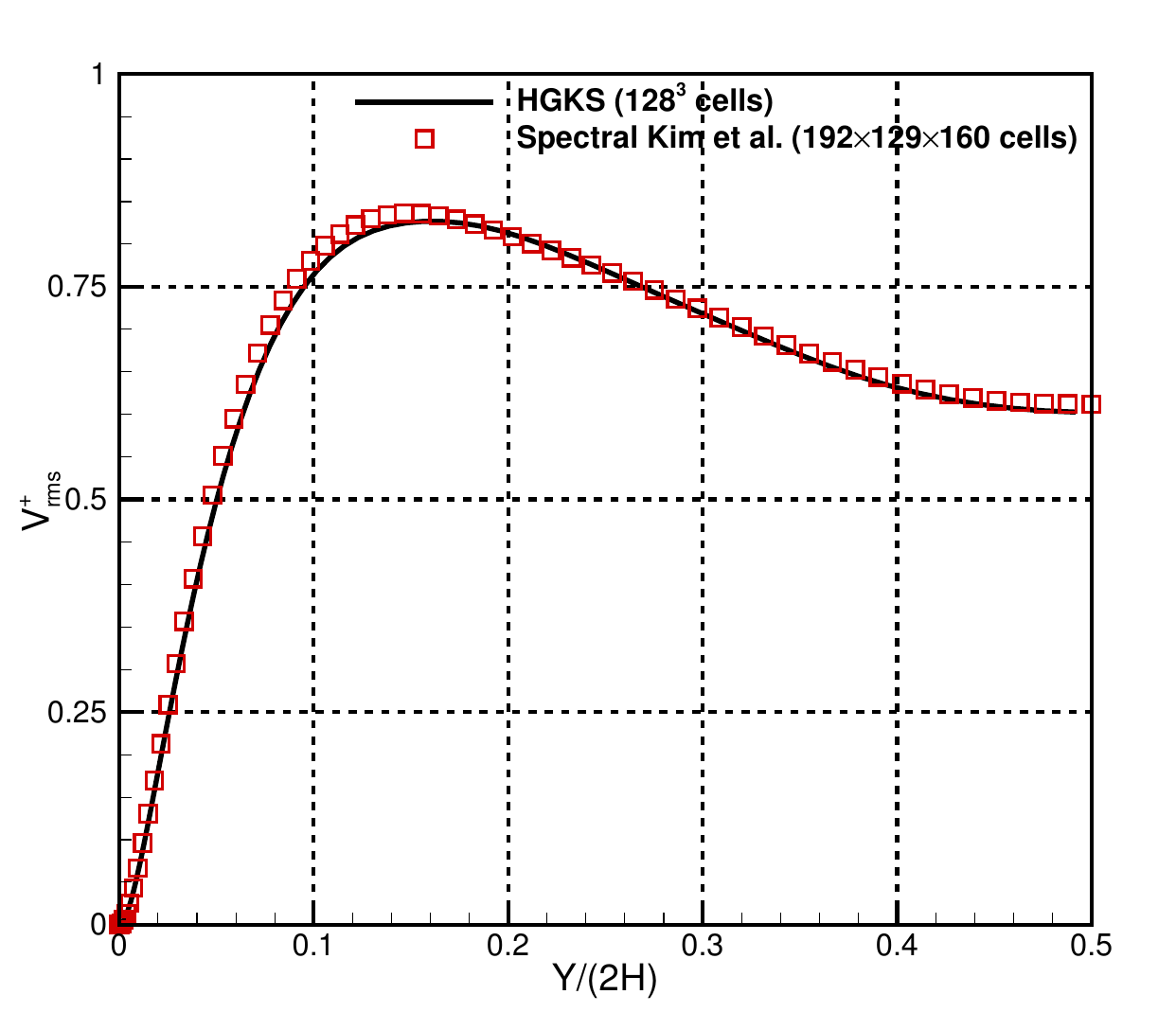}
\includegraphics[width=0.475\linewidth]{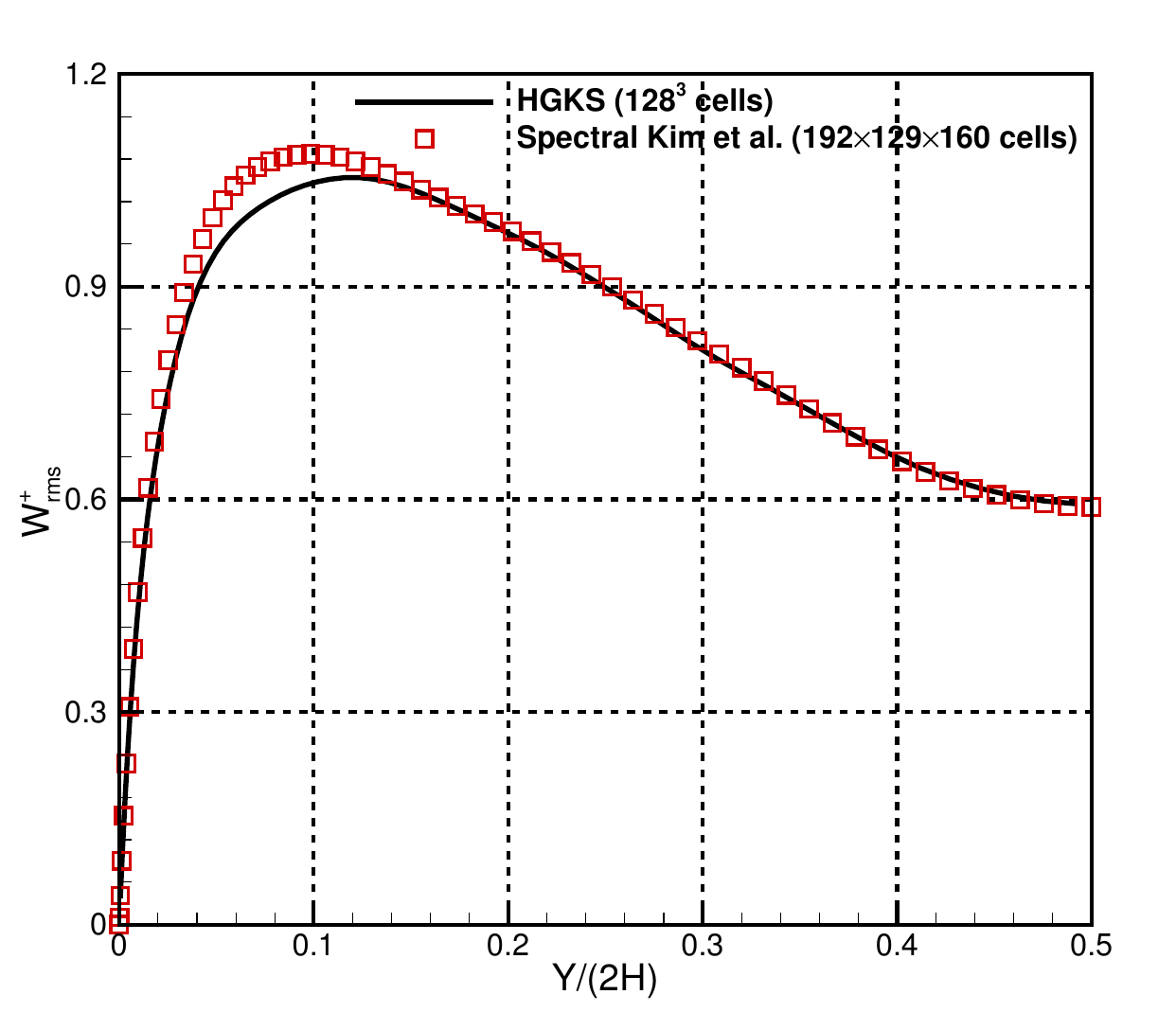}
\includegraphics[width=0.475\linewidth]{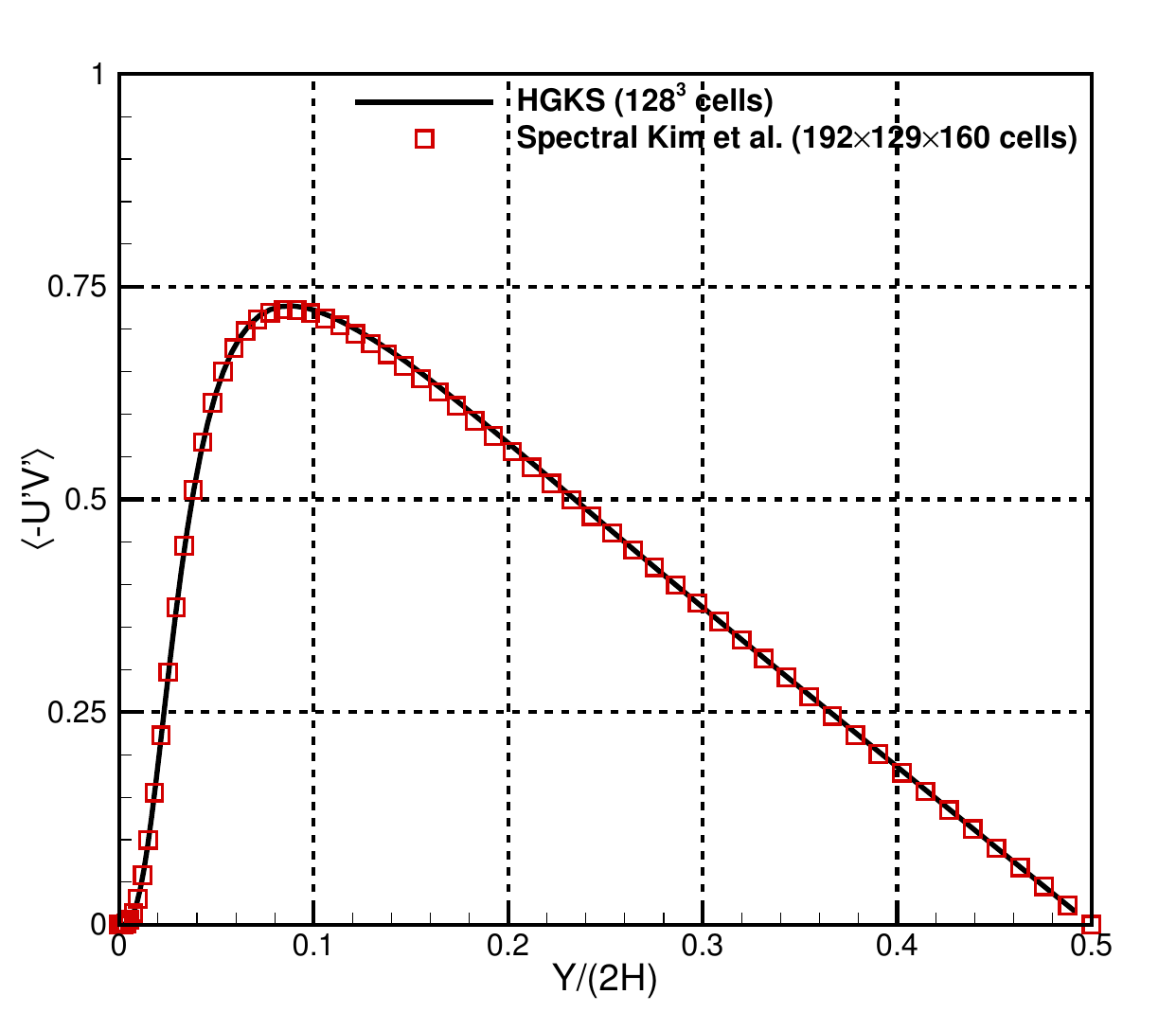}
\caption{\label{channel-flow-2} Turbulent channel flow: the profiles of the root-mean-square fluctuation velocity 
$U_{rms}^+$, $V_{rms}^+$, $W_{rms}^+$ and the Reynolds stress $\langle -U'V' \rangle$.}
\end{figure}

In the current computation, $128\times128\times128$ cells are used, which are uniformly distributed in $(\xi,\eta,\zeta)$ space.  
The flow quantity $\phi$ is decomposed into 
$\phi =\langle \phi \rangle+ \phi^{'}$, where $\langle \cdot \rangle$ represents the mean results over the spanwise direction in space and the statistical periods in time and $\phi^{'}$ is the fluctuation. 
The mean normalized streamwise velocity $\langle U^{+} \rangle$ versus the normalized height $Y^+$ is shown in Fig.\ref{channel-flow-1}, 
where the high-order GKS solution is in good agreement with the reference solution \cite{channel-dns-1}. 
The reference solution is obtained by the spectral method with $192 \times 129 \times 160$ cells. 
The normalized root-mean-square fluctuation velocity profiles $U_{rms}^+$, $V_{rms}^+$, $W_{rms}^+$ 
and normalized Reynolds stress profile $\langle -U'V'\rangle$ are shown in Fig.\ref{channel-flow-2}.
The root-mean-square is defined as $\phi^{+}_{rms}=\langle \sqrt{(\phi-\langle \phi \rangle)^{2}} \rangle$.
High-order GKS solutions are also in good agreements with the reference solution in \cite{channel-dns-1}. 
The high-order GKS provides a reliable tool for the simulation of weakly compressible turbulent flows.

\begin{figure}[!h]
\centering
\includegraphics[width=0.475\linewidth]{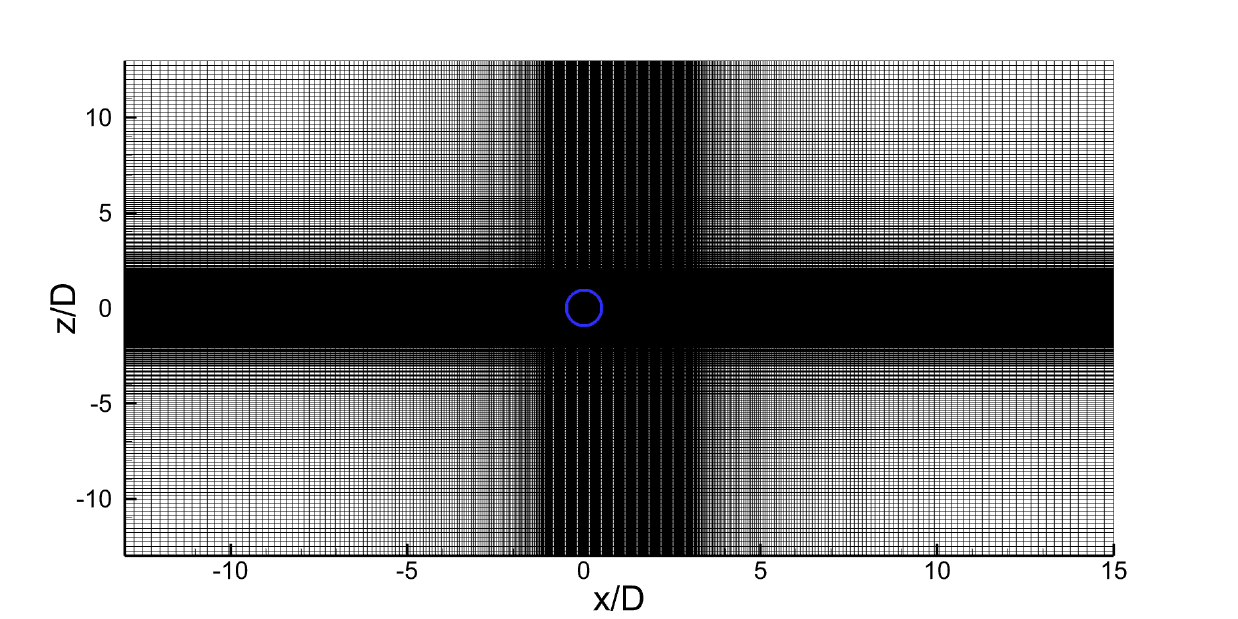}
\includegraphics[width=0.475\linewidth]{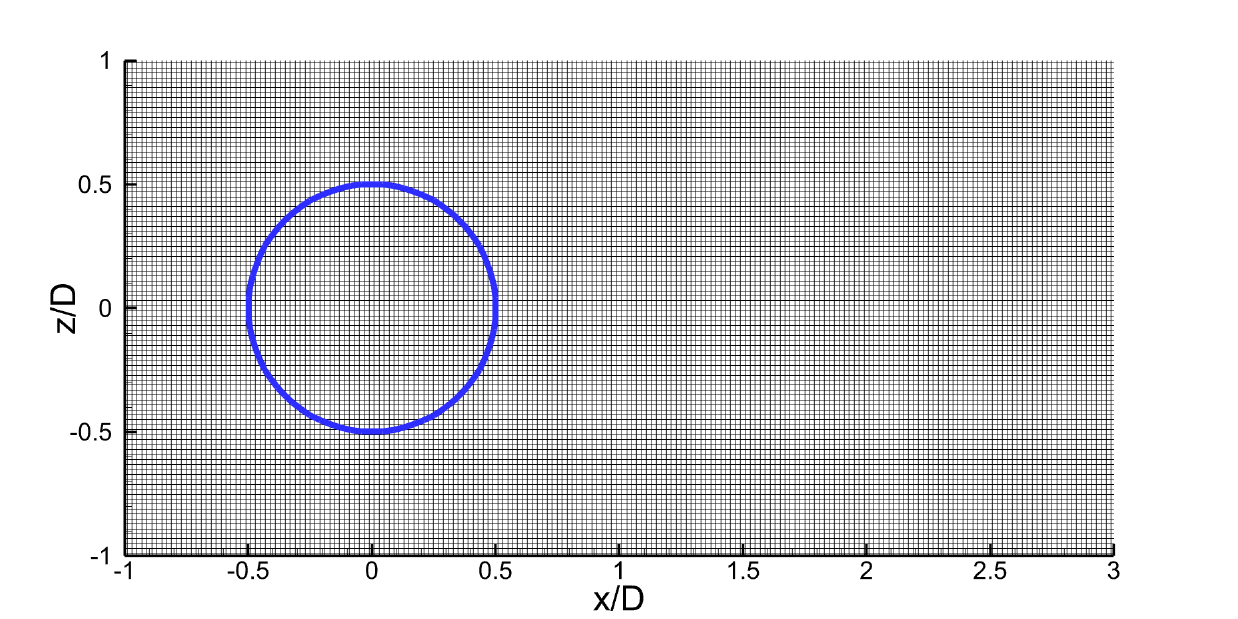}
\caption{\label{cylinder3900-mesh} Turbulent flow past a circular cylinder: the mesh distribution in $x$-$z$ plane (left) and the locally 
refined distribution (right) with Lagrangian points representing the cylinder (blue circle).}
\centering
\includegraphics[width=0.65\linewidth]{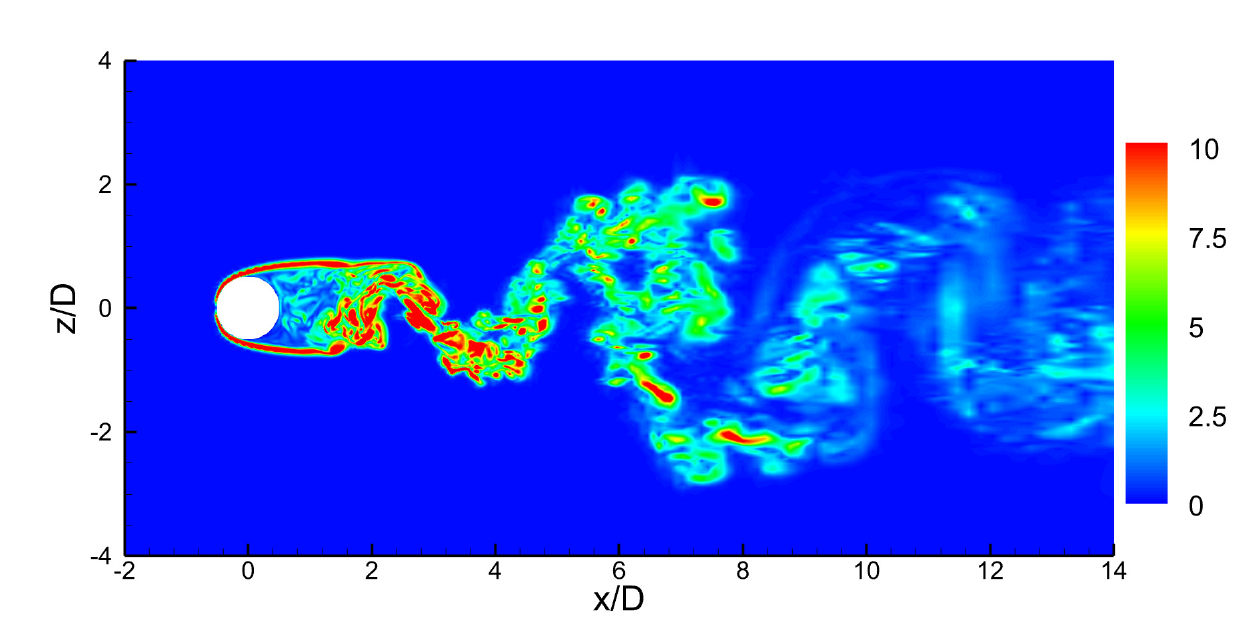}
\caption{\label{cylinder3900-vorticity} Turbulent flow past a circular cylinder: the instantaneous vorticity magnitude in $x$-$z$ plane with $y=\pi D/2$.} 
\end{figure}

\begin{figure}[!h]
\centering
\includegraphics[width=0.475\linewidth]{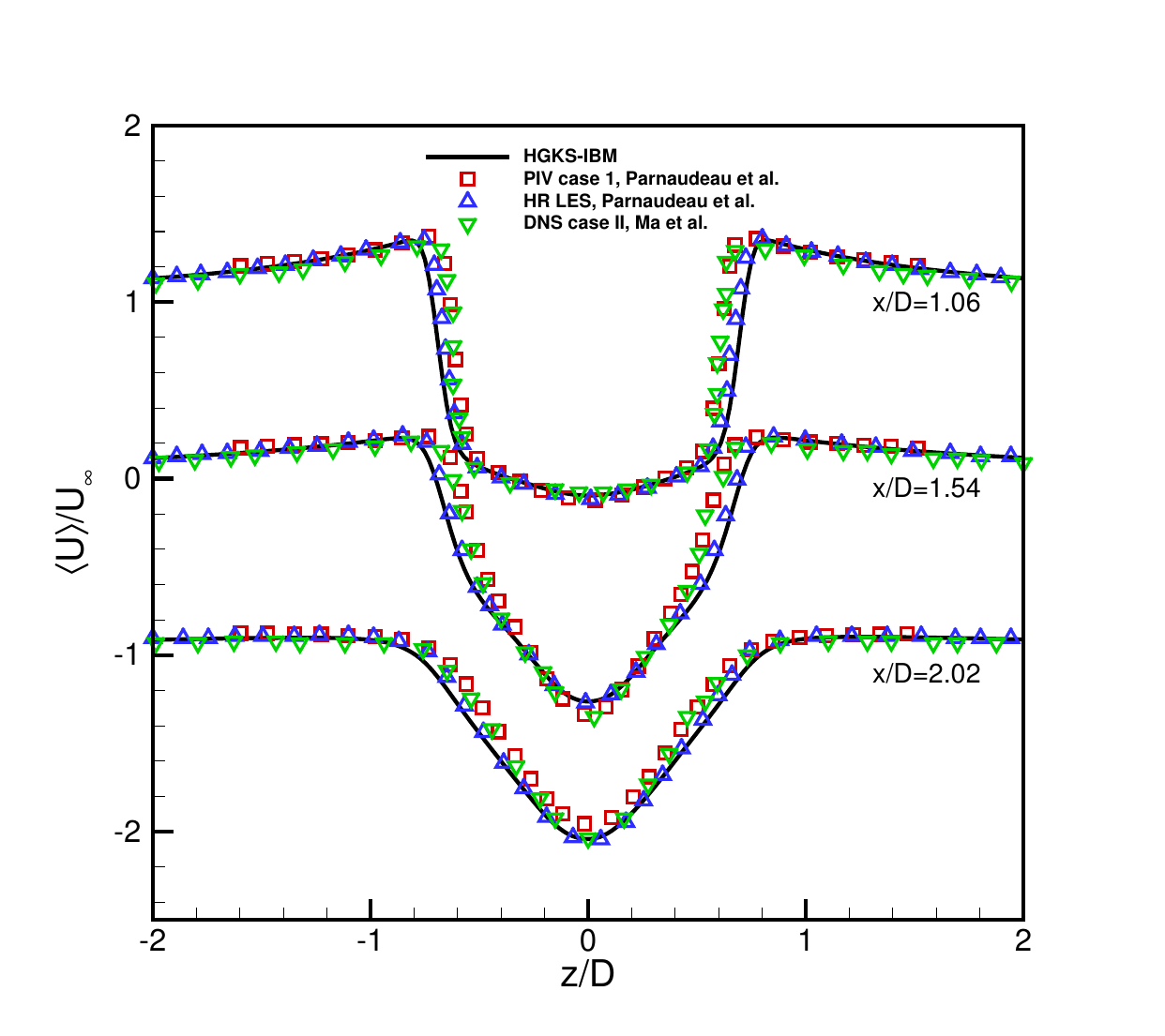}
\includegraphics[width=0.475\linewidth]{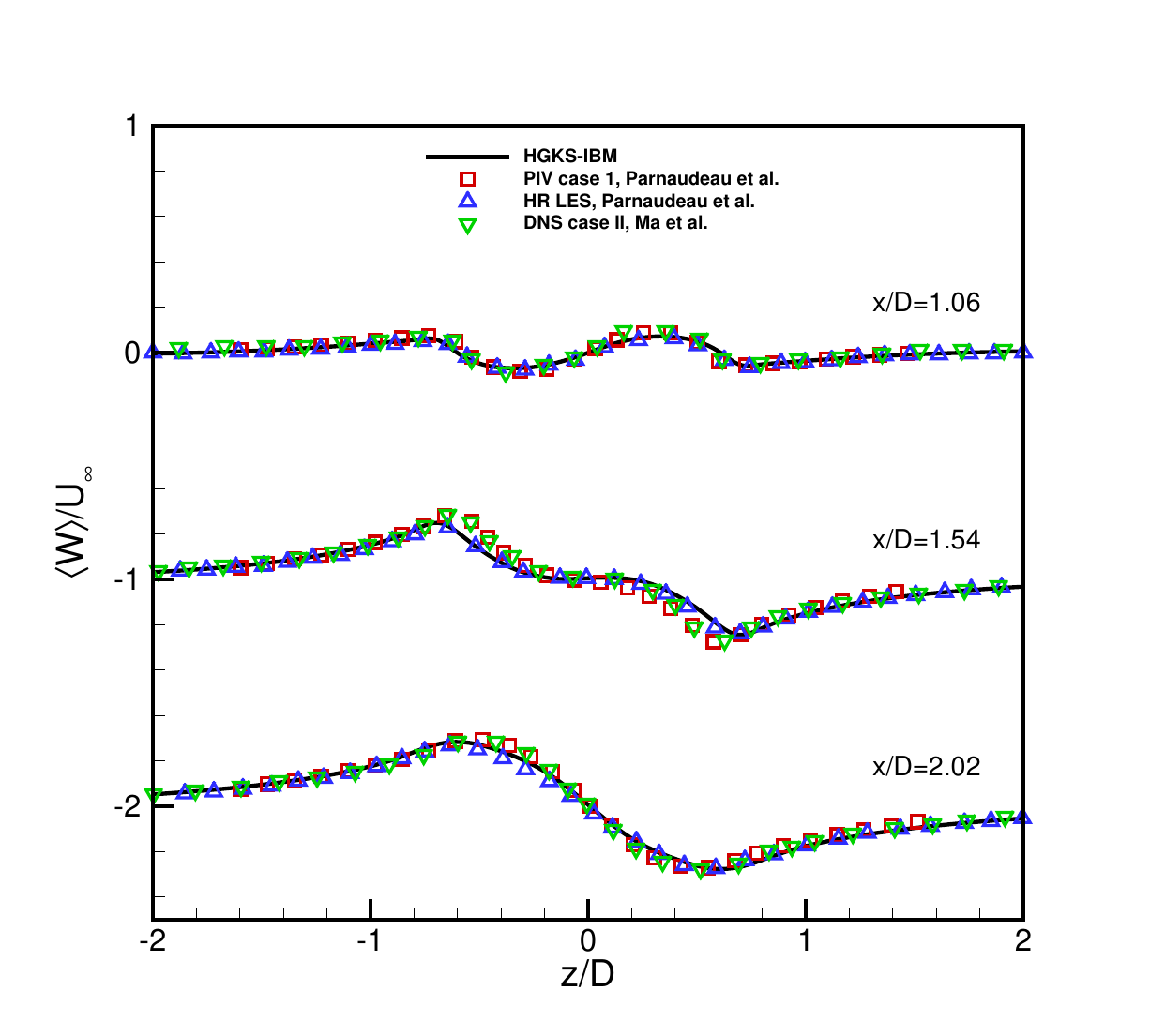}
\caption{\label{cylinder3900-statistics-1} Turbulent flow past a circular cylinder: the mean profiles of streamwise velocity and 
vertical velocity at $x/D=1.06$, $1.54$ and $2.02$. The ordinates have been shifted downward for cases $x/D=1.54$ and $x/D=2.02$.}
\end{figure}

\subsection{Turbulent flow past circular cylinder}
To validate the performance of high-order GKS and immersed boundary method for weakly compressible turbulent flows, 
the turbulent flow past a circular cylinder is tested \cite{cylinder-1,cylinder-2,cylinder-3} .
For the case, the Reynolds number $Re_{D}=\rho U_{\infty}D/\mu=3900$ and Mach number $Ma=0.1$,  
where $U_{\infty}$ is the free streamwise velocity and $D$ is the cylinder diameter. 
The computational domain takes $(x,y,z) \in [-13D,15D]\times[0,\pi D]\times[-13D,13D]$, where the $x$-,
$y$-, $z$-directions represent the streamwise, transverse and
vertical directions, and the velocity components are denoted as $U$, $V$, and $W$, respectively. The cylinder center is located at $(x,z)=(0,0)$, and the mesh with $460\times160\times360$ cells is used.
The grid size is uniform in the spanwise $y$-direction, and the local refined mesh in $x$-$z$ plane is shown in Fig.\ref{cylinder3900-mesh}. 
$57826$ Lagrangian points are used to represent the cylinder surface and  about $240$ Lagrangian points are used to represent the circumference, averagely. 
The inlet boundary and outlet boundary are implemented based on the Riemann invariant. 
An extrapolation boundary condition is used in the vertical $z$-direction. Periodic boundary conditions are imposed
on the boundaries in the spanwise $y$-direction.
The simulation time duration is $T=400 D / U_{\infty}$ and the statistical results are obtained over $T=150$ - $400 D / U_{\infty}$, which is consistent with \cite{cylinder-1}. 

\begin{figure}[!h]
\centering
\includegraphics[width=0.475\linewidth]{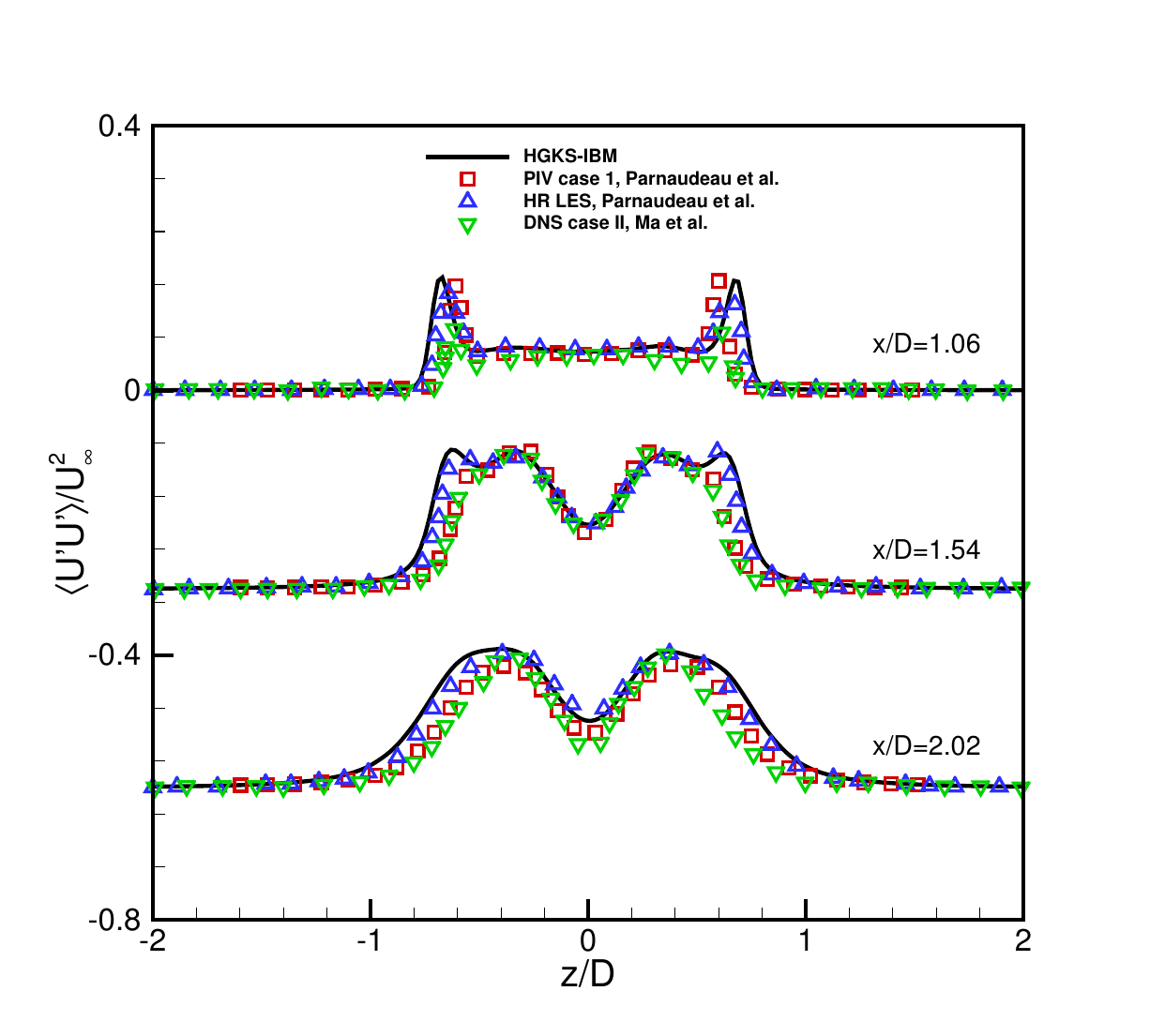}
\includegraphics[width=0.475\linewidth]{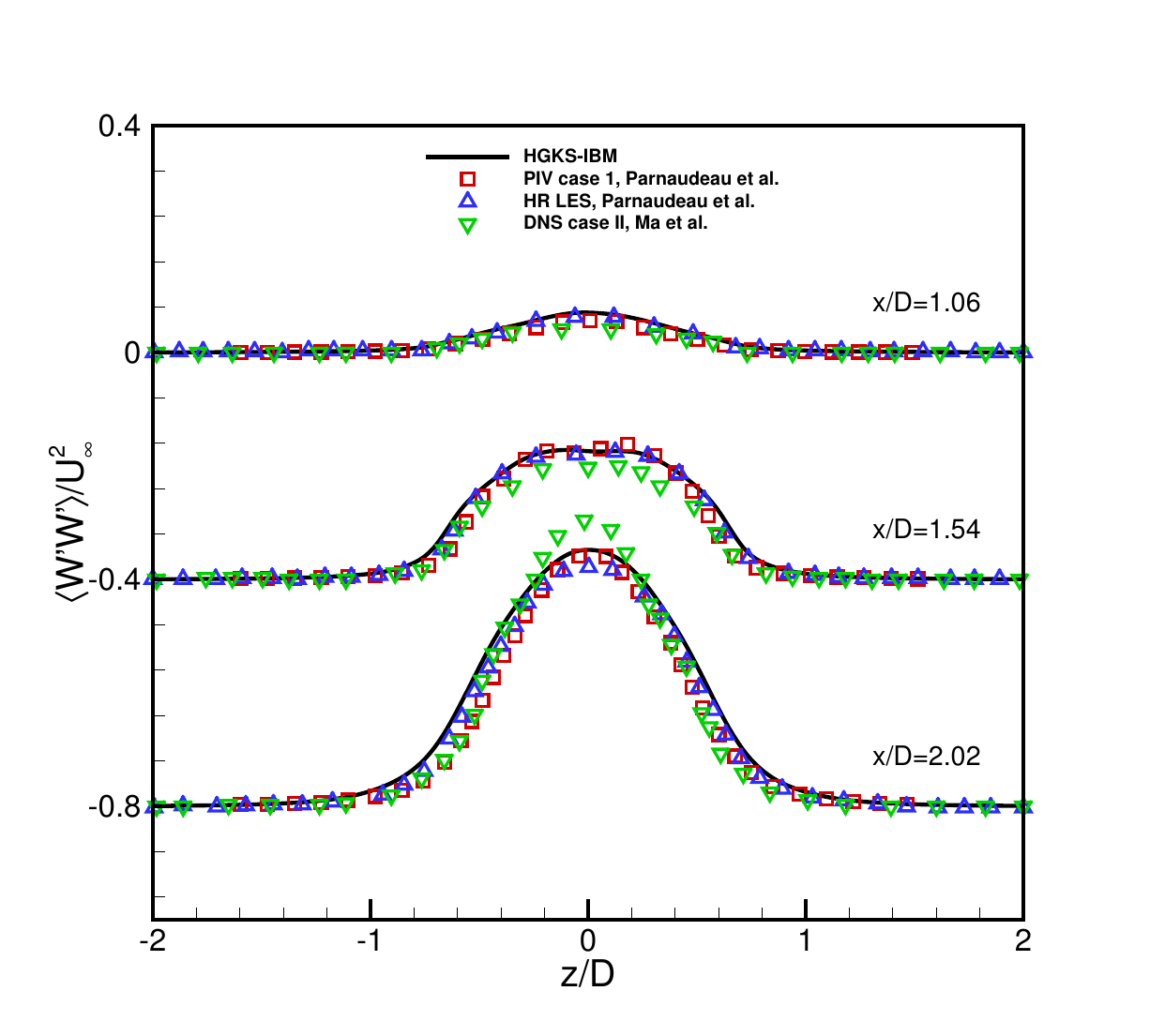}
\includegraphics[width=0.475\linewidth]{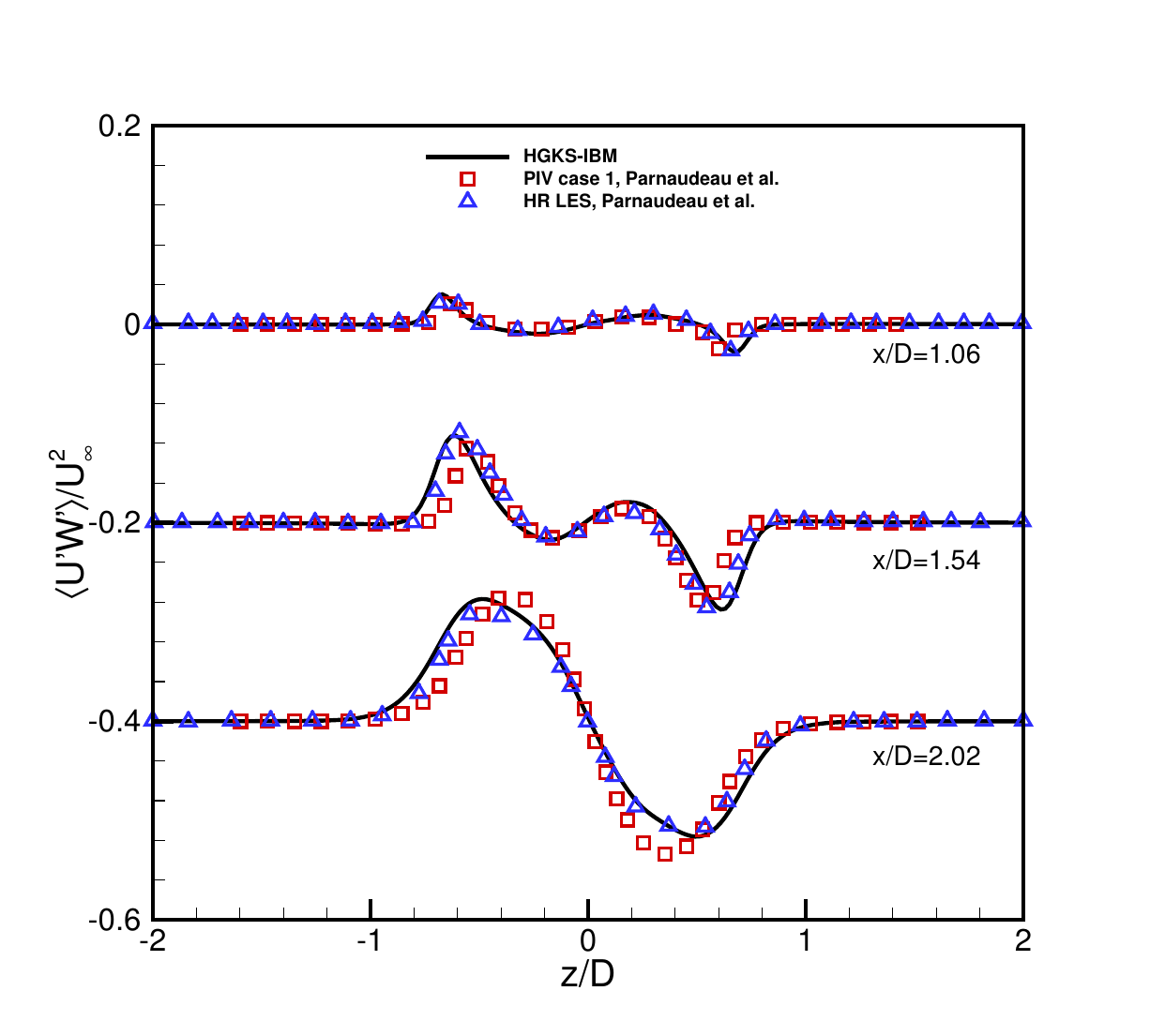} 
\caption{\label{cylinder3900-statistics-2} Turbulent flow past a circular cylinder:  
the mean profiles of streamwise, transverse and shear Reynolds stress at $x/D=1.06$, $1.54$ and $2.02$. 
The ordinates have been shifted downward for cases $x/D=1.54$ and $x/D=2.02$. The $\langle U'W' \rangle /U^{2}_{\infty}$ data is not provided in DNS study \cite{cylinder-2}.}
\end{figure}

\begin{figure}[!h]
\centering 
\includegraphics[width=0.475\linewidth]{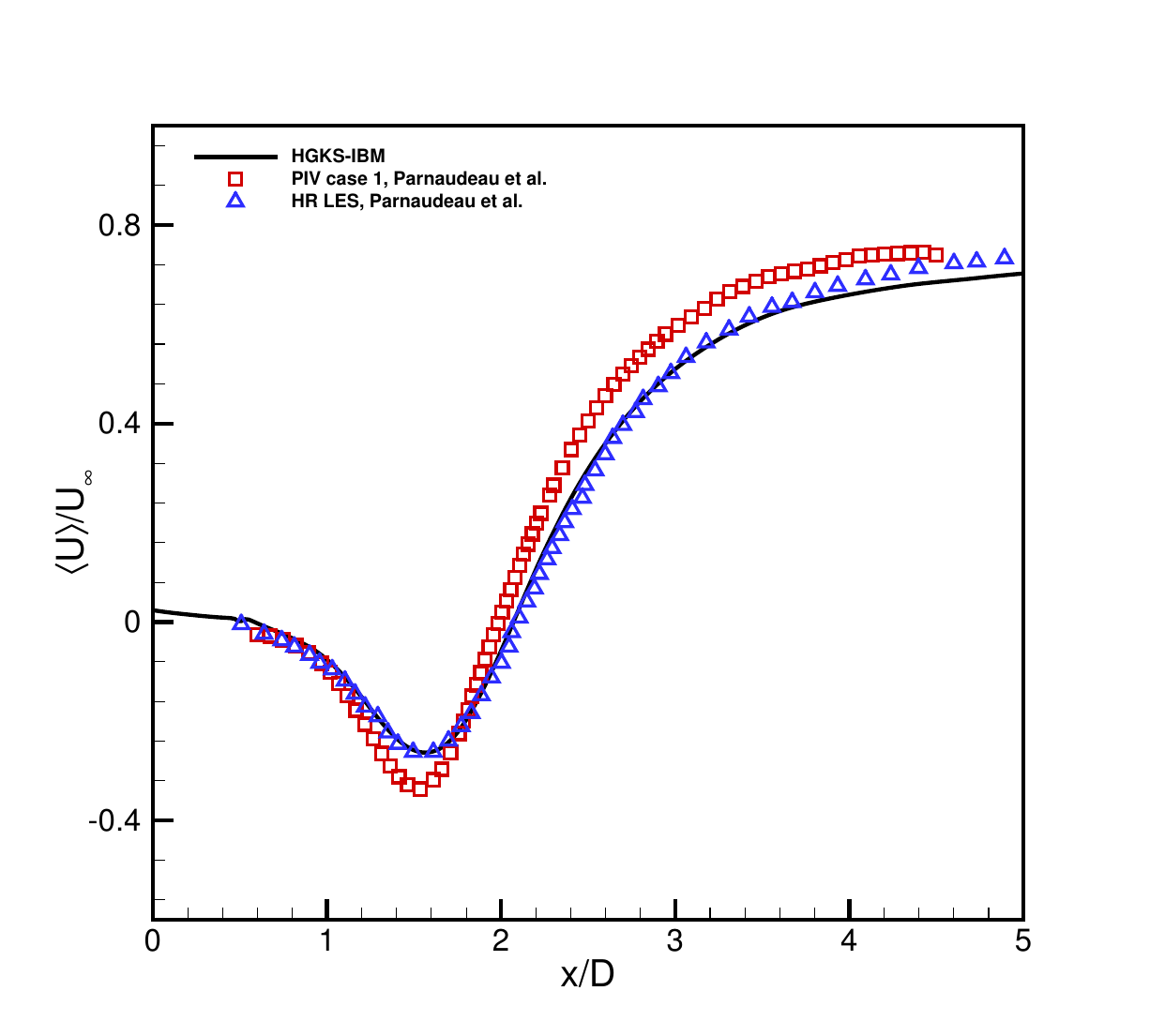}
\caption{\label{cylinder3900-statistics-3} Turbulent flow past a circular cylinder:  
the mean profile of streamwise velocity along the wake centerline.}
\end{figure}

The vorticity magnitude in $x$-$z$ plane with $y=\pi D/2$ is shown in Fig.\ref{cylinder3900-vorticity}, 
and the high-order GKS and immersed boundary method well resolves the large vortex shedding behind the cylinder.  
The mean streamwise velocity $\langle U \rangle /U_{\infty}$ and vertical velocity $\langle W \rangle /U_{\infty}$ profiles are given in Fig.\ref{cylinder3900-statistics-1}, 
and the mean streamwise, transverse and shear Reynolds stress $\langle U'U'\rangle$, $\langle W'W'\rangle$ and $\langle U'W'\rangle$ profiles 
are given in Fig.\ref{cylinder3900-statistics-2} at $x/D=1.06$, $1.54$ and $2.02$.
For streamwise velocity, a strong velocity deficit occurs in the recirculation region with  $U$-shape close to
the cylinder which evolves toward $V$-shape further downstream. 
Meanwhile, for vertical velocity, the anti-symmetry with respect to $z=0$ plane is obtained.
The profile of mean streamwise velocity along the wake centerline is shown in Fig.\ref{cylinder3900-statistics-3}.
The streamwise velocity is zero at $x=0.5D$, as the non-slip boundary condition is satisfied. 
The minimum value is reached at around $x=1.6D$ in the recirculation zone and then the cylinder wake begins to recover. 
The solutions of high-order GKS and  immersed boundary method are in good agreement with the reference solutions, 
where the PIV experimental data \cite{cylinder-1}, large eddy simulation (LES) \cite{cylinder-1} and direct numerical simulation (DNS) solutions \cite{cylinder-2} are given as reference.

\begin{figure}[!h]
\centering
\includegraphics[trim=80 100 100 100, clip = true, width=1.0\linewidth]{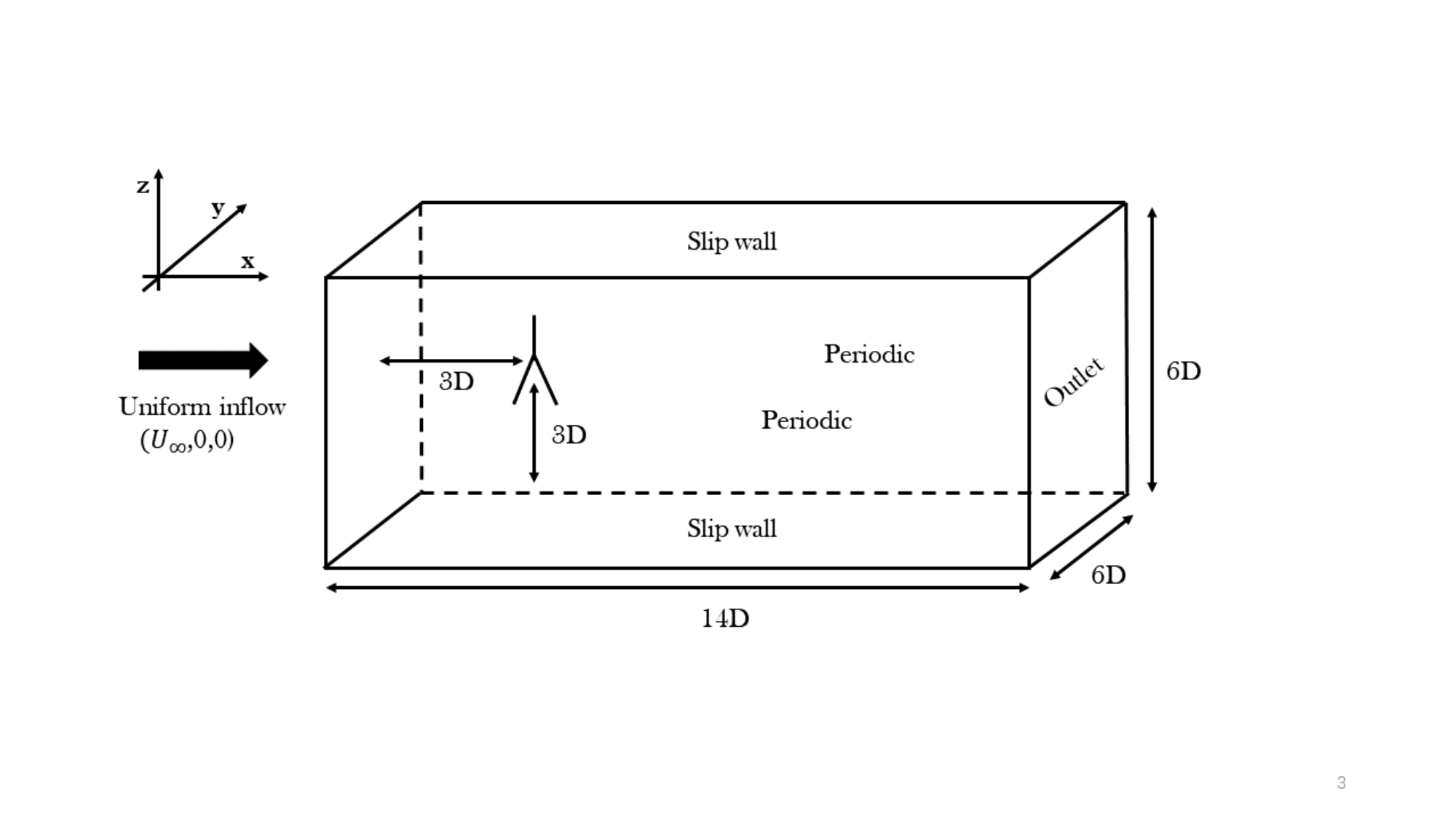}
\caption{\label{Schematic-2} NREL $5$MW wind turbine wake: schematic of the computation domain and boundary conditions.}
\end{figure}

\subsection{NREL $5$MW wind turbine wake}
The NREL $5$ MW reference wind turbine has three blades and a rotor diameter of $D=126$ meters, and
the detailed airfoil data of the NREL $5$ MW reference wind turbine is provided by \cite{NREL-turbine}. 
The computational domain and boundary conditions are shown in Fig.\ref{Schematic-2} for the simulation of the NREL $5$ MW reference wind turbine. 
The computational domain $(x,y,z)\in[-3D,11D]\times[-3D,3D]\times[-3D,3D]$, where the $x$-,
$y$-, $z$-directions represent the streamwise, transverse and
vertical directions, and the velocity components are denoted as $U$, $V$, and $W$, respectively. The center of the wind turbine is
located at $(0D, 0D, 0D)$. The computational domain length upstream
of the wind turbine is $3D$, and the downstream length is $11D$.
In the current ALM, $100$ actuator points are evenly distributed to
represent each blade. No tower or nacelle models are implemented.

\begin{figure}[!h]
\centering
\includegraphics[width=0.475\linewidth]{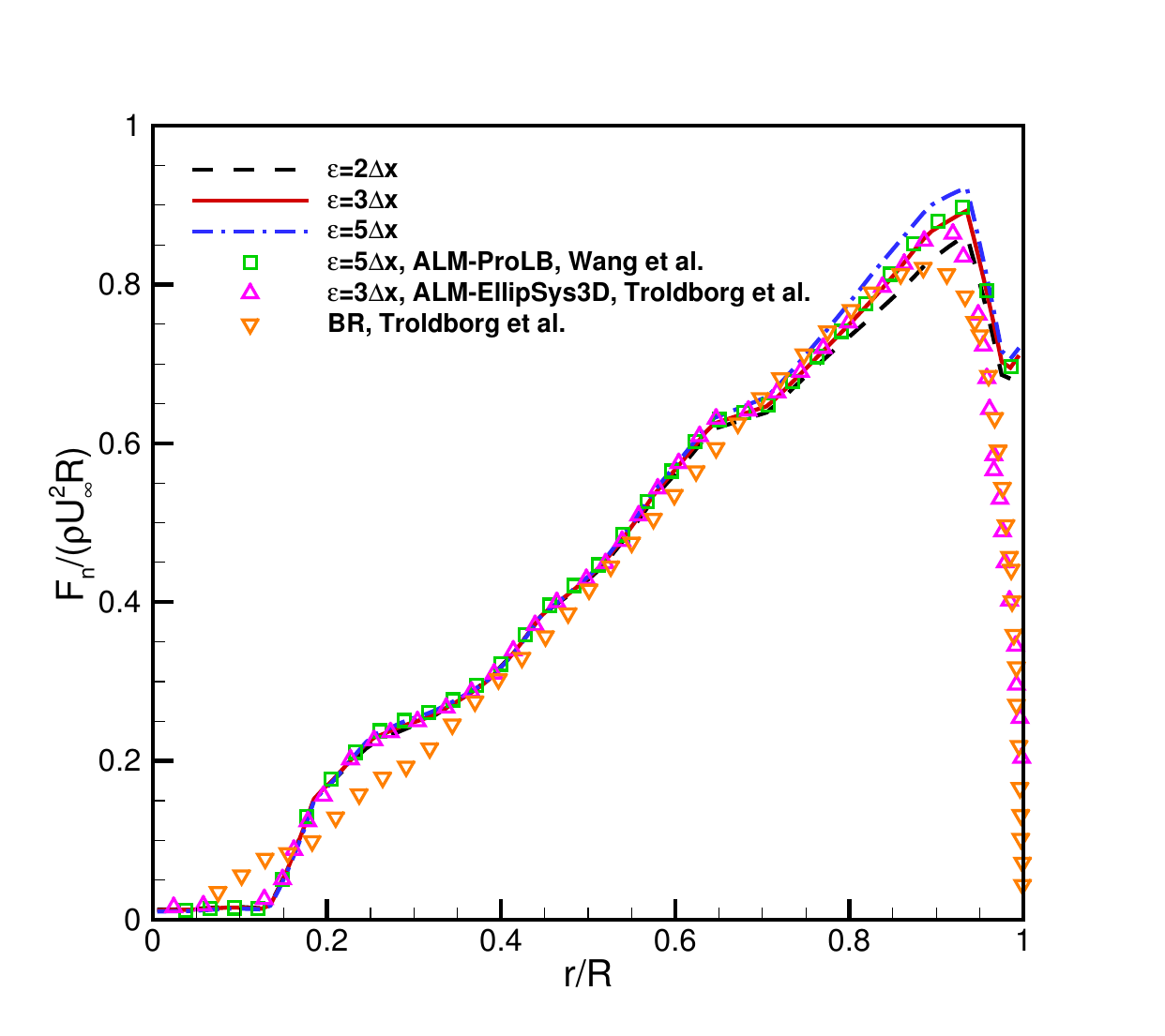}
\includegraphics[width=0.475\linewidth]{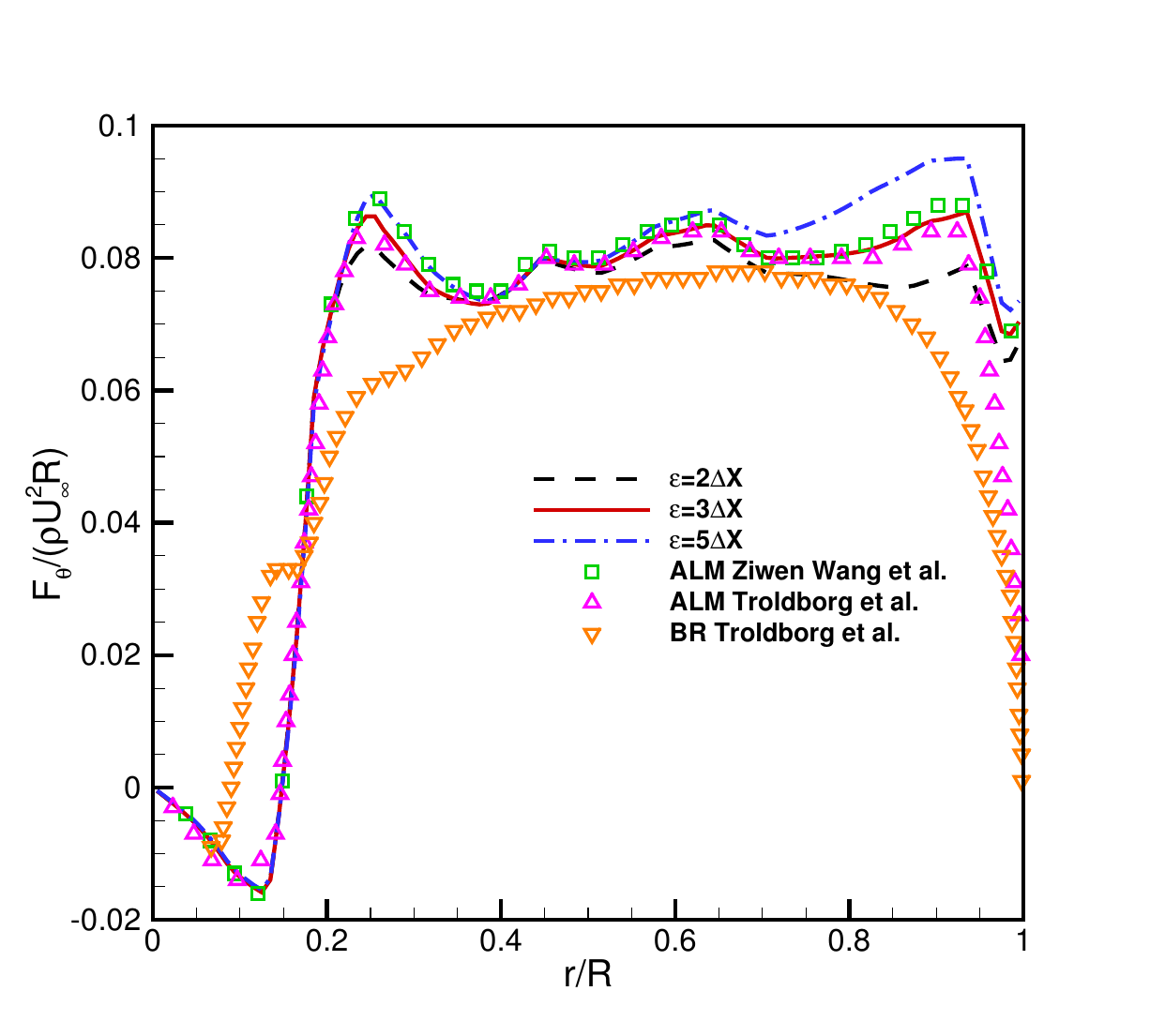}
\caption{\label{along-blade-eps-2} NREL $5$MW wind turbine wake:  
the time-averaged normal (left) and tangential (right) forces along turbine blade with different  $\varepsilon$.}
\includegraphics[width=0.475\linewidth]{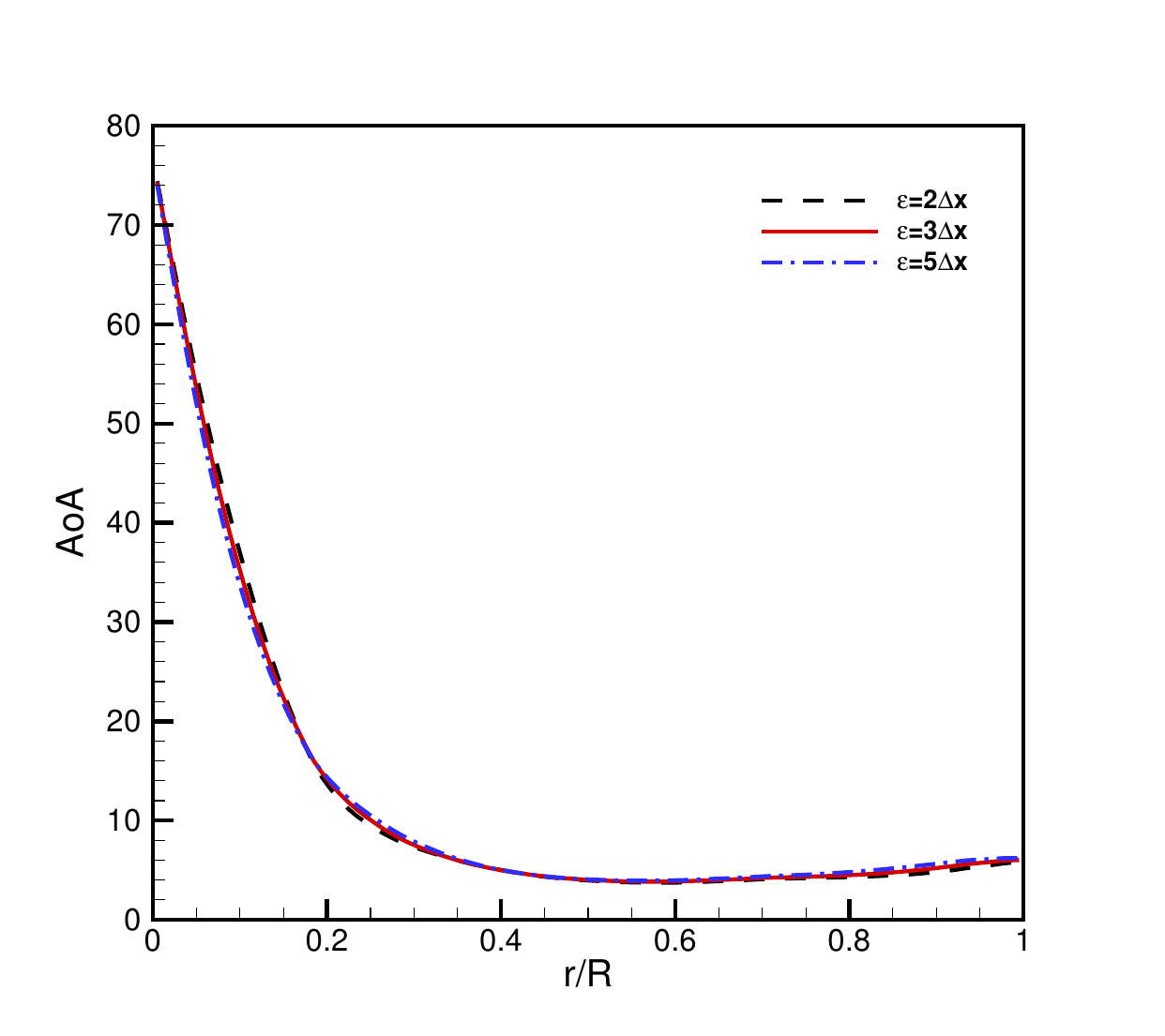}
\includegraphics[width=0.475\linewidth]{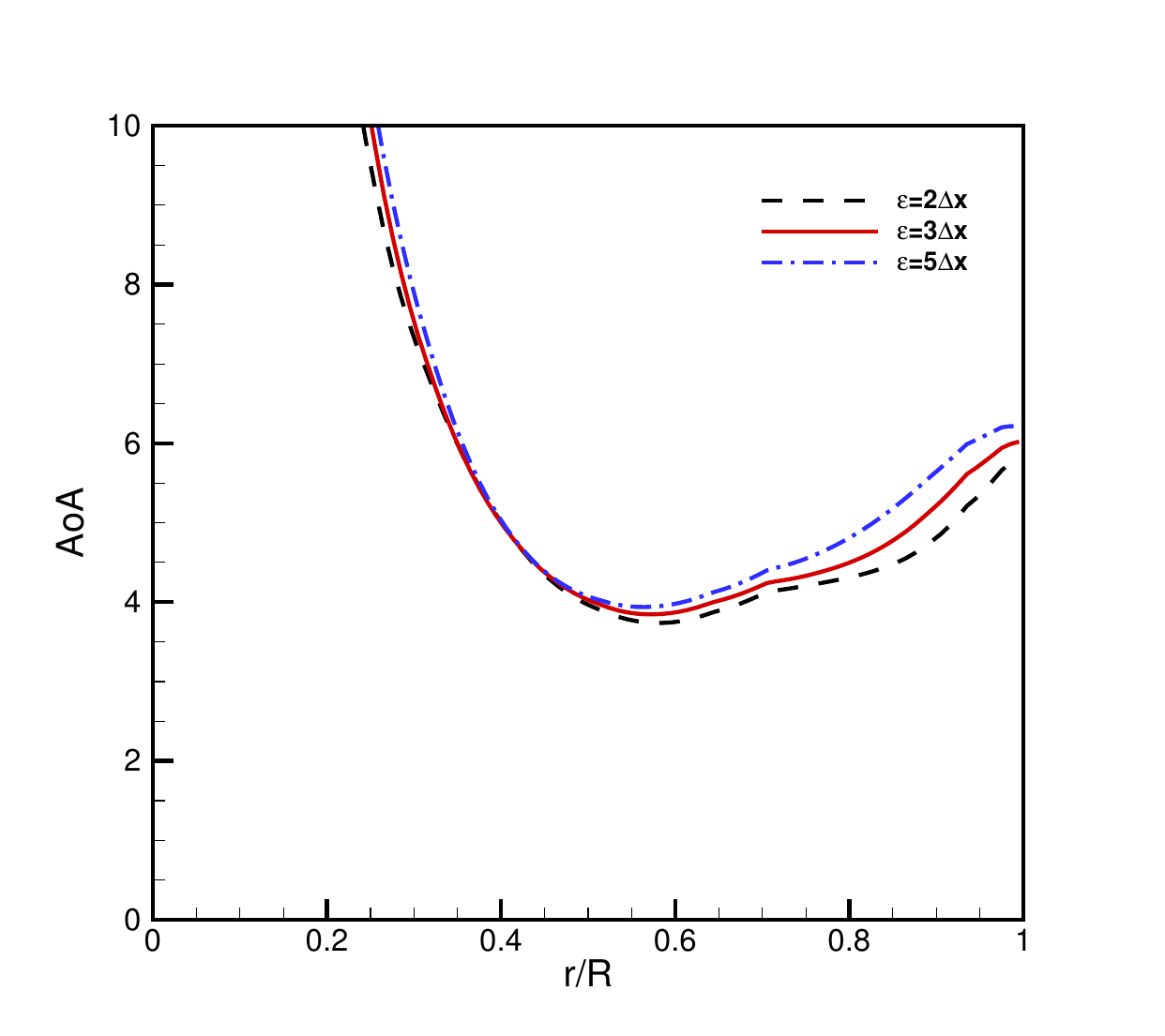}
\caption{\label{along-blade-eps-1} NREL $5$MW wind turbine wake: 
the time-averaged angle of attack along the turbine blade with different $\varepsilon$.}
\end{figure}

The rotational speed of the rotor is $9.155$ revolutions per minute
at the free-stream wind speed $U_{inlet} = 8$ m/s, giving a tip
speed ratio of $TSR=7.55$. The Mach number $Ma=0.15$ and the
Reynolds number $Re=\rho_{\infty} U_{\infty} R/\mu\approx 3.38\times
10^7$, with the characteristic density $\rho_{\infty} = 1.2$
kg/m$^{3}$, the radius $R = D/2$ and viscosity coefficient $\mu
=1.7891\times10^{-5}$ kg/m/s. 
The computation is non-dimensionalized with the characteristic density $\rho_{\infty}$, 
the radius $R = D/2$ and the free-stream wind speed $U_{inlet}.$ 
In the streamwise $x$-direction, a
uniform and non-turbulent free-stream inflow
$(U_{\infty},V_{\infty},W_{\infty})=(1,0,0)$ is applied in the inlet
boundary. The Riemann invariant boundary
condition is used for the density at the inlet boundary. An
extrapolation boundary condition is used in the outlet boundary.
Periodic boundary conditions are imposed on the boundaries in the
transverse $y$-direction. A slip wall is used on the top and bottom
boundaries in the vertical $z$-direction.
In the computation, a uniform grid with $700\times300\times300$
cells is used, and the cell size is $\Delta x=\Delta y=\Delta
z=D/50$. The characteristic time step takes a constant of $\Delta
t=4\times10^{-3}$.

\begin{figure}[!h]
\centering
\includegraphics[width=0.475\linewidth]{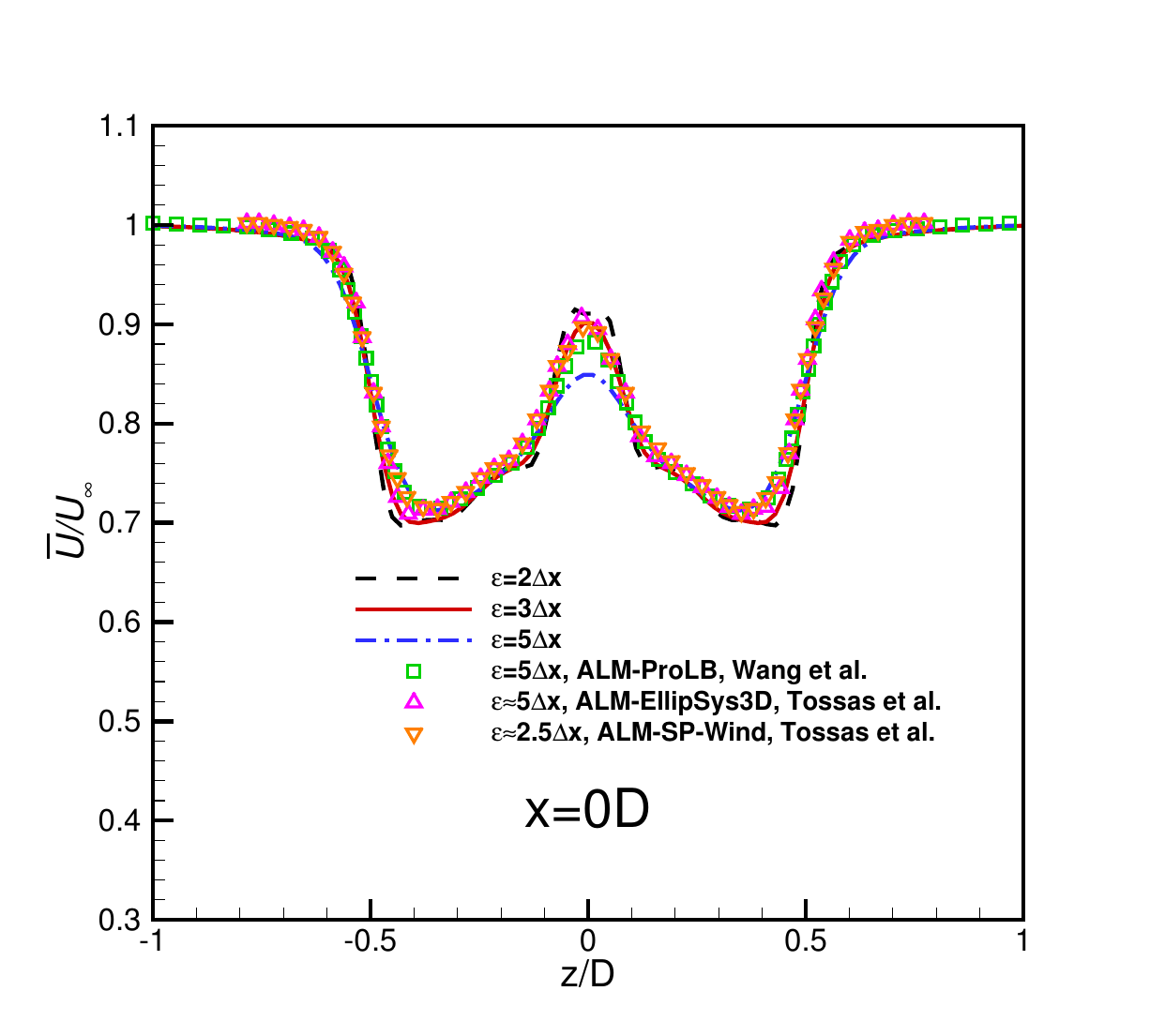}
\includegraphics[width=0.475\linewidth]{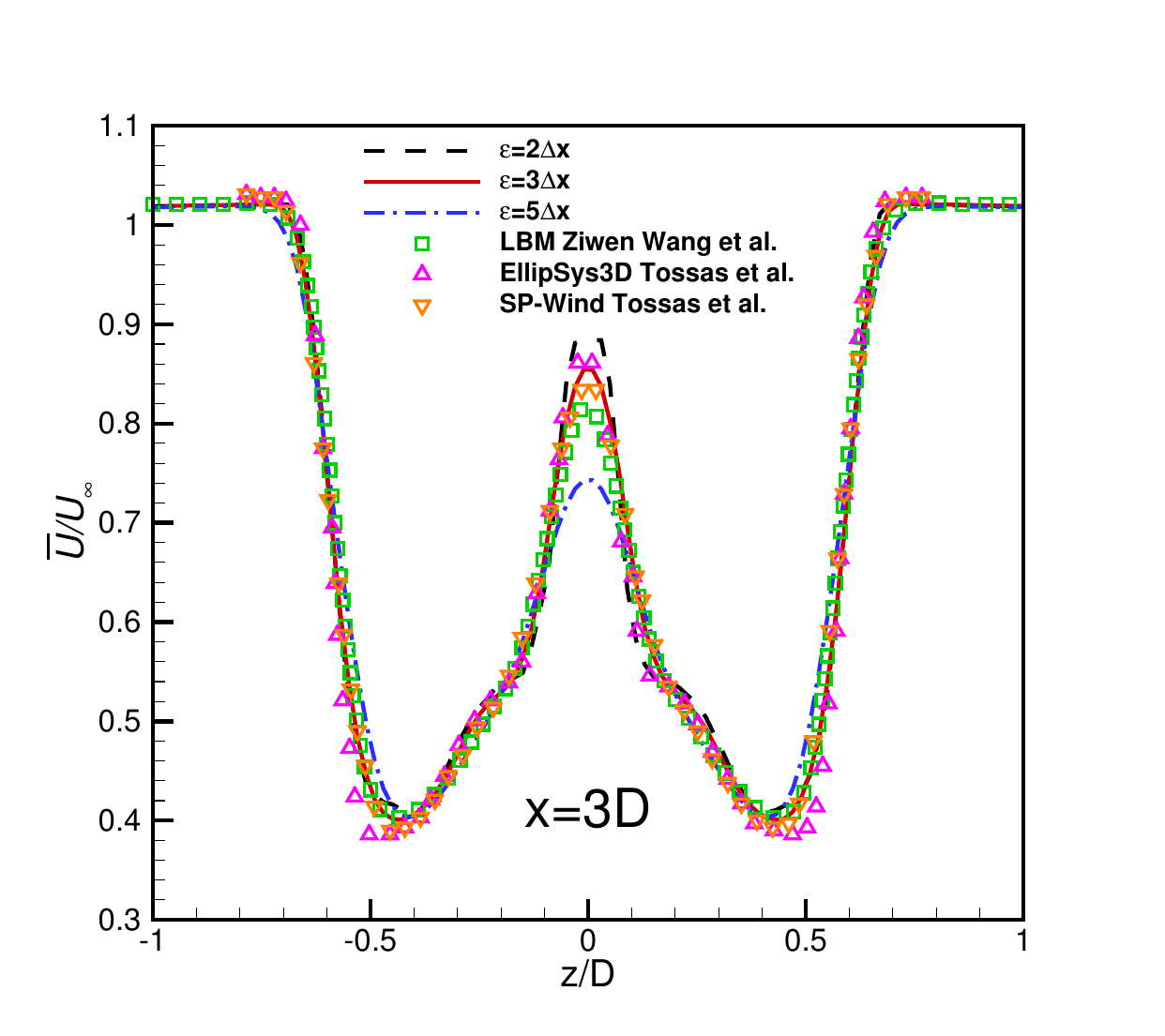}
\includegraphics[width=0.475\linewidth]{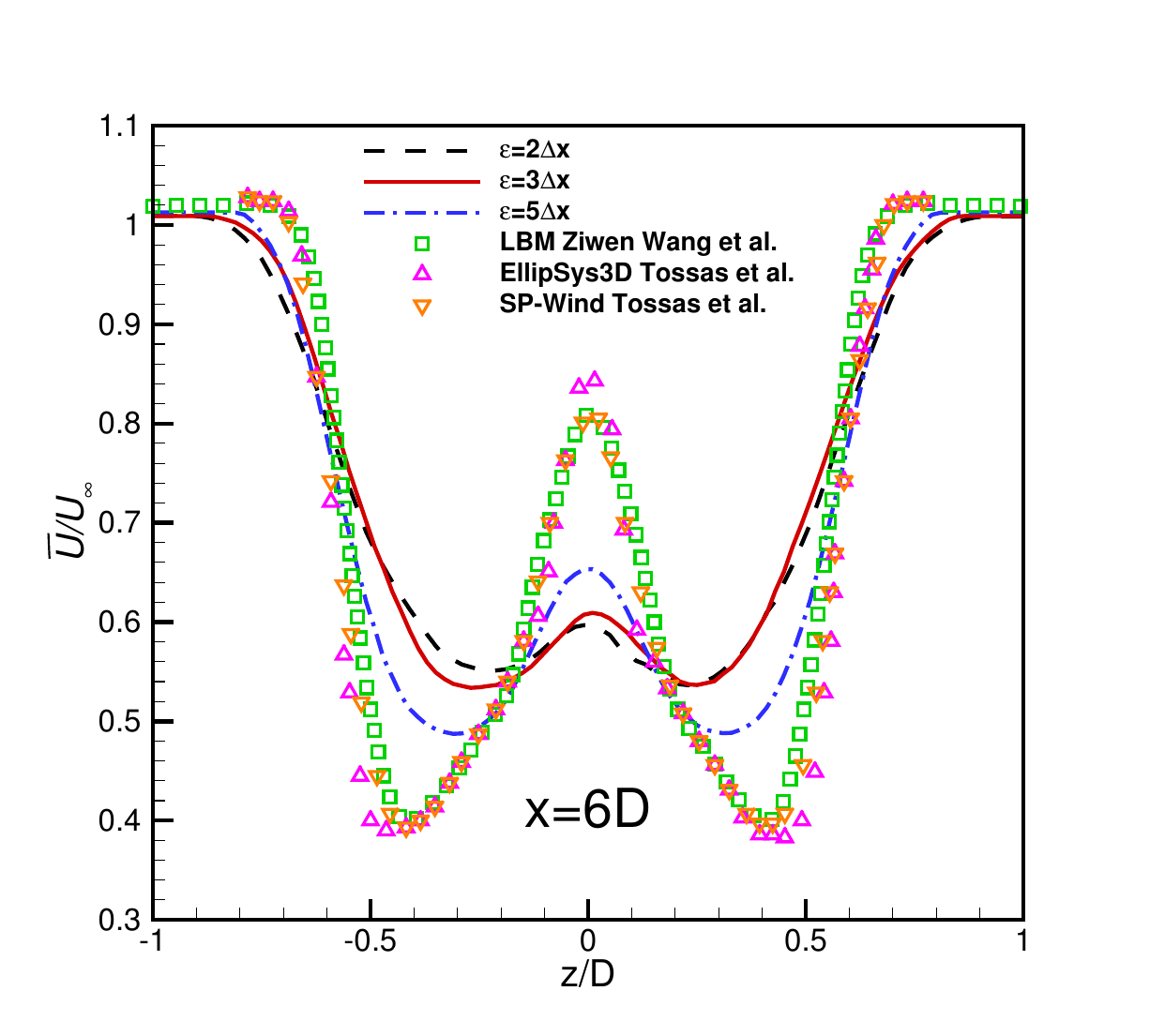}
\includegraphics[width=0.475\linewidth]{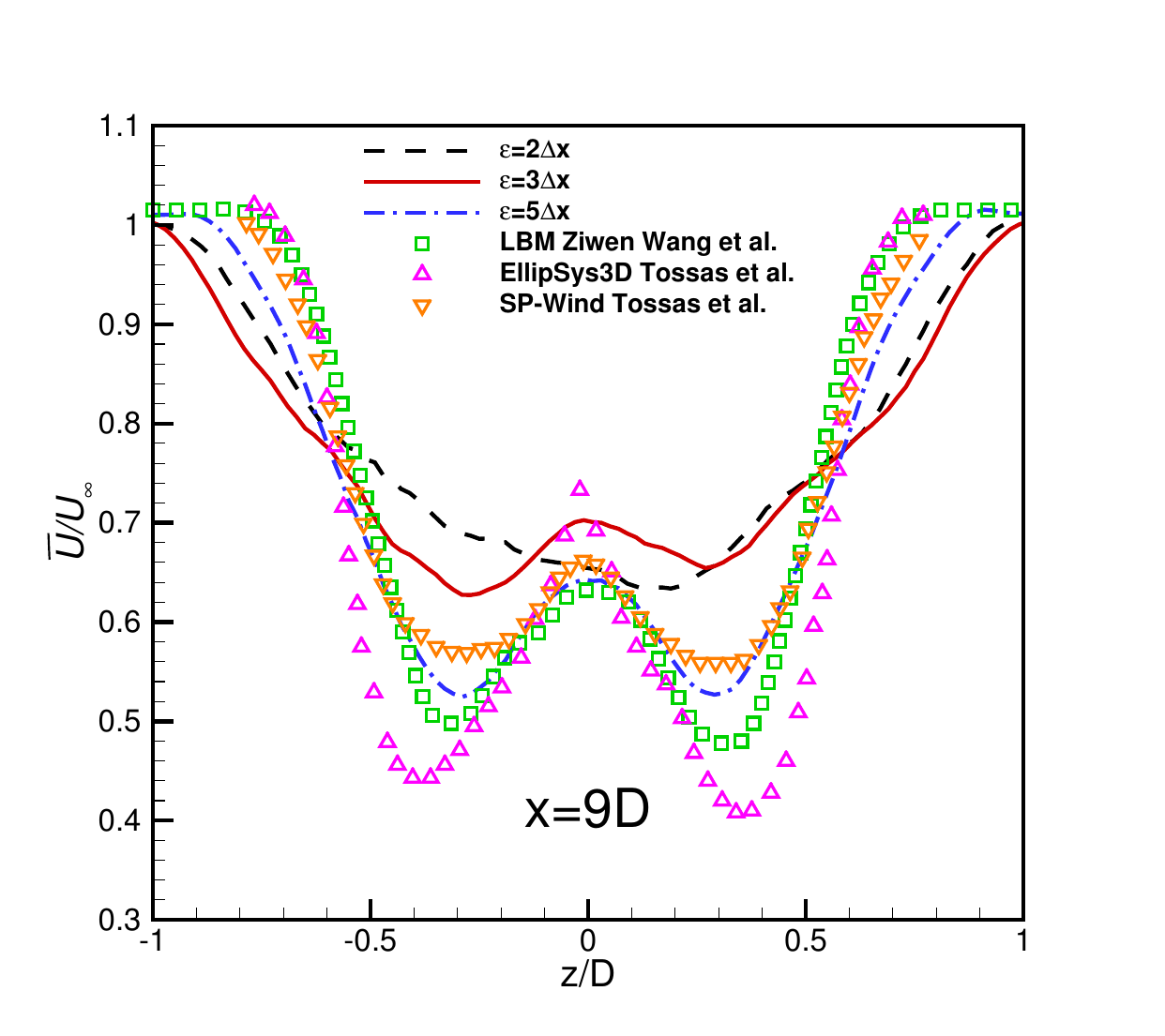}
\caption{\label{velocity-deficit-eps} NREL $5$MW wind turbine wake: the time-averaged streamwise velocity profiles 
at $x=0D, 3D, 6D$ and $9D$ at $y=0$ plane with$\varepsilon=2\Delta x$, $3\Delta x$, $5\Delta x$.}
\end{figure}

\subsubsection{Effect of smearing kernel width}
In the framework of high-order GKS and ALM, the effect of smearing kernel width $\varepsilon$ in Gaussian function in Eq.\eqref{Gaussian} is tested. 
The parameter $\varepsilon$ determines how the point forces are projected onto the surrounding grids as body forces, which plays
an important role in predicting along-blade quantities and affects the wake of wind turbine. 
To avoid the numerical instabilities \cite{ALM-5}, it is suggested that $\varepsilon \ge 2\Delta x$ should be satisfied. 
The value of  Gaussian function decays to $1\%$ of its initial value at a distance of $r = 2.15\varepsilon$. 
To span the force through the blade area, the value of $\varepsilon = c/4.3$ would be required, where $c$ is the chord length \cite{wind-turbine-modeling}. 
To determine the optimal value of $\varepsilon $, the value of $\varepsilon =0.2c$ is suggested \cite{Optimal-eps}.

In the current computation, the effect of smearing kernel width is investigated with $\varepsilon=2\Delta x$, $3\Delta x$ and $5\Delta x$
implemented with high-order GKS, and the Gaussian function is truncated at the distance of $r = 3\varepsilon$.
The reference solutions are given by proLB \cite{ALM-LBM-3},  EllipSys3D \cite{4-les-ALM}, SP-Wind \cite{4-les-ALM} and 
RANS-based blade-resolved (BR) simulation \cite{ALM-4}, where
the grid size $\Delta x=D/63$ with fixed smearing kernel width $\varepsilon=5\Delta x$ is chosen in  ProLB, 
the grid size $\Delta x=D/64$ with fixed smearing kernel width $\varepsilon =5D/63 \approx 5\Delta x$ is chosen in EllipSys3D,
and  the grid size  $\Delta x=\Delta y=D/32$ and $\Delta z=D/64$  with fixed smearing kernel width $\varepsilon =5D/63 \approx 2.5\Delta x$ is chosen in SP-Wind.
The time-averaged normal and tangential forces along the blade  are given in Fig.\ref{along-blade-eps-2}.
The BR simulation gives a relatively smaller normal and tangential forces.  The time-averaged angle of attack
(AoA) along the blade is shown in Fig.\ref{along-blade-eps-1}. In the center region and outer region of the
blade, the increase of $\varepsilon$ gives larger AoA which increases the normal and tangential forces as shown in Fig.\ref{along-blade-eps-2}. 
The time-averaged streamwise velocity profiles $\overline{U}/U_{\infty}$ are shown in Fig.\ref{velocity-deficit-eps}  at $x=0D, 3D, 6D$ and $9D$ behind the turbine, where $\overline{(\cdot)}$ represents the time-averaged quantities. 
In the near wake region ($x\in [0D, 3D]$), the flow is laminar and the streamwise velocity profiles are very similar with different $\varepsilon$. 
This region of the flow is governed by the inviscid equations, and turbulence is not triggered until a later stage in the wake. 
However, in the far wake region ($x\in [6D,9D]$), the profiles vary significantly due to different locations where the wake becomes turbulent. 
With smaller $\varepsilon$, the turbulence is triggered earlier and the streamwise velocity of the wake recovers faster. 
For the ALM with high-order GKS, $\varepsilon =3\Delta x$ is adopted, whose solutions agree well with the reference ALM solutions.

\begin{figure}[!h]
\centering
\includegraphics[width=0.6\linewidth]{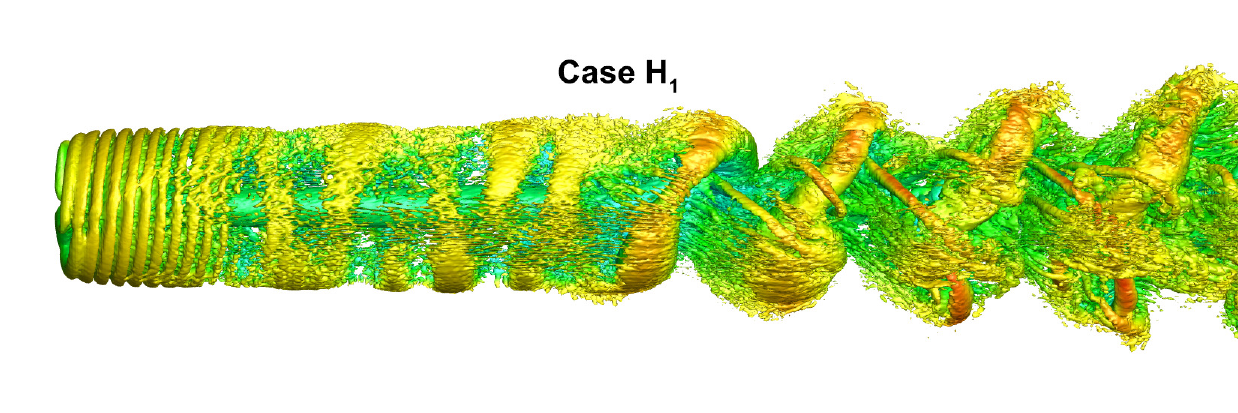}  
\includegraphics[width=0.6\linewidth]{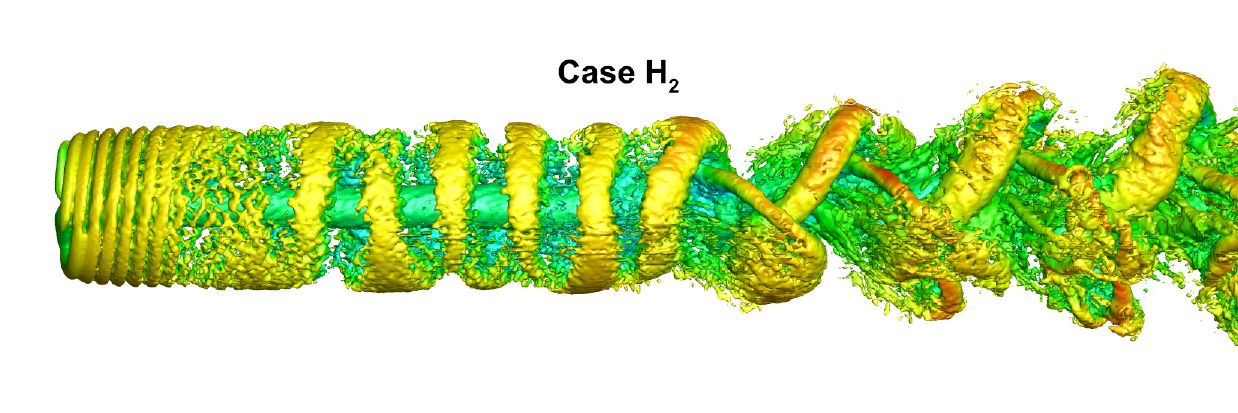} 
\includegraphics[width=0.6\linewidth]{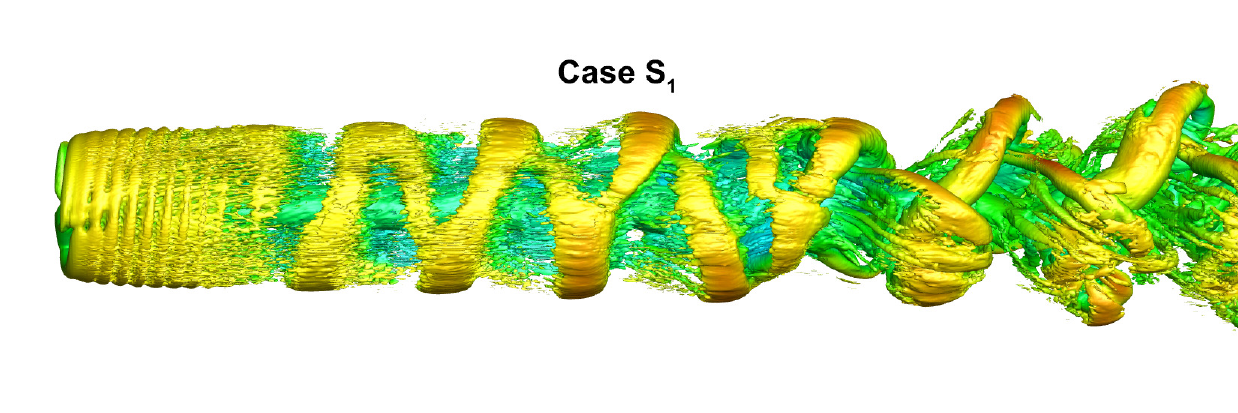} 
\includegraphics[width=0.6\linewidth]{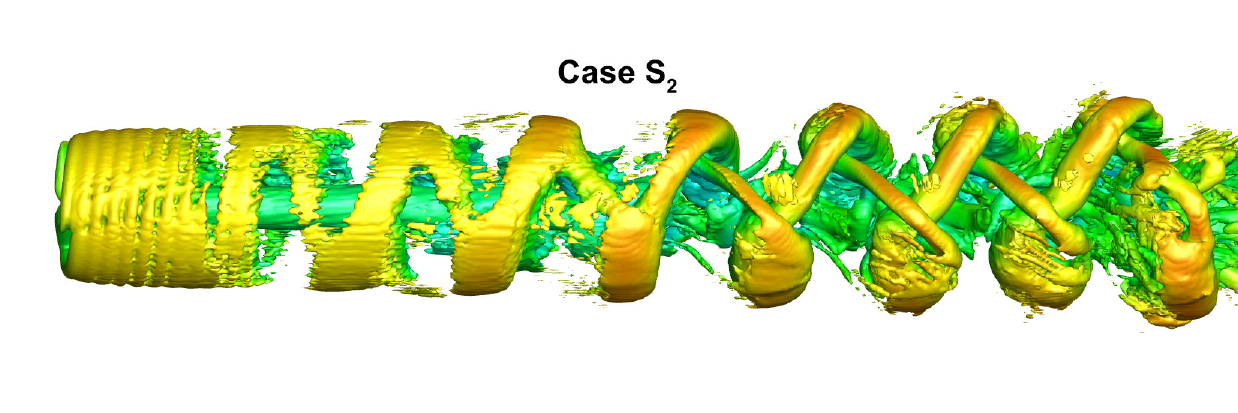} 
\caption{\label{refined-iso-1} NREL $5$MW wind turbine wake: the instantaneous contours of Q-criterion colored by magnitude of velocity for four cases in Tab.\ref{accuracy_case}.}
\end{figure}

\begin{figure}[!h]
\centering
\includegraphics[width=0.9\linewidth]{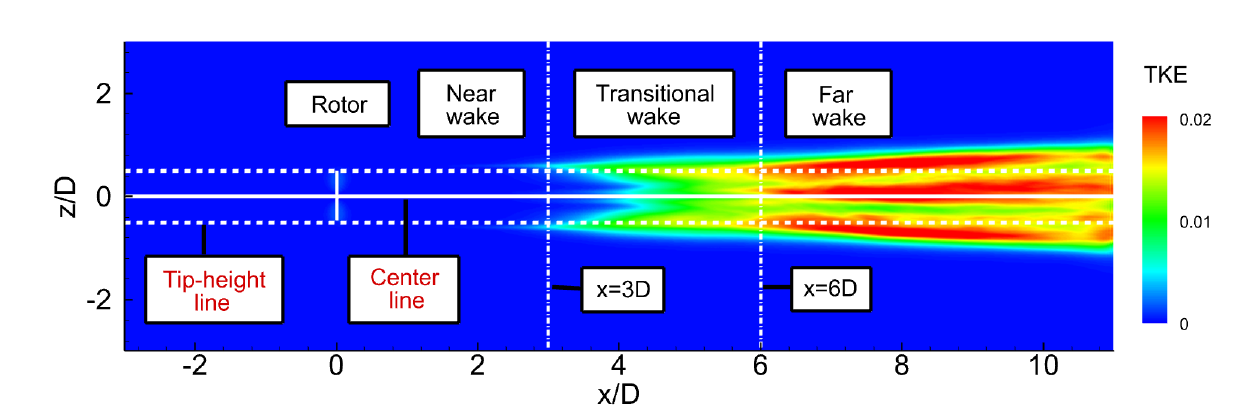}
\caption{\label{tke-schematic-accuracy} NREL $5$MW wind turbine wake: the time-averaged normalized TKE at $y = 0$ plane for Case $H_1$.}
\end{figure}

\begin{table}
\begin{center}
\def\temptablewidth{1.0\textwidth}
{\rule{\temptablewidth}{1.0pt}}
\begin{tabular*}{\temptablewidth}{@{\extracolsep{\fill}}ccc}
Case &Numerical schemes &Grid settings  \\
\hline
H$_1$ &High-order GKS   &$\Delta x=D/50, \Delta y=\Delta z=D/100$\\
\hline
H$_2$ &High-order GKS   &$\Delta x=\Delta y=\Delta z=D/50$\\
\hline
S$_1$ &Second-order GKS   &$\Delta x=D/50, \Delta y=\Delta z=D/100$\\
\hline
S$_2$ &Second-order GKS   &$\Delta x=\Delta y=\Delta z=D/50$ \\
\end{tabular*}
{\rule{\temptablewidth}{1.0pt}}
\caption{\label{accuracy_case}NREL $5$MW wind turbine wake: the grid settings for high-order and second-order GKS. }
\end{center}
\end{table}

\begin{figure}[!h]
\centering
\includegraphics[width=0.475\linewidth]{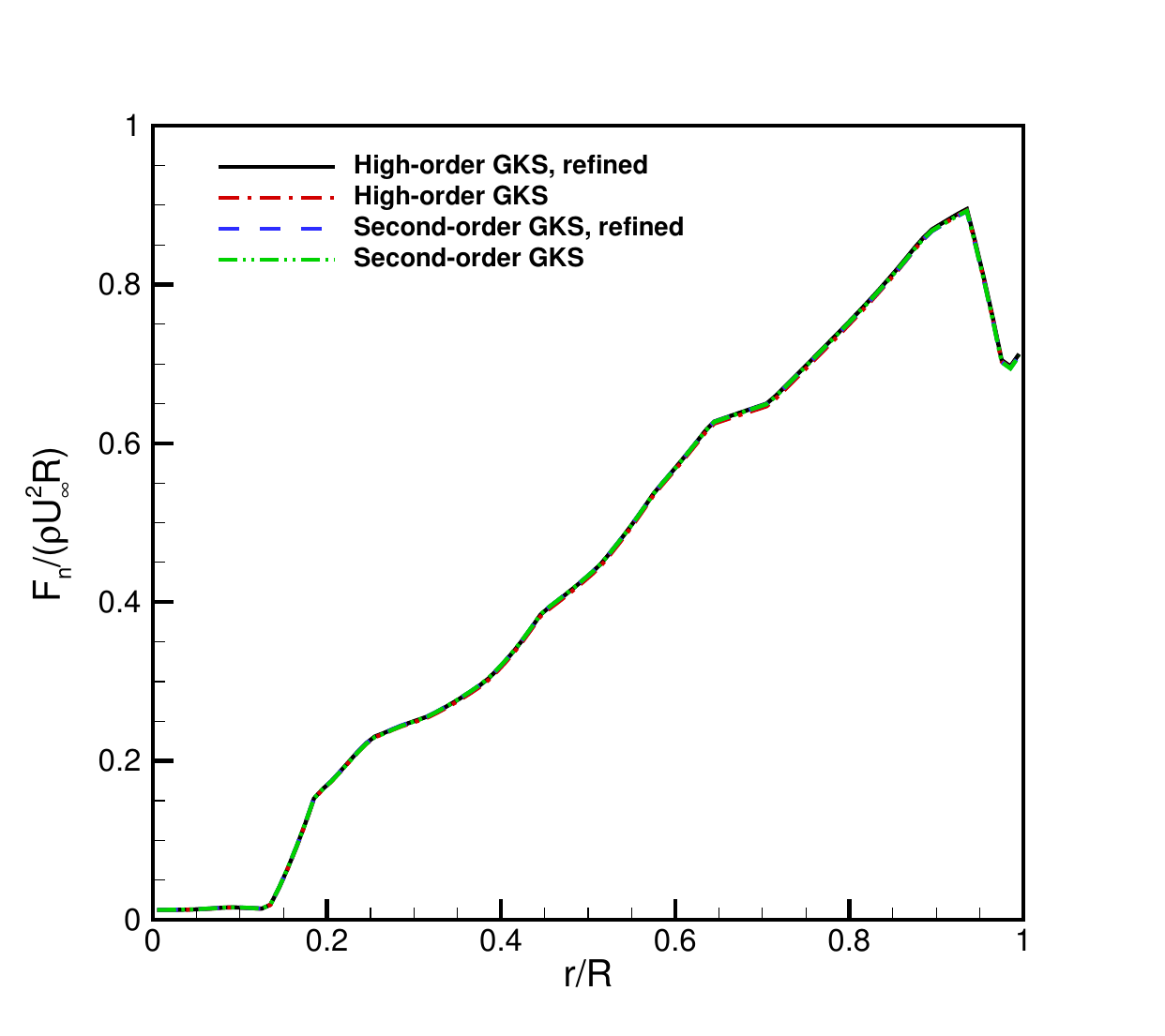}
\includegraphics[width=0.475\linewidth]{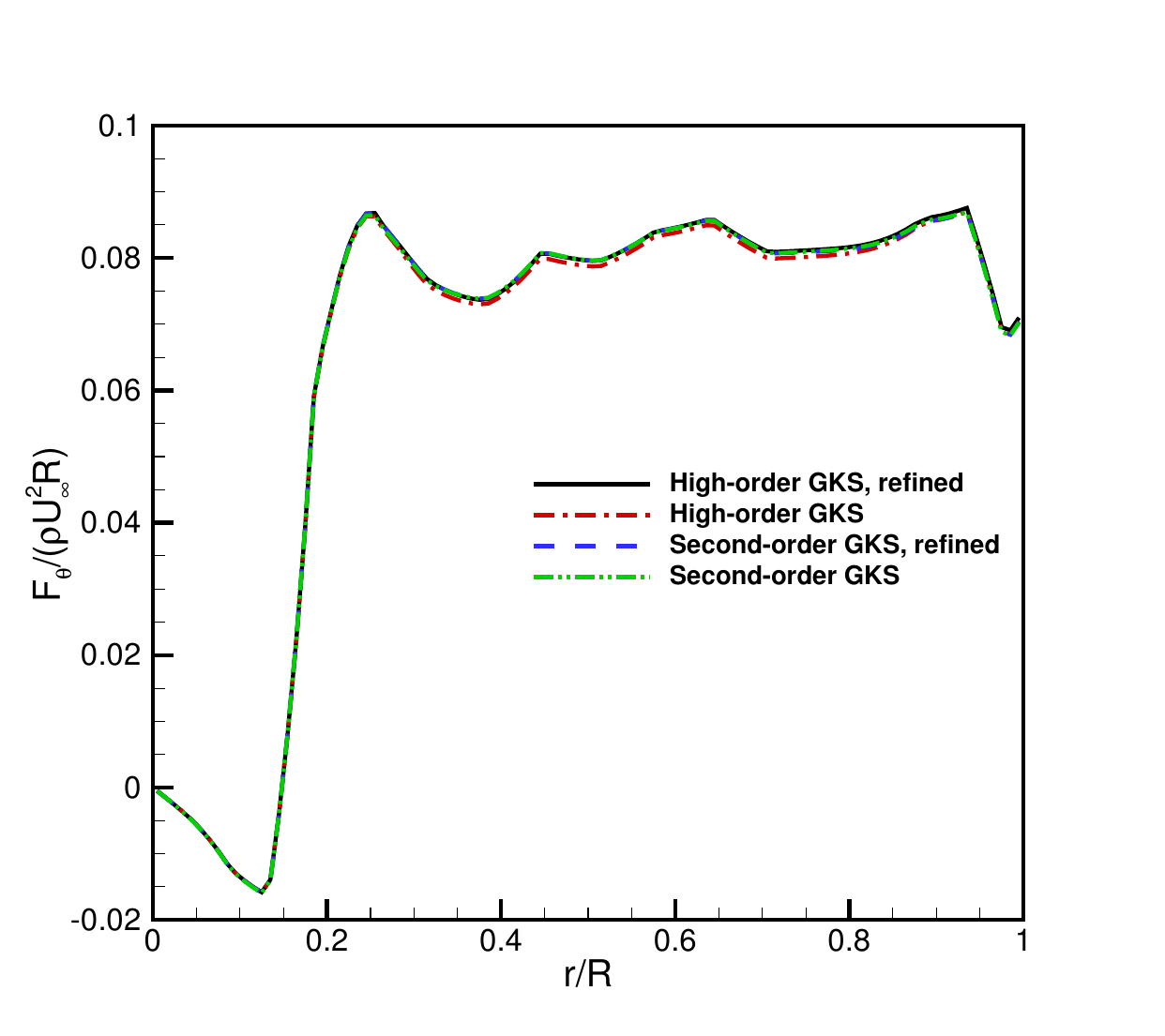}
\caption{\label{along-blade-force-accuracy} NREL $5$MW wind turbine wake: the time-averaged normal (left) and 
tangential forces (right) along the blade with the high-order and second-order GKS using locally-refined and uniform grids.}
\end{figure}

\begin{figure}[!h]
\centering
\includegraphics[width=0.475\linewidth]{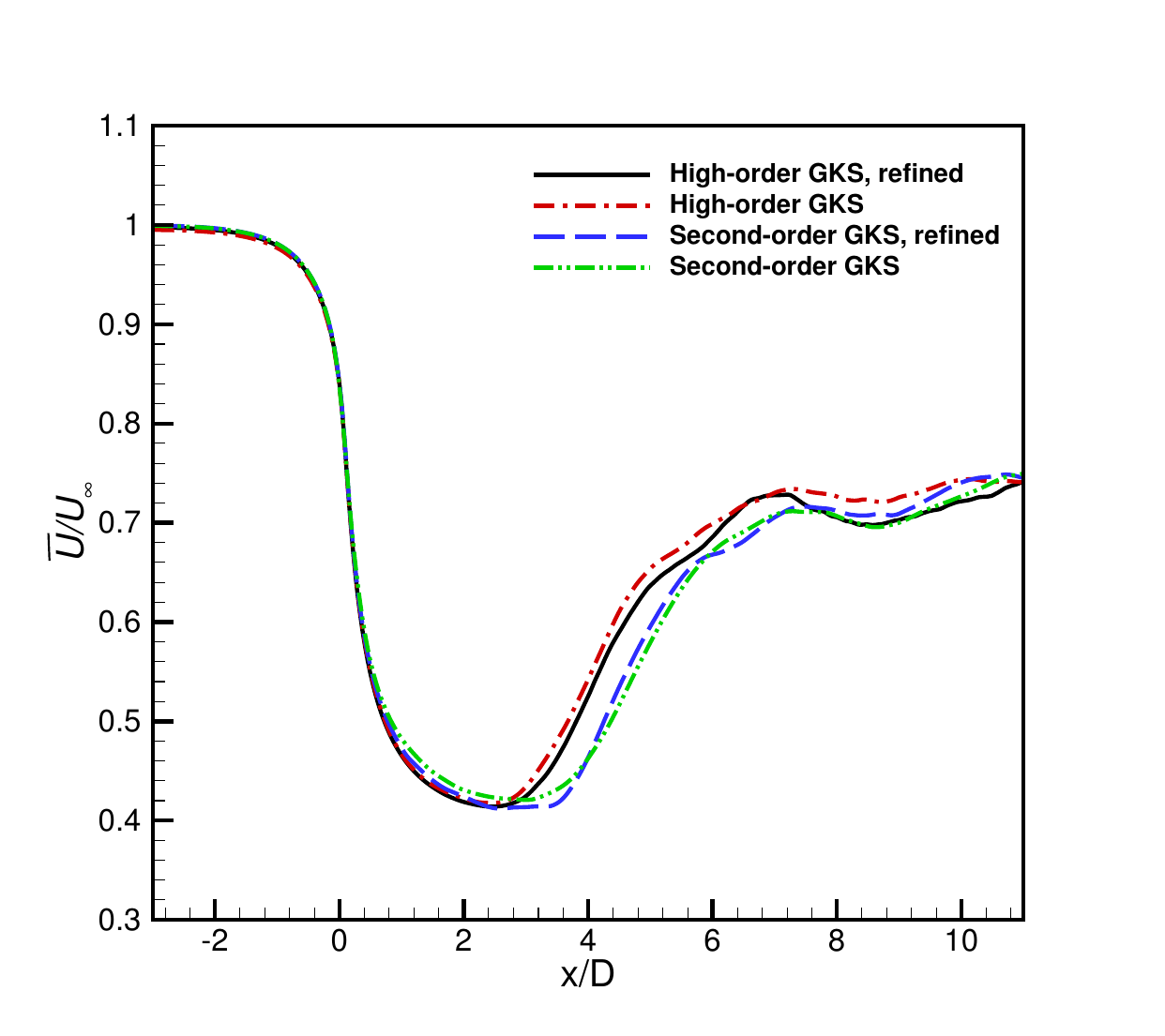}
\includegraphics[width=0.475\linewidth]{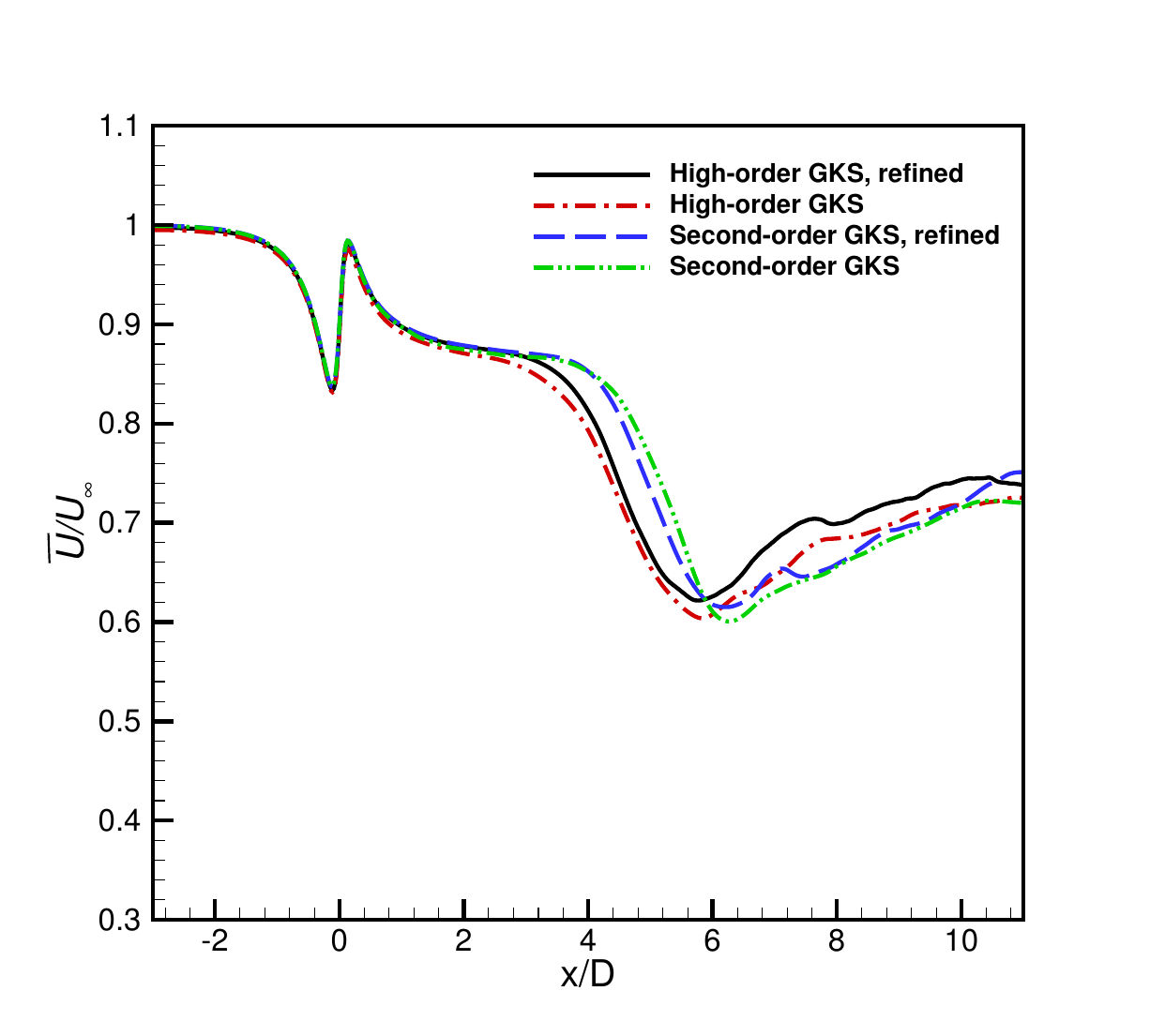}
\includegraphics[width=0.475\linewidth]{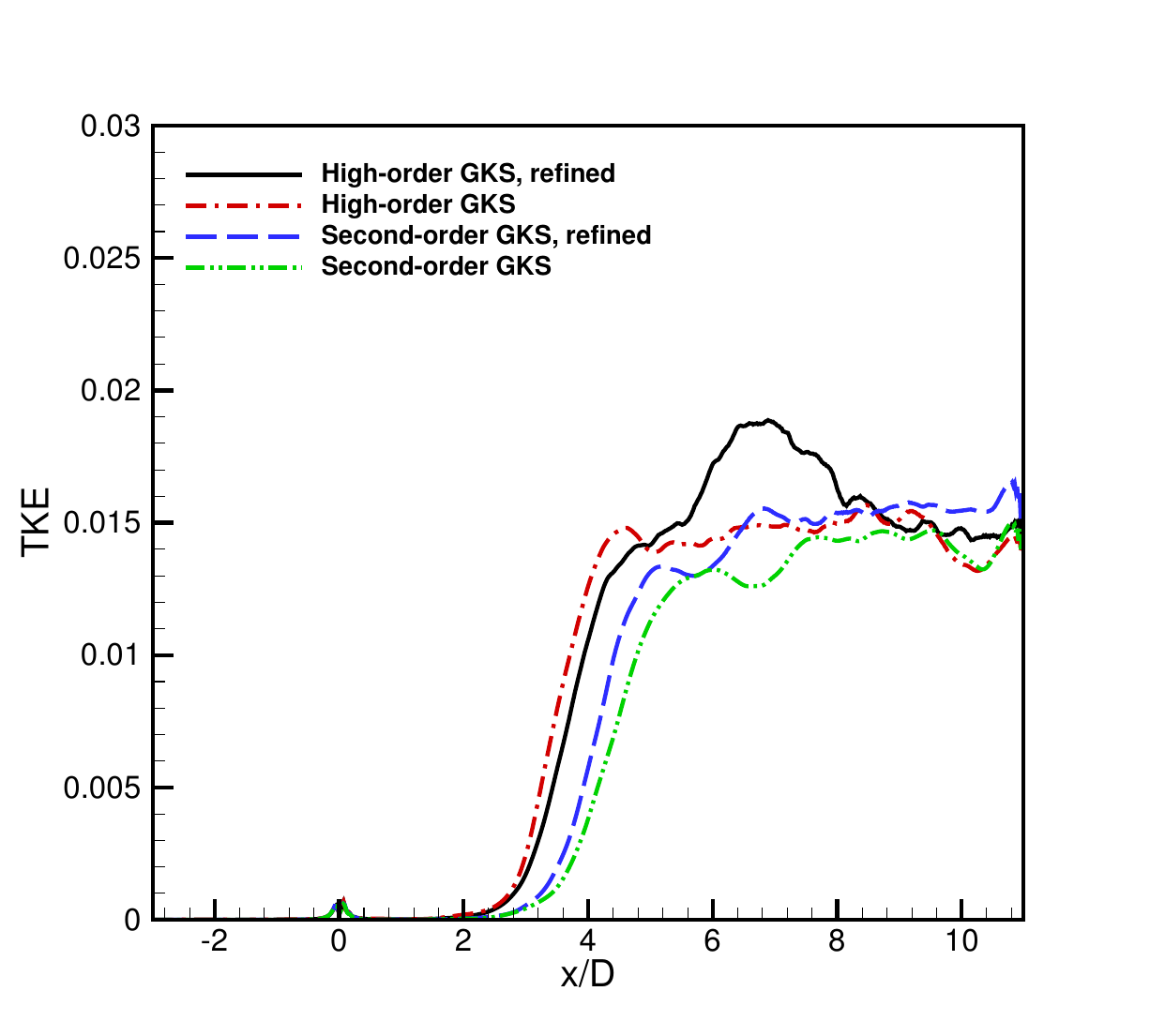}
\includegraphics[width=0.475\linewidth]{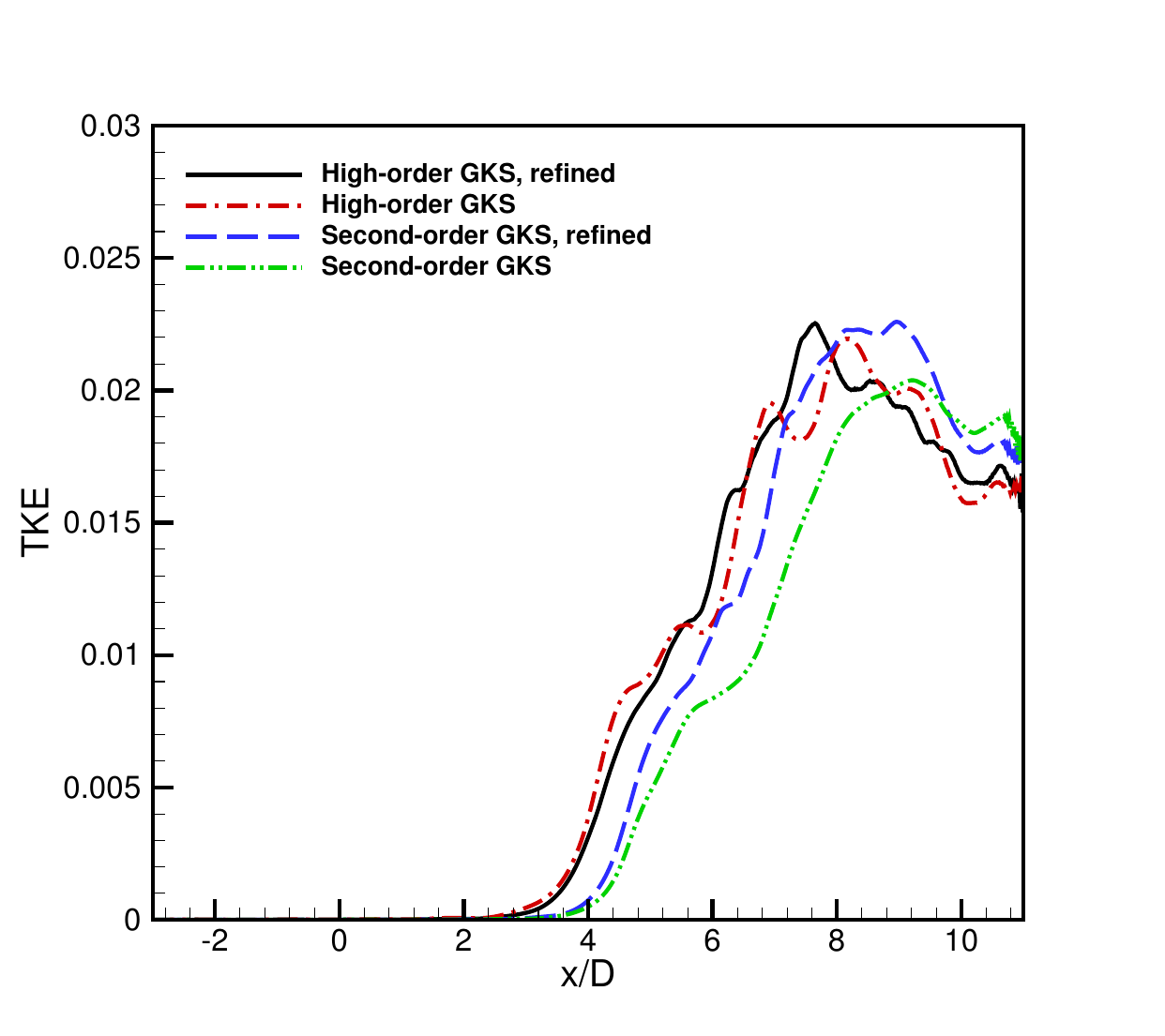}
\caption{\label{axial-line-accuracy} NREL $5$MW wind turbine wake: The time-averaged streamwise
velocity (top) and TKE (bottom) profiles at the tip height (left) and the center (right) line at the $y = 0$ plane.}
\end{figure}

\subsubsection{Numerical accuracy analysis}
To investigate the effect of numerical order, the comparisons on the second-order GKS and high-order GKS with two set of grids are studied. 
Tab.\ref{accuracy_case} shows four setups varying the numerical orders of the GKS and the grid settings. A locally-refined
grid is used, remaining the $700\times300\times300$ cells and  the center domain$[-3D,11D]\times[-D,D]\times[-D,D]$ is 
locally refined with $\Delta x=D/50$ and $\Delta y=\Delta z=D/100$. The smearing kernel width takes a constant value of 
$\varepsilon =3\varDelta x=3D/50$ for the four cases. Simulations with the high-order GKS under the locally-refined grid are regarded
as the highest-fidelity reference solutions. 
The instantaneous Q-criterion iso-surface of four cases in Tab.\ref{accuracy_case} are shown in Fig.\ref{refined-iso-1} at the fifth flow-through time. The contours
are colored by the magnitude of local streamwise velocity. 
With the same grid, the high-order GKS shows higher resolution in vortex structures
than the second-order GKS, with resolving much smaller turbulent structures.
Using the same numerical scheme, much smaller vortex structures can be resolved using
the locally-refined grid.

The time-averaged normalized turbulent kinetic energy (TKE) for Case 
$H_1$ is also shown in Fig.\ref{tke-schematic-accuracy} at the $y=0$ plane. 
The normalized TKE is defined as \begin{align*}
TKE=\frac{1}{2}(\overline{{U^{'}}^{2}}+\overline{{V^{'}}^{2}}+\overline{{W^{'}}^{2}})/U_{\infty}^2,
\end{align*}
where $U^{'}=U-\overline{U}, V^{'}=V-\overline{V}$ and $W^{'}=W-\overline{W}$.
Based on the magnitude of TKE at the top and bottom tip height $z=\pm
0.5D$, we define three wake regions for the following quantitative
analysis. Near wake is defined from $x=0D$ to $x=3D$, which is
laminar region. The turbulence roughly begins to be triggered from
$x=3D$ and transits to $x=6D$ where the magnitude of TKE reaches to
the plateau, which is defined as the transitional wake. In the
transitional wake, the rapid increase of TKE results from the tip
vortex breakdown into turbulence. The far wake starts from $x=6D$,
which is the fully turbulent region. For the center mixing, we find
the almost same far wake region as the magnitude of TKE reaches to
the plateau after $x = 6D$. Numerical-order impact is analyzed for the rotor loading and wake
evolution.

\begin{figure}[!h]
\centering
\includegraphics[width=0.475\linewidth]{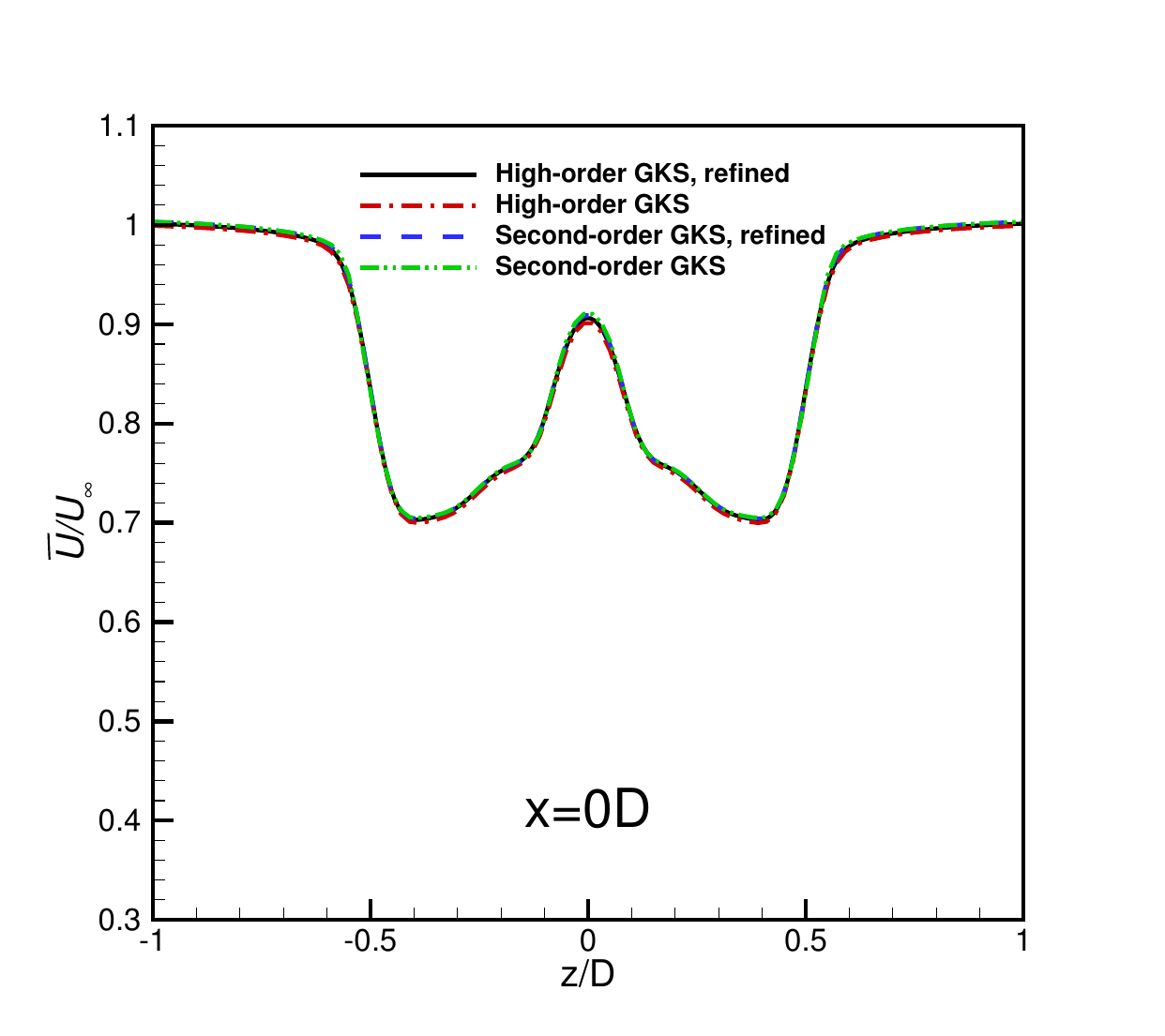}
\includegraphics[width=0.475\linewidth]{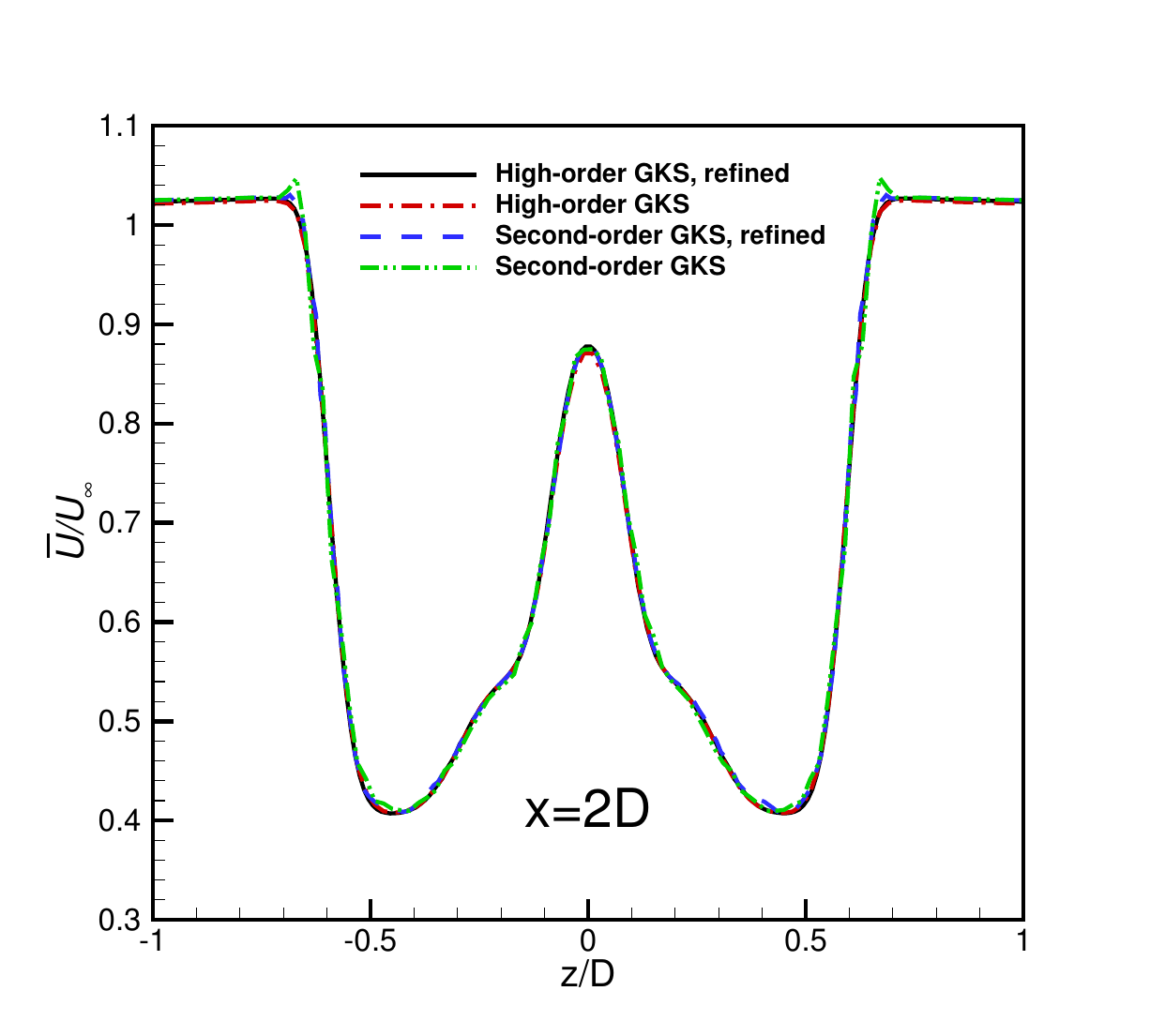}
\includegraphics[width=0.475\linewidth]{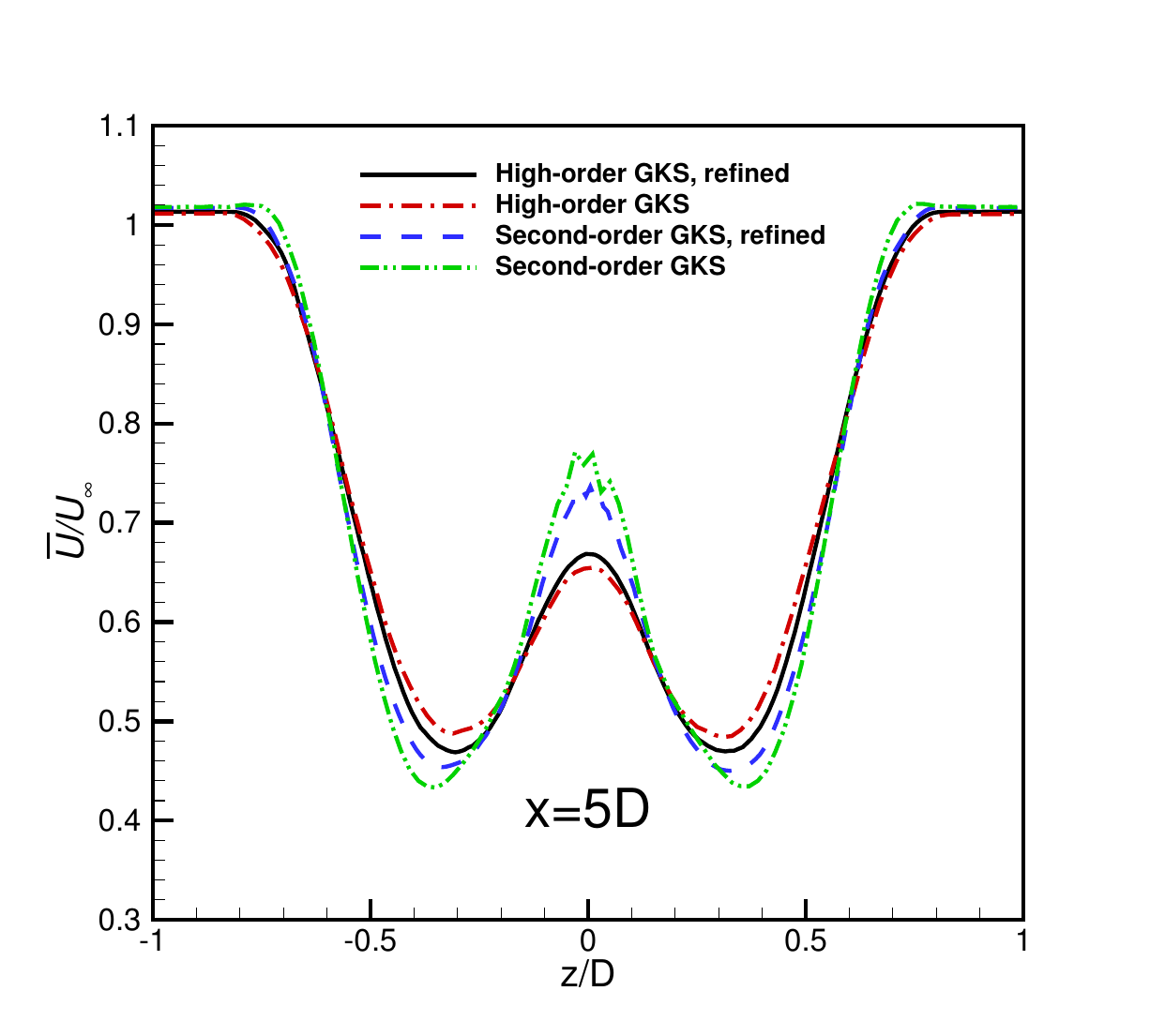}
\includegraphics[width=0.475\linewidth]{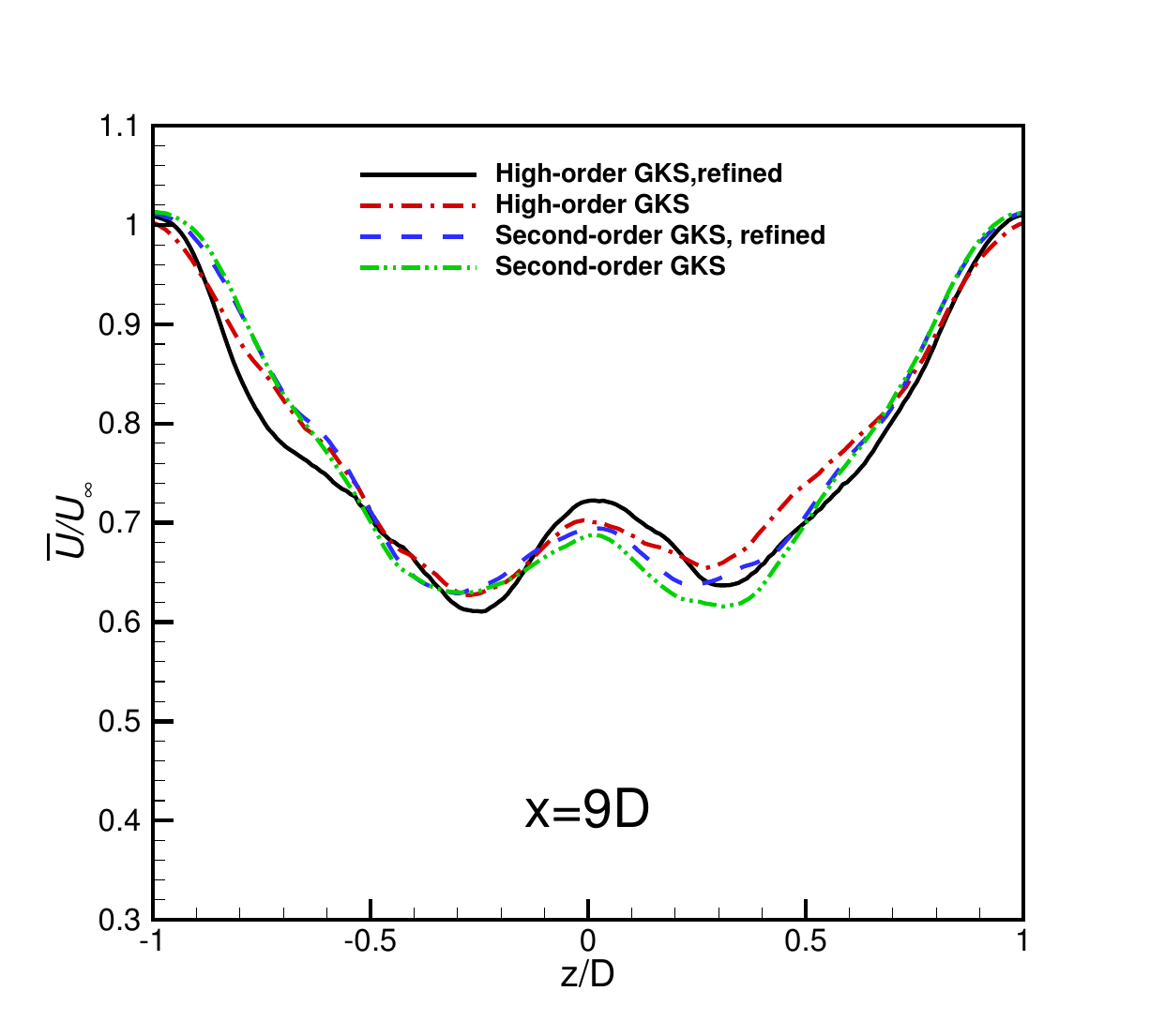}
\caption{\label{velocity-deficit-accuracy} NREL $5$MW wind turbine wake: 
the time-averaged streamwise velocity profiles at $x=0D, 2D, 5D$ and $9D$ at the $y=0$ plane.}
\end{figure}

\begin{figure}[!h]
\centering
\includegraphics[width=0.475\linewidth]{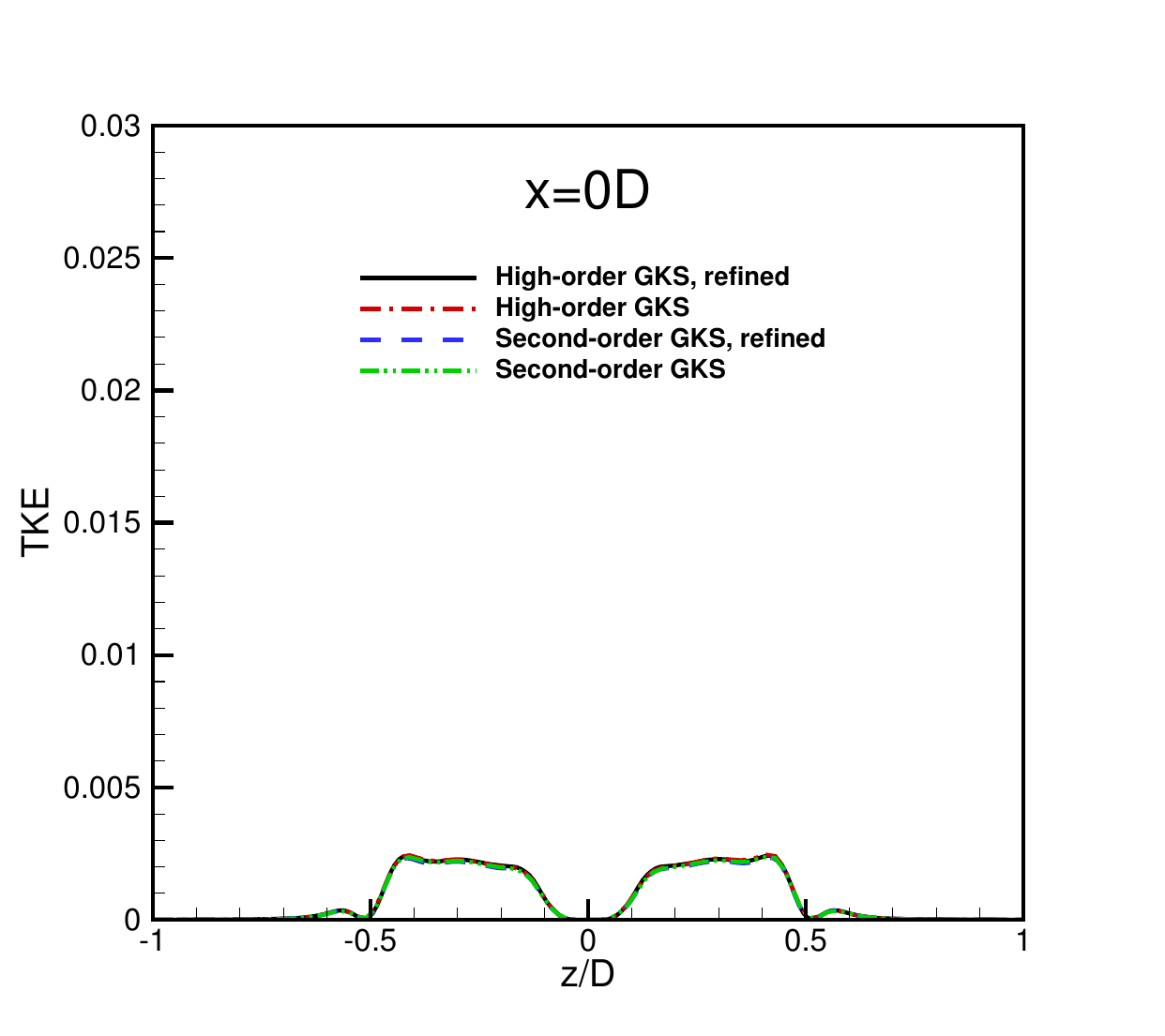}
\includegraphics[width=0.475\linewidth]{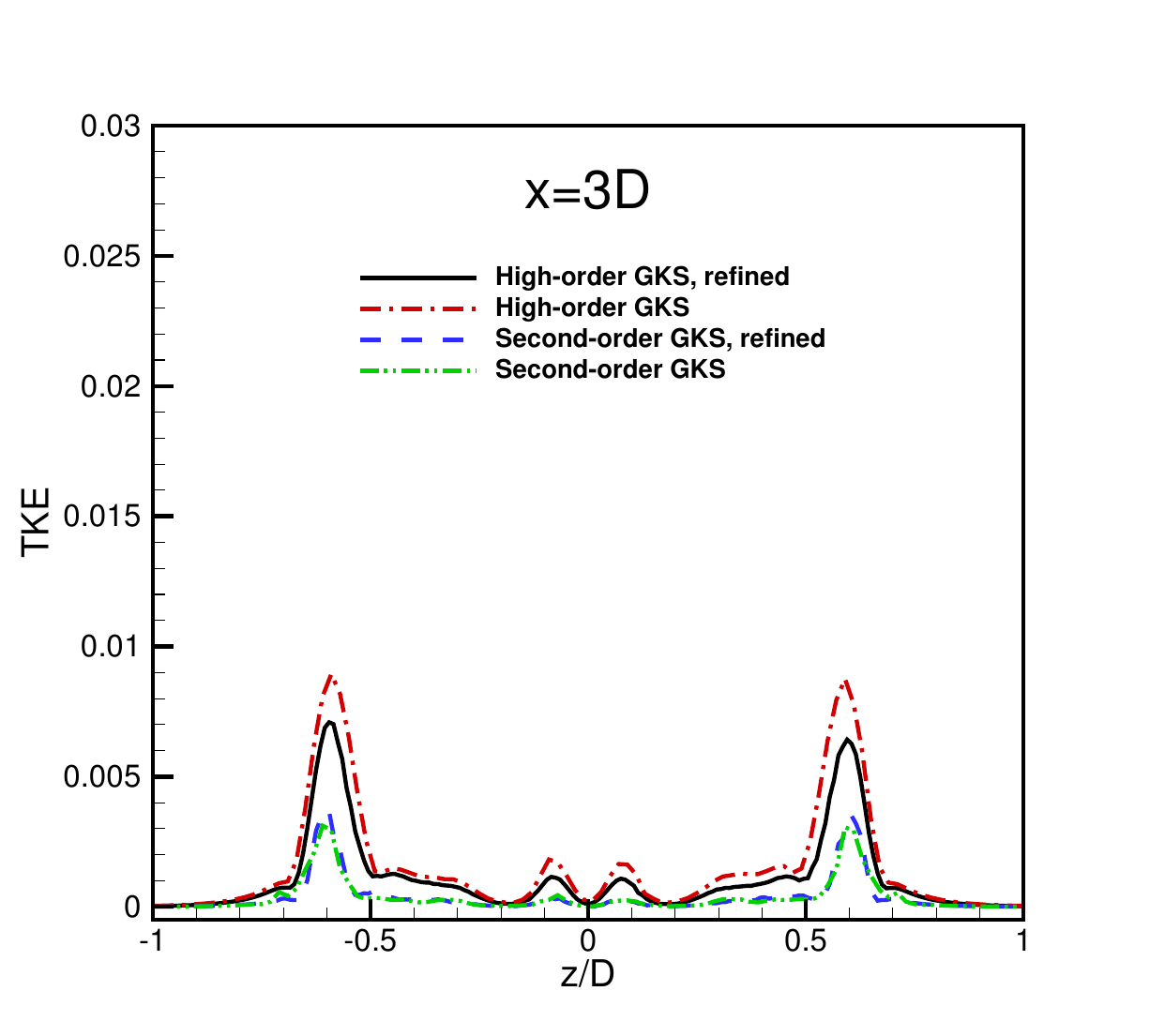}
\includegraphics[width=0.475\linewidth]{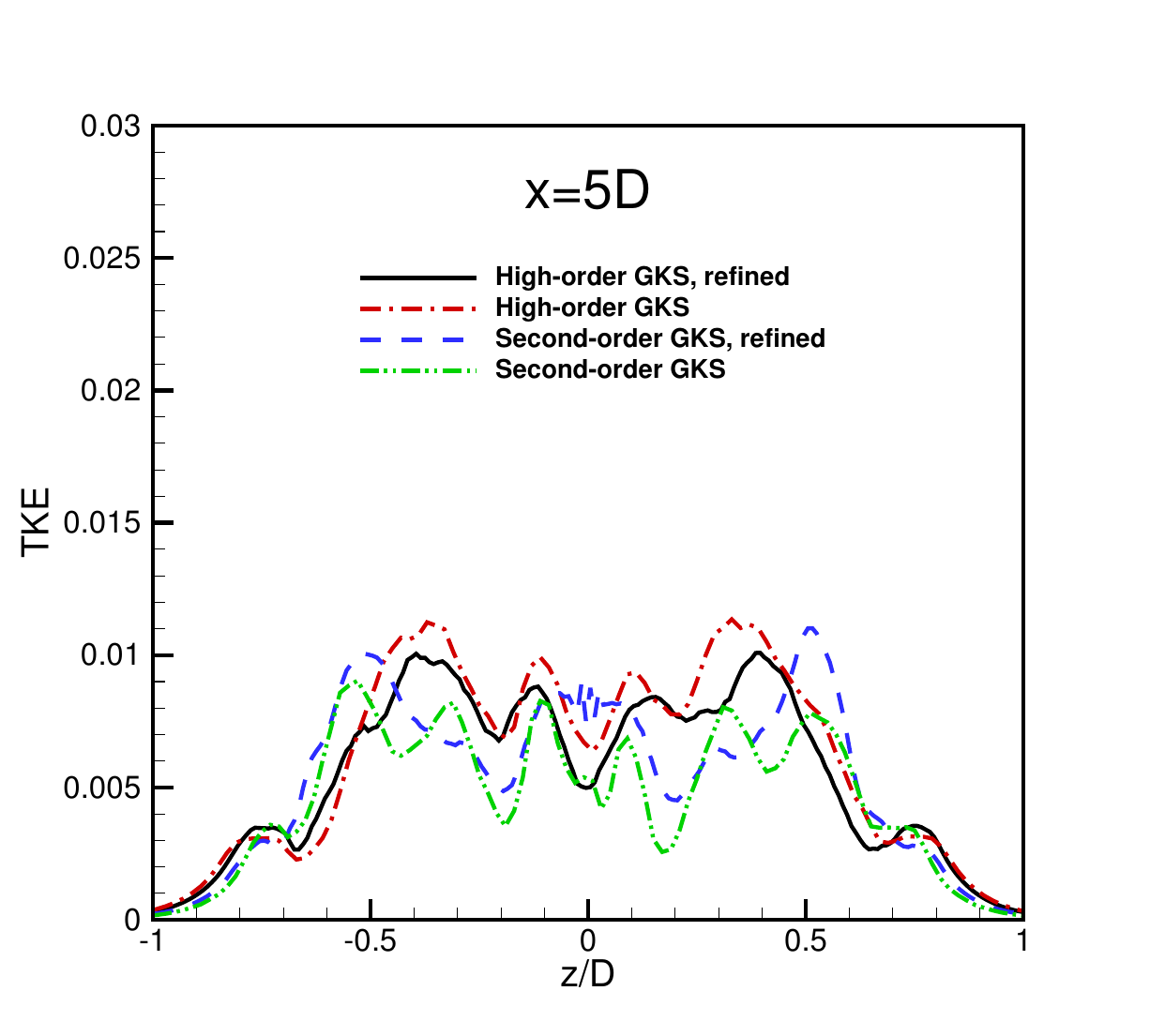}
\includegraphics[width=0.475\linewidth]{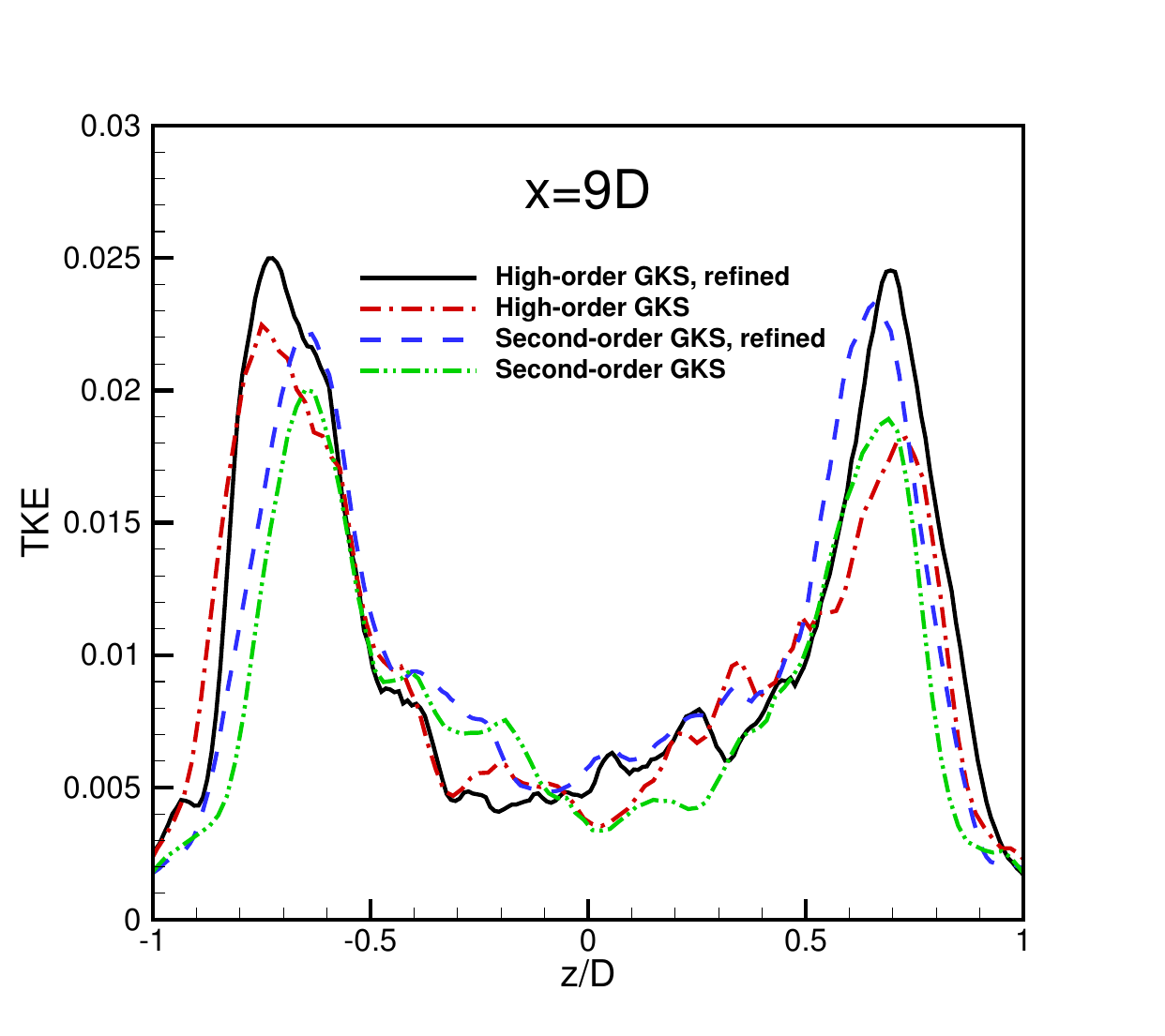}
\caption{\label{tke-profile-accuracy} NREL $5$MW wind turbine wake: 
the time-averaged normalized turbulent kinetic energy at $x=0D, 3D, 5D$ and $9D$ at the $y=0$ plane.}
\end{figure}

The detailed analysis in wake is implemented based on the
three wake regions. The time-averaged normal and tangential forces along the blade are
given in Fig.\ref{along-blade-force-accuracy}. The spanwise blade
quantities agree well among the GKS solutions with different
numerical orders and grid settings, indicating that the numerical
discretization error has a very limited impact on the blade force
predictions.

The time-averaged normalized streamwise velocity and TKE are shown
in Fig.\ref{axial-line-accuracy} along the center and tip height
line at the $y = 0$ plane. The tip height line solution is the mean
of solutions at the top tip height ($z= 0.5D$) and the bottom tip
height ($z= -0.5D$). In the near wake region, all solutions collapse
well. At around $x=3D$, the magnitude of normalized streamwise
velocity changes rapidly, and the TKE increases significantly both
at the center line and at the tip height line. The time-averaged
normalized streamwise velocity and TKE profiles agree reasonably
well for the high-order GKS regardless of the grid refinement. The
high-order GKS can preserve better high wave-number structures that
earlier trigger the breakdown of the tip vortex and center mixing.
Solutions from the second-order GKS with locally-refined grids
approach to those from the high-order GKS, confirming that the
high-order numerical scheme improves the accuracy of transitional
wake. For the far-wake region, the discrepancies to the reference
solution scatters.  

The time-averaged streamwise velocity profiles are shown in
Fig.\ref{velocity-deficit-accuracy} at $x=0D, 2D, 5D$ and $9D$ at
the $y =0$ plane. In the near-wake region, all solutions agree well
among different numerical orders and grid settings, confirming that
the near-wake laminar structures are well resolved. In the
transitional-wake region, the dissipative scale of large Reynolds
number wake turbulence requires high-resolution simulations. At
$x=5D$, the streamwise velocity of high-order GKS recovers faster
than that of the second-order GKS. The velocity profiles from the
second-order GKS with locally-refined grids are closer to those from
the high-order GKS, demonstrating that the high-order numerical
scheme improves the resolution of the transitional-wake turbulence.
For far wake at $x=9D$, we observe the relative larger discrepancies
compared to the reference solution. 

The TKE profiles are shown in Fig.\ref{tke-profile-accuracy} at
$x=0D, 3D, 5D $ and $9D$. At $x=0D$, the TKE arises from the rotor
revolution. In the transitional-wake region, all cases show the wake
expansion as the peak positions of the TKE are off the $z \pm 0.5
D$. Compared to the reference solution, the high-order GKS gives the
much closer peak TKE in the center mixing and tip vortex regions.
With the local-refined grid, the solution from the second-order GKS
gives the higher peak in the tip height region than that from the
uniform grid. At $x=9D$, the magnitude of TKE increases downstream
to the far-wake region, and we observe the asymmetric TKE profile.
Compared to the reference solution, the high-order GKS with uniform
grids gives the closer solution in the $z/D \in [-1, 0.5]$ region,
while the second-order GKS with local-refined grid agrees better in
the $z/D \in [0.5, 1]$ region.

\begin{figure}[!h]
\centering
\includegraphics[trim=80 100 100 100, clip = true, width=0.80\linewidth]{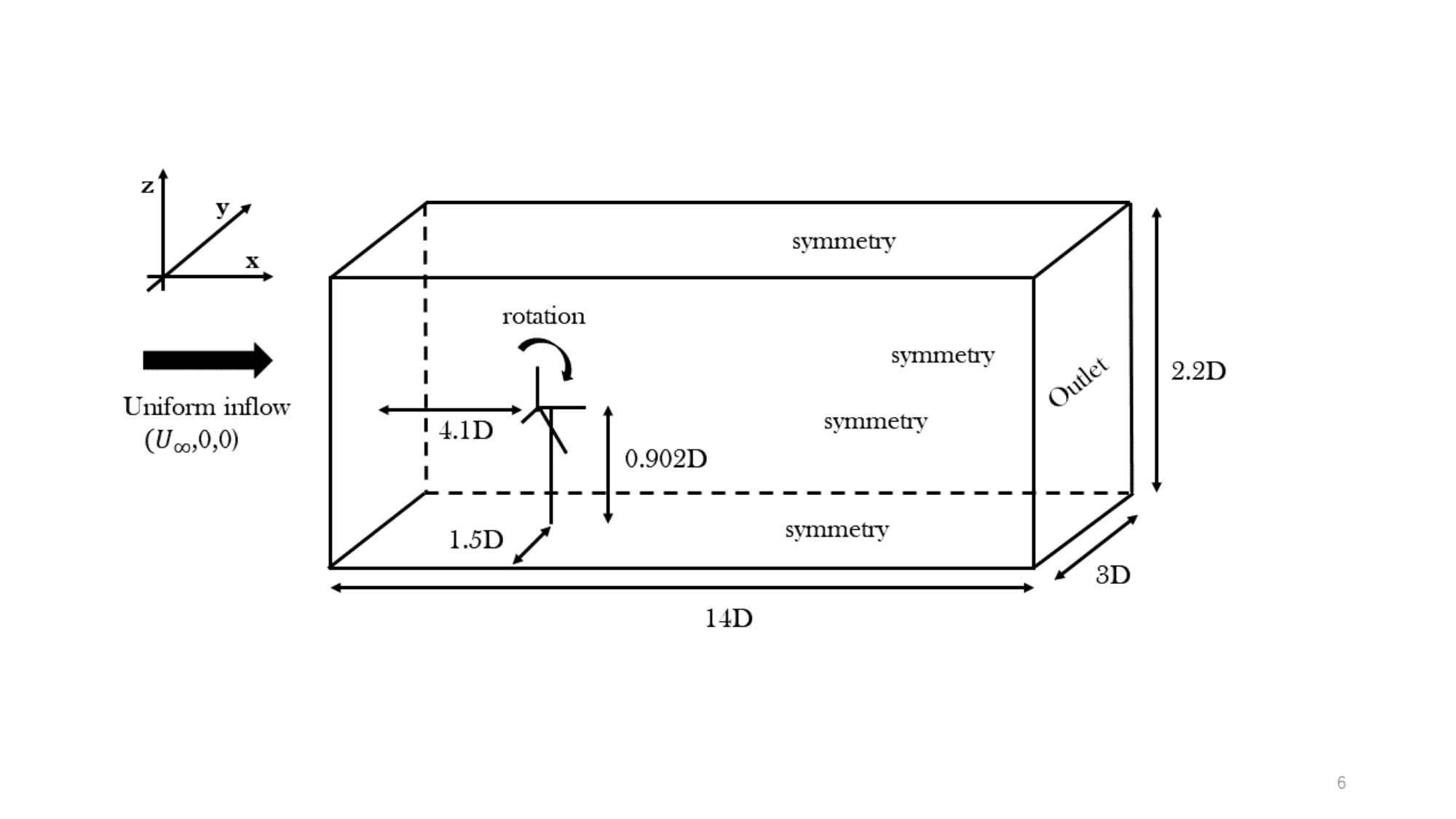}
\caption{\label{NTNU-domain} NTNU rotor simulation: the schematic of computational domain and boundary conditions.}
\vspace{3mm}
\centering
\includegraphics[trim=80 100 100 100, clip = true, width=0.80\linewidth]{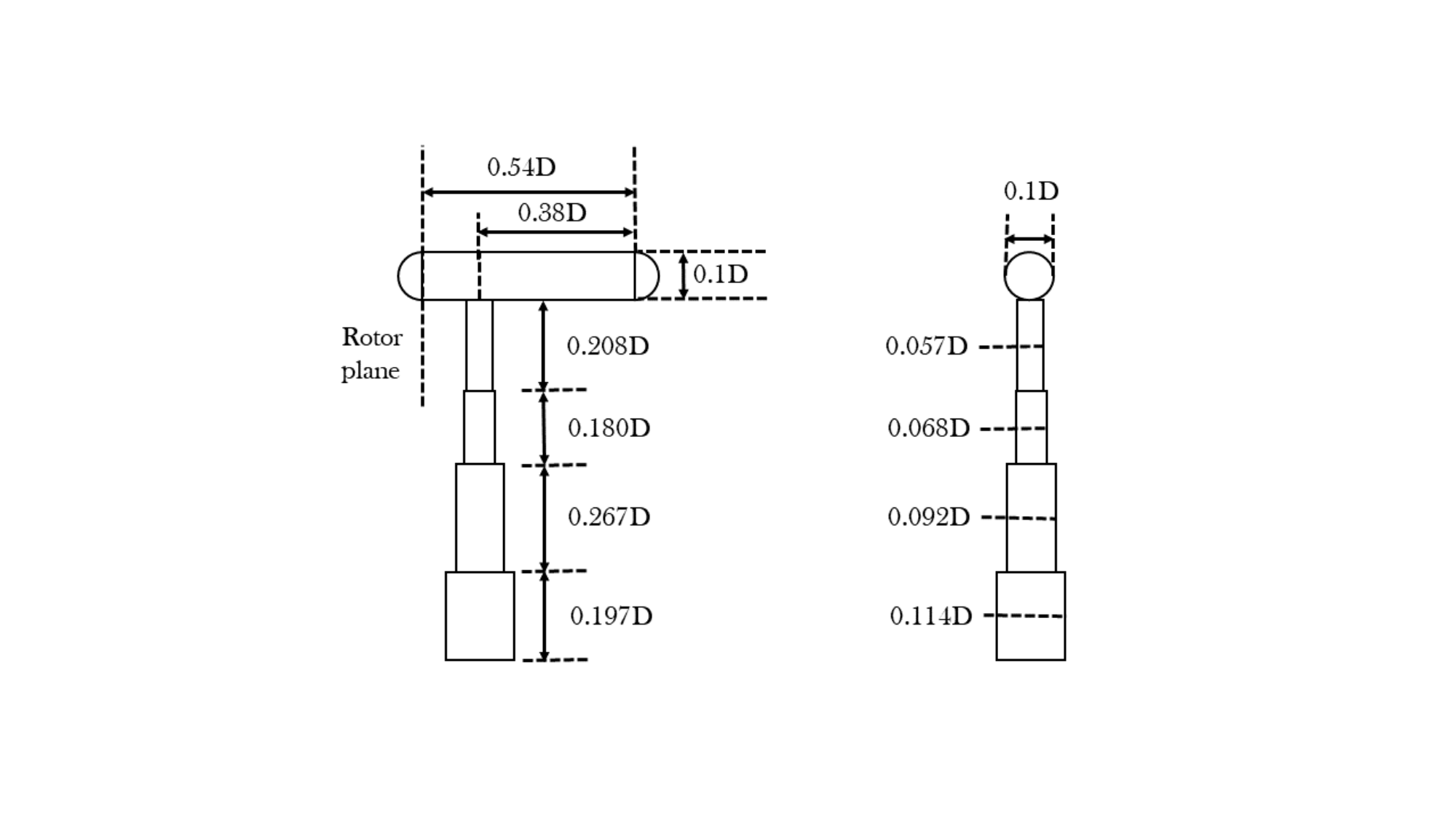}
\caption{\label{NTNU-detail} NTNU rotor simulation: the detailed geometry of the tower and the nacelle. }
\end{figure}

\subsection{NTNU rotor with nacelle and tower}
The high-order GKS with actuator line model and immersed boundary method is implemented to simulate the NTNU wind turbine with nacelle and tower \cite{NTNU-turbine}.
The computational domain and boundary conditions are shown in Fig.\ref{NTNU-domain}, 
and the rotor is rotating clockwise from the direction of inlet. 
The computational domain is $ (x,y,z) \in [-4.1D,9.9D]\times[-1.5D,1.5D]\times[0D,2.2D]$, where the $x$-,
$y$-, $z$-directions represent the streamwise, transverse and
vertical directions, and the velocity components are denoted as $U$, $V$, and $W$, respectively.
The wind turbine rotor center is located at $(0D, 0D, 0.902D)$ and the symmetry boundary conditions are imposed on the walls. 
The uniform grid with $1400\times 300\times 220$ cells is used, giving a grid size of $\Delta x=\Delta y=\Delta z=D/100$. 
The detailed geometry of the nacelle and tower is shown in Fig.\ref{NTNU-detail},  which follows the settings in \cite{NTNU-IBM-Santoni}. 
The tower consists of $4$ cylinders of different diameters and lengths. 
From top to bottom, the cylinder diameters are $d_1 = 0.057D, d_2 = 0.068D, d_3 = 0.092D$, and $d_4 = 0.114D$, 
and the lengths of the cylinders are $L_1 = 0.208D, L_2 = 0.180D, L_3 = 0.267D$, and $L_4 = 0.197D$. 
The nacelle consists of a cylinder and two hemispheres, the cylinder has a diameter of $0.1D$ and a total length of $0.54D$ 
and the radius of the hemisphere is $0.05D$. The overhang takes $0.16D$. 

In the computation, both nacelle and tower effects are modeled using immersed boundary method. 
For the actuator line model, $200$ actuator points are distributed to represent each blade, 
and the actuator points are locally refined  at the tip and root of the blade.
The smearing kernel width $\varepsilon=3\Delta x=3D/100$ in the ALM. 
The line sampling method proposed by \cite{ALM-2}  is implemented for velocity sampling,
and the tip correction factor proposed by \cite{tip-correction-shen}  is used, which is defined as 
\begin{align}\label{tip-correction}
\begin{split}
f_{tip}&=\frac{2}{\pi}\cos^{-1} \big[\exp\big(-g_{1}\frac{N(R-r)}{2rsin\phi}\big)\big],\\
g_{1}&=\exp(-c_{1}(N\lambda-c_{2}))+0.1,
\end{split}
\end{align}
where $N$ represents the number of blades, $R$ is the rotor radius, $\lambda$ is the tip speed ratio $(TSR)$, and 
$c_1$ and $c_2$ are empirical coefficients. 
In the computation, 
$c_1=0.1219$, $c_2=21.52$ are used in Eq.\eqref{tip-correction} to calculate the tip correction factor for axial force, 
which is denoted as $f_{tip}^{axi}$ and $c_1=0.0984$, $c_2=13.026$ are used in Eq.\eqref{tip-correction} to calculate the tip correction factor for tangential force, which is denoted as $f_{tip}^{tan}$  \cite{tip-correction-W&W}. 
The modified axial and tangential forces are given as follows
\begin{align*}
F_x     &= f_{tip}^{axi}  (L \cos \phi + D \sin \phi),\\
F_\theta&= f_{tip}^{tan} (L\sin \phi-D\cos \phi),
\end{align*}
where $L$ and $D$ are given by Eq.\eqref{axial-tangential}.

\begin{figure}[!h]
\centering
\includegraphics[width=0.475\linewidth]{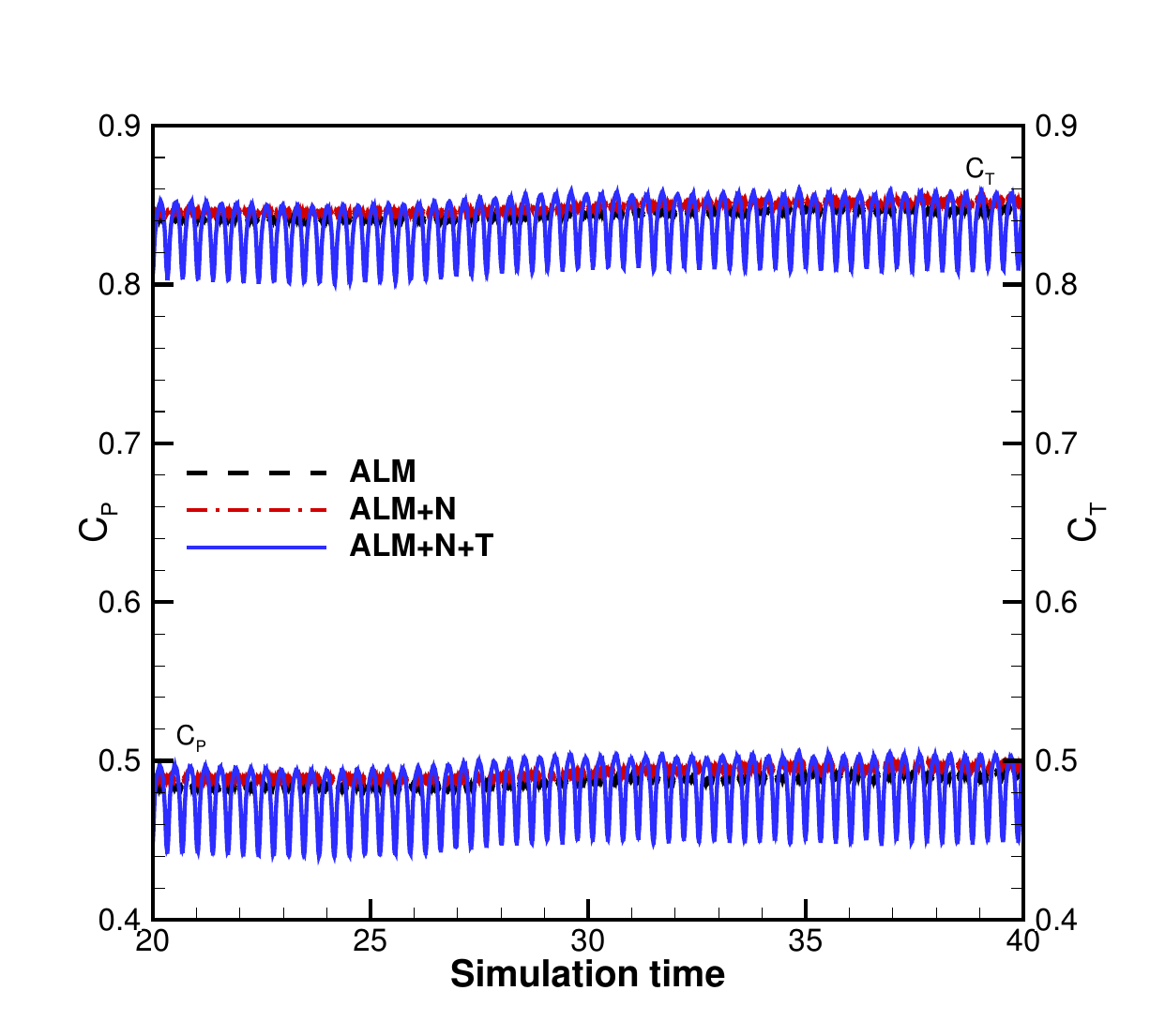}
\includegraphics[width=0.475\linewidth]{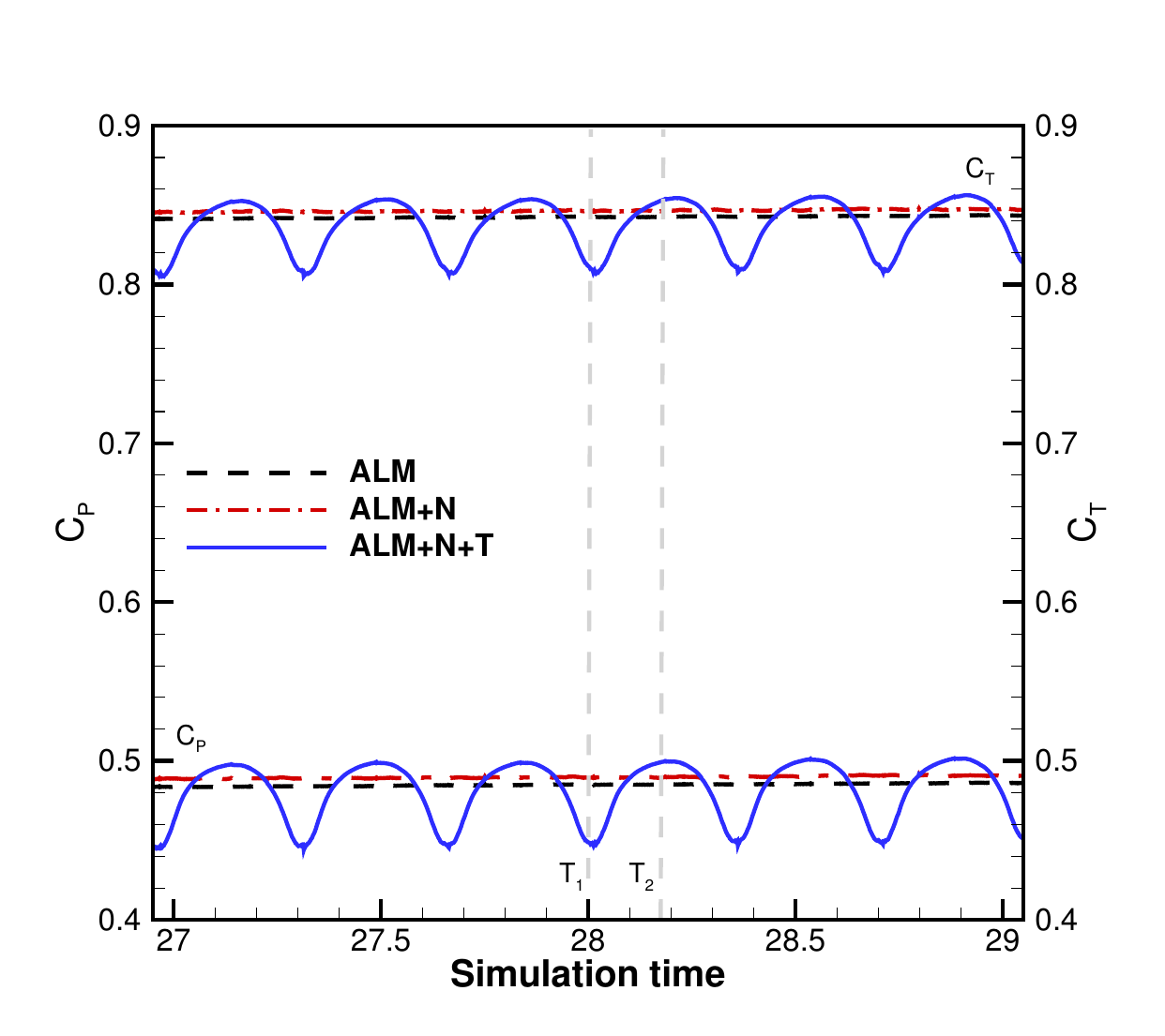}
\caption{\label{NTNU-coefficients-1} NTNU rotor with nacelle and tower: the time histories of $C_P$ and $C_T$ for three cases.}
\end{figure}

\begin{figure}[!h]
\centering
\includegraphics[width=0.475\linewidth]{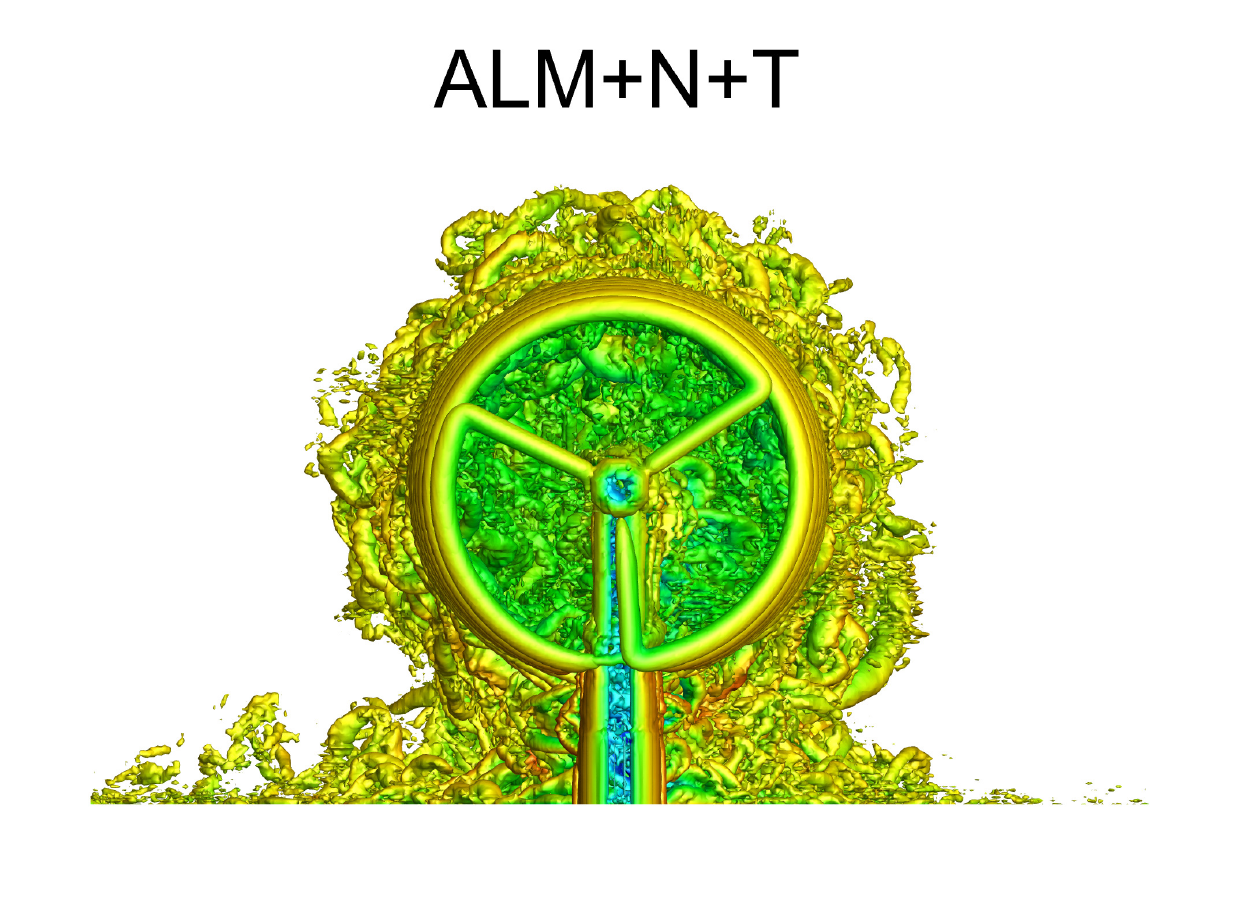}
\includegraphics[width=0.475\linewidth]{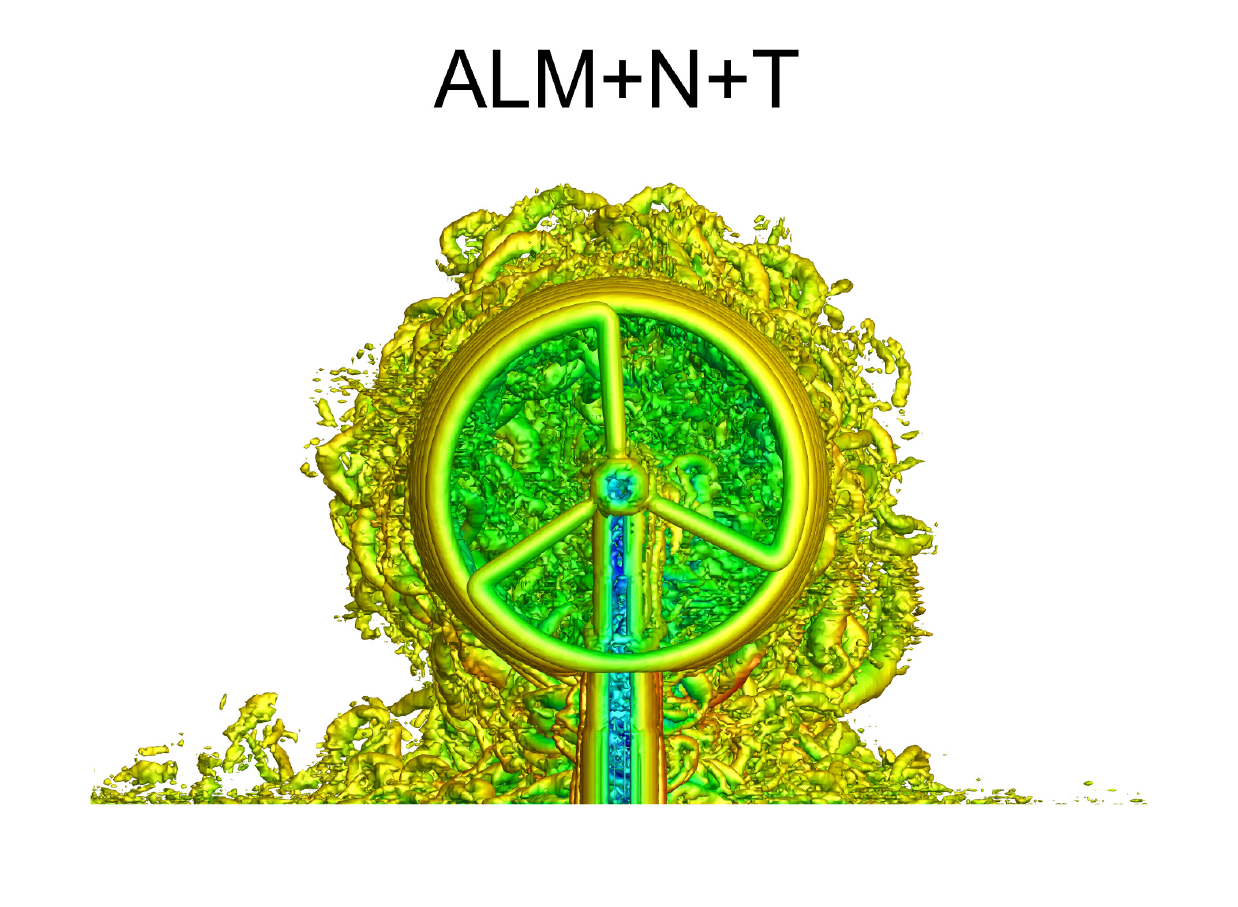}
\caption{\label{NTNU-coefficients-2} NTNU rotor with nacelle and tower: the instantaneous Q-criterion iso-surface for ALM+N+T case at $T_1$ (left) and $T_2$ (right).}
\end{figure}


\begin{table}[!h]
\begin{center}
\def\temptablewidth{1.0\textwidth}
{\rule{\temptablewidth}{1.0pt}}
\begin{tabular*}{\temptablewidth}{@{\extracolsep{\fill}}cccccc}
Case &Blade model &Nacelle model &Tower model & $C_P$ & $C_T$ \\
\hline
ALM &ALM &None   &None & 0.487& 0.844\\
\hline
ALM+N &ALM &IBM &None & 0.492& 0.849\\
\hline
ALM+N+T &ALM &IBM &IBM & 0.481& 0.837\\
\hline
ALM-LES  \cite{Zormpa-phd}  &ALM & cell-blocking   &None & 0.482 & 0.846\\
\hline
Experiment  \cite{NTNU-turbine} & & & & 0.447 & 0.906\\
\end{tabular*}
{\rule{\temptablewidth}{1.0pt}}
\caption{\label{NTNU_predicted}NTNU rotor with nacelle and tower: the computational models and $C_P$, $C_T$ coefficients predicted at $TSR=6$ for three cases. }
\end{center}
\end{table}

\subsubsection{Power and thrust coefficients}
Three cases are simulated. The first case only simulates the blade effect using ALM 
and no nacelle or tower models are used, which is denoted as "ALM". The second case simulates the blade using ALM 
and only the nacelle is modeled using IBM with $4724$ Lagrangian points, which is denoted as "ALM+N". 
In the third case, ALM is used for blade and IBM is used for nacelle and tower with $10730$ Lagrangian points, 
which is denoted as "ALM+N+T". The power and thrust coefficients are defined by
\begin{align*}
C_P=\frac{2\omega T^{'}}{ \rho\pi R^{2} U_{\infty}^{3}},~C_T=\frac{2D^{'}}{\rho\pi R^{2} U_{\infty}^{2}},
\end{align*}
where $T^{'}$ is the torque generated by the rotor, $\rho$ is the air density and $D^{'}$ is the drag force. 
The numerical results of $C_P, C_T$ are presented in Tab.\ref{NTNU_predicted} for three cases, 
where the experimental data  \cite{NTNU-turbine} and the ALM-LES data \cite{Zormpa-phd}
are given as reference. The results agree well with the experiment data. 
The time histories for $C_P$ and $C_T$ over  $T\in[20, 40]$ (about $19$ rotor cycles) are shown in Fig.\ref{NTNU-coefficients-1}. 
The ALM+N case predicts the highest value, while the ALM+N+T case predicts the lowest result. 
Meanwhile, the ALM+N+T case yields periodic results while the other two cases predict stable results. 
The time histories for $C_P$ and $C_T$ over $2$ rotor cycles are shown in Fig.\ref{NTNU-coefficients-1}. 
Each period represents a $120^{\circ}$ rotation of rotor. 
For ALM+N+T case, the Q-criterion iso-surface colored by streamwise velocity are presented in Fig.\ref{NTNU-coefficients-2} at $T_1$ and $T_2$ denoted in Fig.\ref{NTNU-coefficients-1}. 
The $C_P$ and $C_T$ reach the lowest value when the blade passes the tower,  and reach the highest value when the blades are farthest from the tower.

\begin{figure}[!h]
\centering
\includegraphics[width=0.95\linewidth]{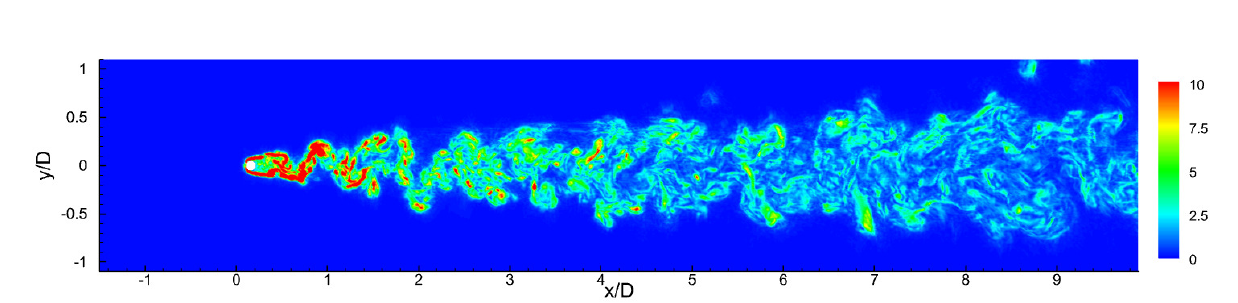}
\includegraphics[width=0.95\linewidth]{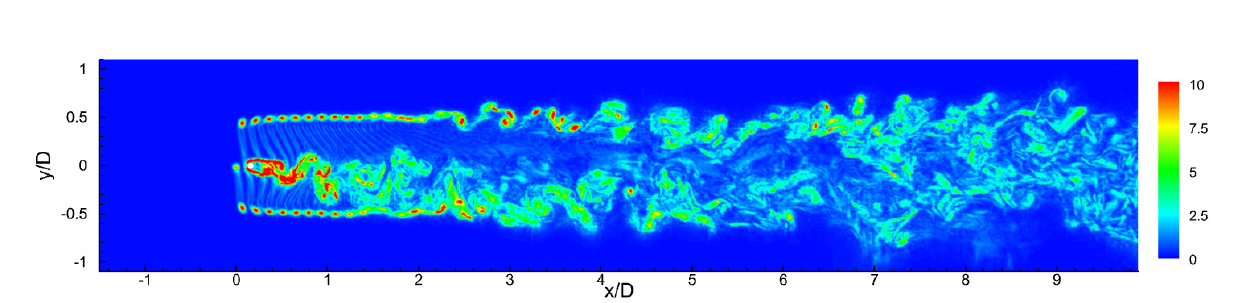}
\includegraphics[width=0.95\linewidth]{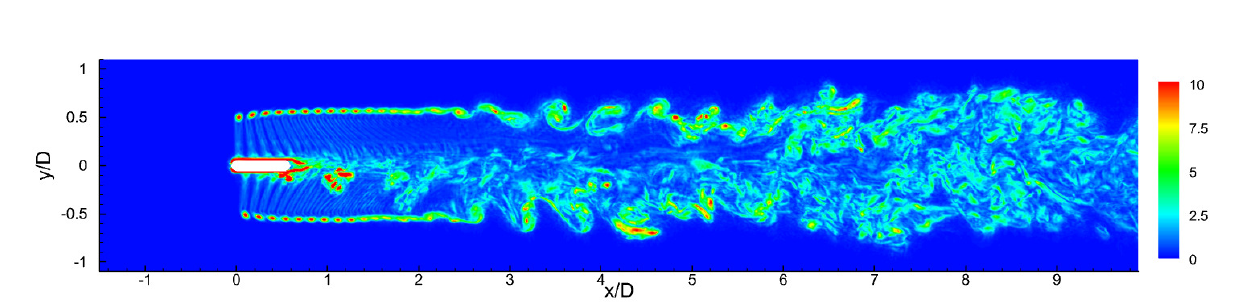}
\caption{\label{ALM+N+T-vorticity-z} NTNU rotor with nacelle and tower: the instantaneous vorticity magnitude at three horizontal planes $z=0.3D, 0.65D$ and $0.902D$ for ALM+N+T case.}
\centering
\includegraphics[width=0.325\linewidth]{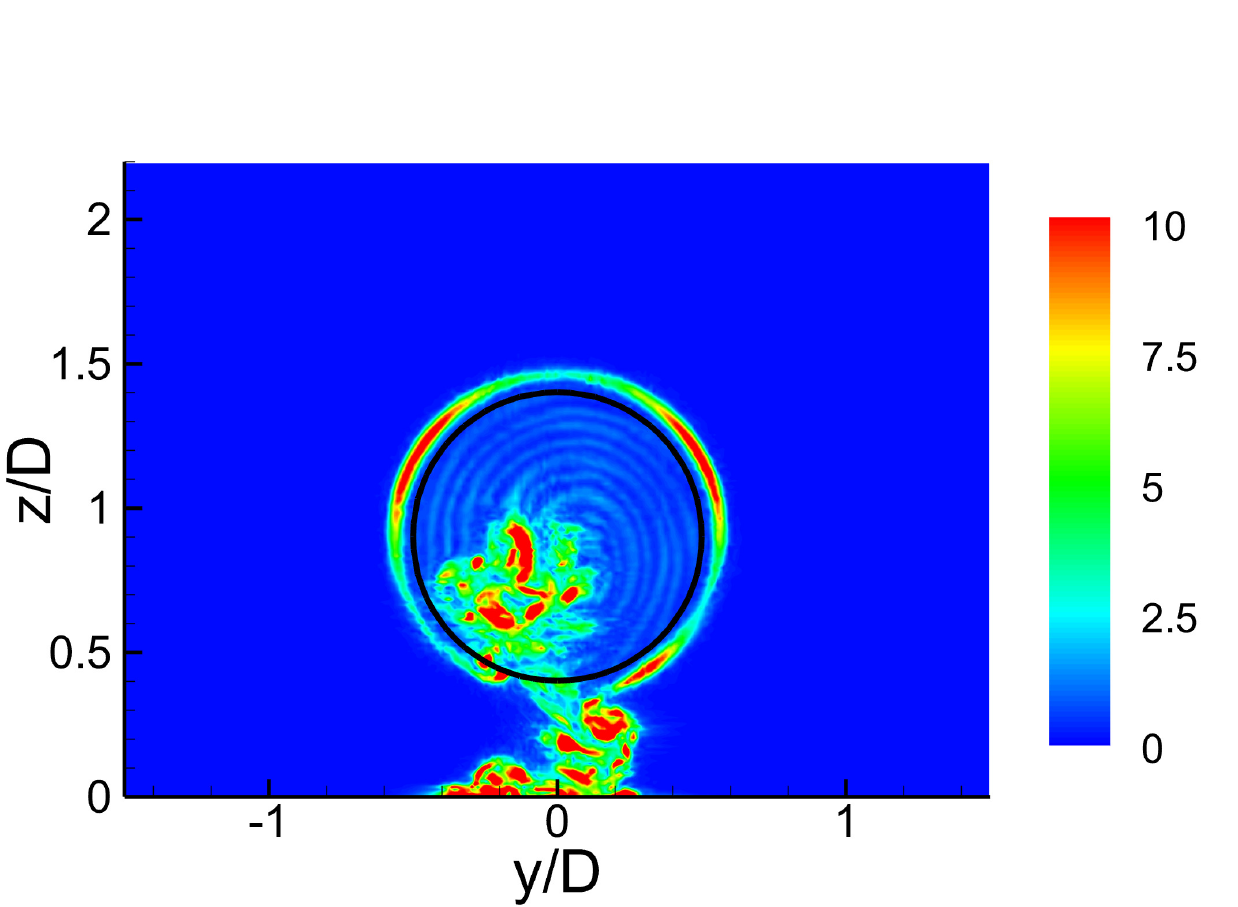}
\includegraphics[width=0.325\linewidth]{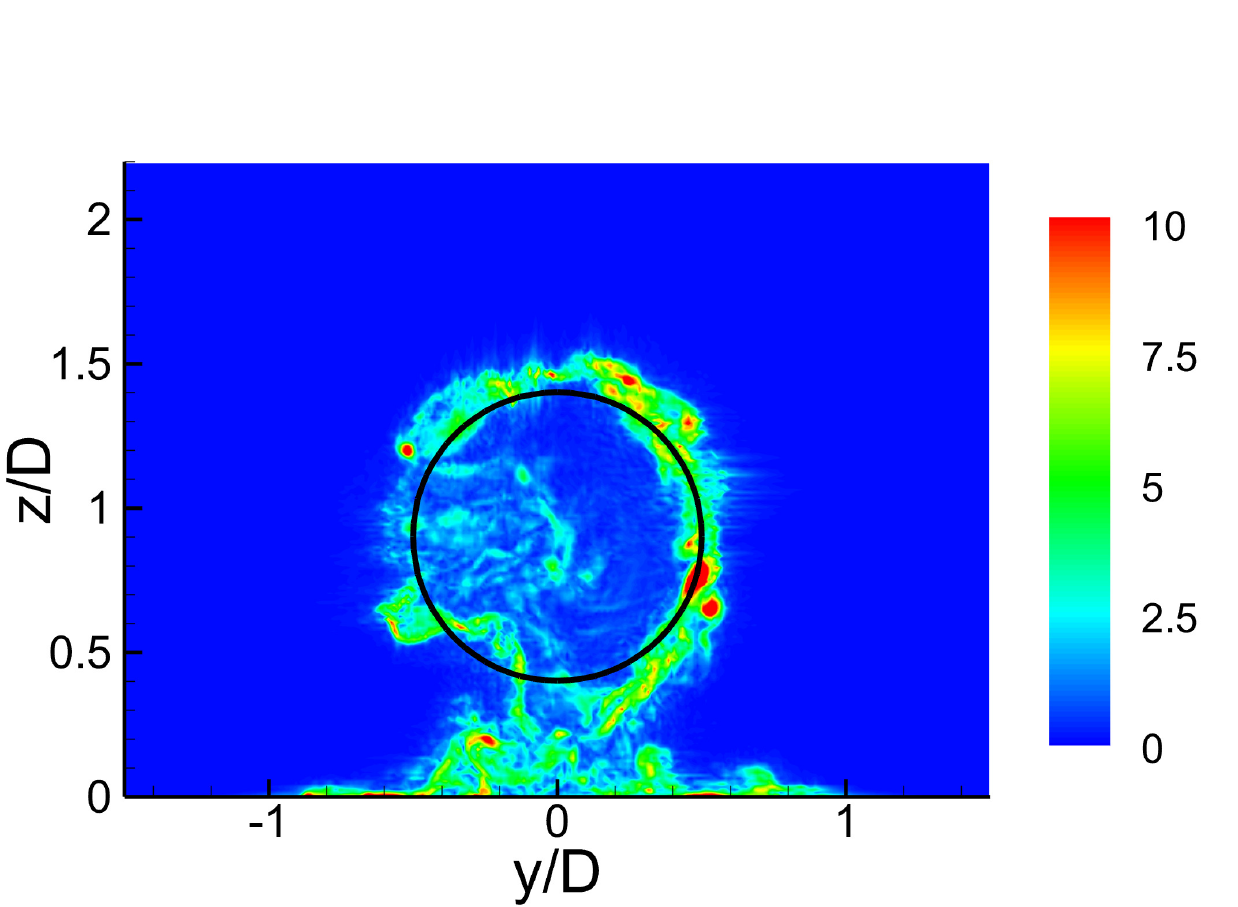}
\includegraphics[width=0.325\linewidth]{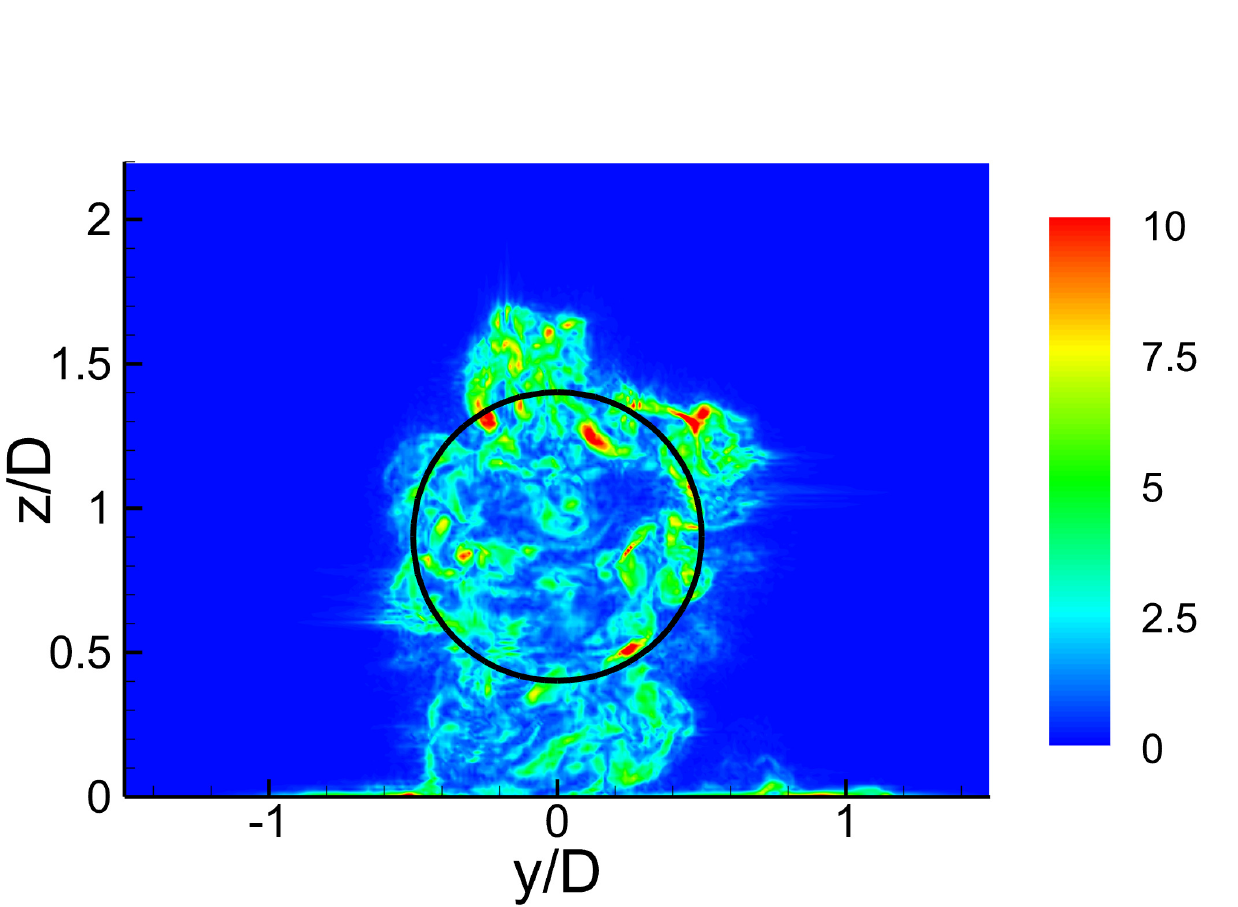}
\caption{\label{ALM+N+T-vorticity-x} NTNU rotor with nacelle and tower: the instanneous vorticity magnitude at three $x$-normal planes $x=1D, 3D$ and $5D$ for ALM+N+T case. The black circle represents the tip-swept region.}
\end{figure}

\begin{figure}[!h]
\centering
\includegraphics[width=0.24\linewidth]{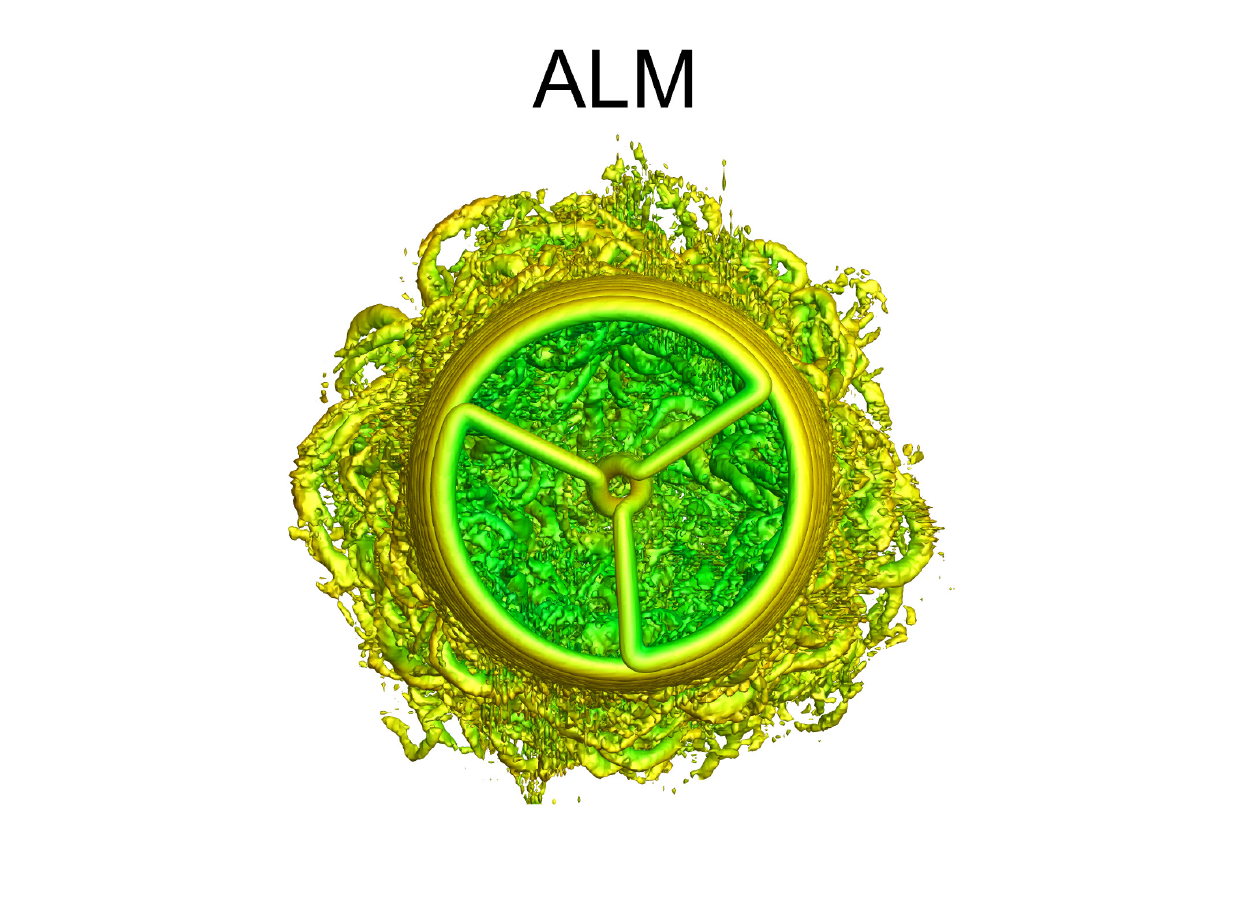}
\includegraphics[width=0.72\linewidth]{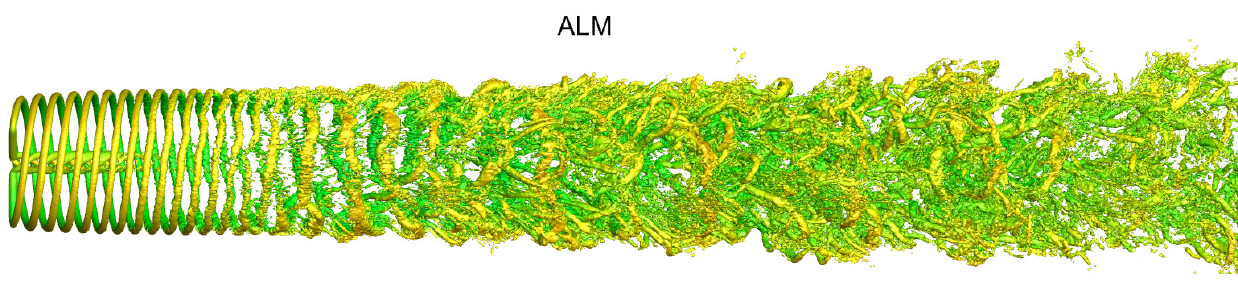}
\includegraphics[width=0.24\linewidth]{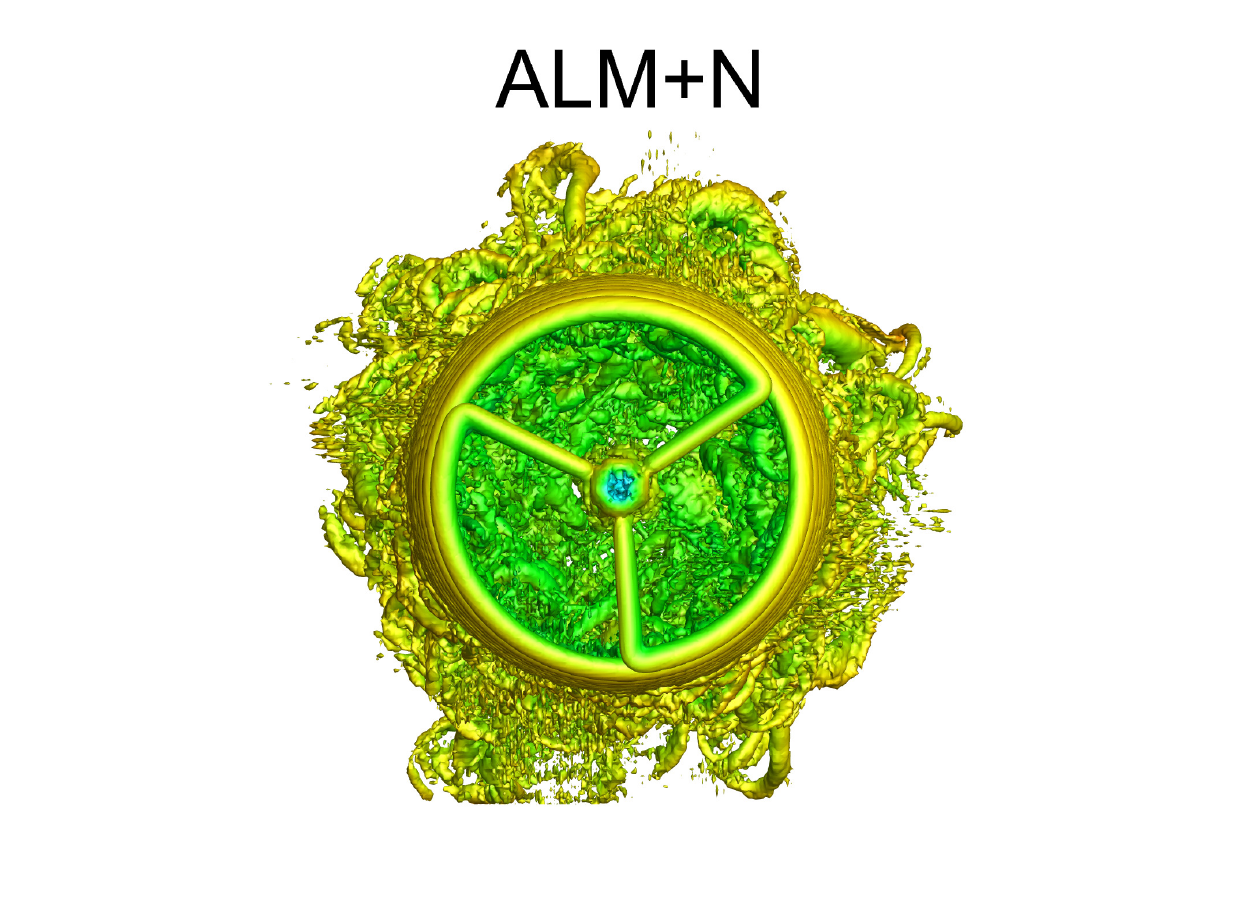}
\includegraphics[width=0.72\linewidth]{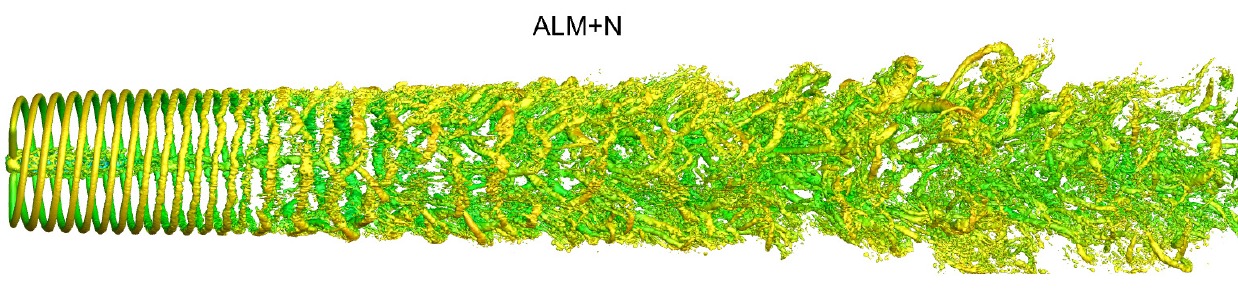}
\includegraphics[width=0.24\linewidth]{ALM+N+T-Q-iso-yz-low-eps-converted-to.pdf}
\includegraphics[width=0.72\linewidth]{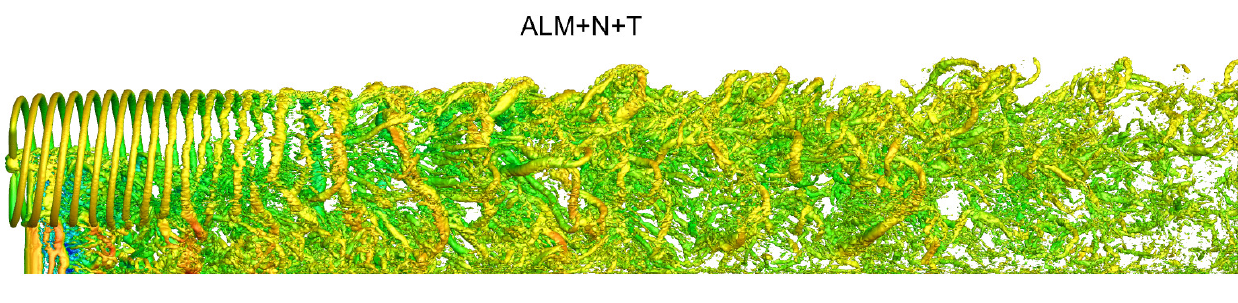}
\caption{\label{NTNU-Q-iso} NTNU rotor with nacelle and tower: the instantaneous Q-criterion iso-surface at $x$-normal plane (left) and $y$-normal plane (right) for three cases.}
\end{figure}

\begin{figure}[!h]
\centering
\includegraphics[width=0.95\linewidth]{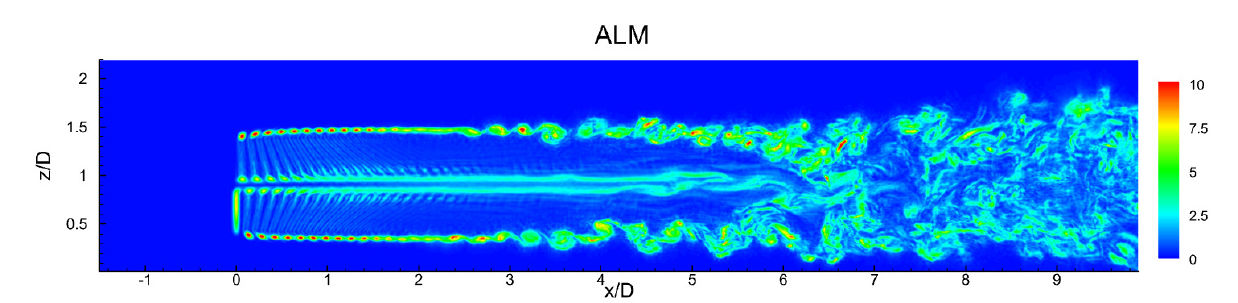}
\includegraphics[width=0.95\linewidth]{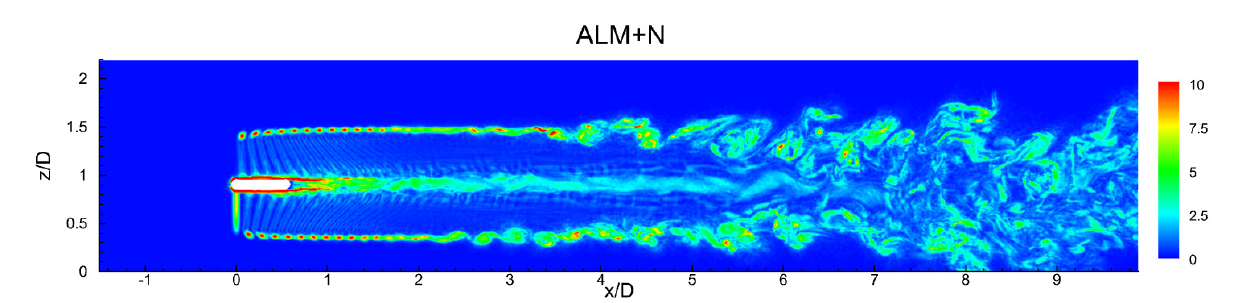}
\includegraphics[width=0.95\linewidth]{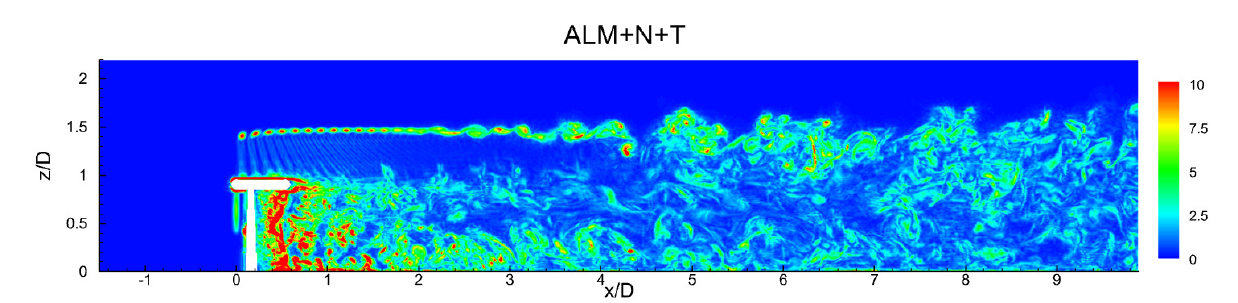}
\caption{\label{NTNU-vorticity} NTNU rotor with nacelle and tower: the instantaneous vorticity magnitude at $y=0$ plane for three cases.}
\end{figure}

\subsubsection{Instantaneous turbulent turbine wakes}
For the ALM+N+T case, the vorticity magnitudes at $T_1$ of three horizontal plane located at 
$z=0.3D, z=0.65D$ and $z=0.902D$ (hub height) are presented in Fig.\ref{ALM+N+T-vorticity-z}. 
At $z=0.3D$, the tower wake is similar with that of the turbulent circular cylinder flow. 
The downstream vortex shed from the tower acts like a K\'{a}rm\'{a}n street, and the blade tip vortex does not effect the tower wake.
The tower wake is symmetrical with respect to the centerline. 
At $z=0.65D$, the tip vortex can be seen, and the vortex shed from the tower can be observed as well. 
However, the tower wake becomes asymmetric and deflects to the $y<0$ side. 
The tower wake interacts with the tip vortex earlier, which is caused by the rotation of blade. 
At $z=0.902D$ (hub height), the vortex street disappears due to the interaction with the nacelle wake and the tower wake is asymmetric. 
Meanwhile, the vorticity magnitudes at $T_1$ of three $x$-normal plane with $x=1D, 3D$ and $5D$ behind the rotor are shown in Fig.\ref{ALM+N+T-vorticity-x}. 
At $x=1D$, the tip vortex can be clearly seen in the tip-swept circular region,
and the tower wake behind the rotor shifts to the $y<0$ region due to the blade rotation. 
At $x=3D$, the tip vortex becomes thicker and spreads out. And the asymmetric phenomenon 
can be still observed. At $x=5D$, the turbulence is developed and the shape of the wake 
becomes irregular.

The iso-surface of Q-criterion colored by streamwise velocity are presented in Fig.\ref{NTNU-Q-iso} for three cases at $T_1$. 
The result of Q-criterion for ALM case is similar with that of ALM+N case, except for the nacelle region. 
In the ALM case, the root vortex can be clearly seen in the near wake region and then breaks down. 
In the ALM+N case, the nacelle vortex is triggered and interact with the root vortex. 
For the ALM+N+T case, the upper side is similar to the other two cases. 
In the lower side, the vortex triggered by the tower is similar to the cylinder flow. 
The tower vortex interact with the tip vortex and the nacelle vortex, 
causing an earlier transition into turbulence. 
The two dimensional vorticity magnitudes is shown in Fig.\ref{NTNU-vorticity} for three cases at $T_1$ in the $y=0$ plane. 
For the ALM case, the tip vortex and root vortex propagate in the wake region.
The mixing begins at around $x=6D$, where the wake turns into turbulent flow. 
The ALM+N case is similar to the ALM case, except for the region behind the nacelle as well. 
The interaction between the root vortex and nacelle vortex leads to a more disordered wake behavior. 
For the ALM+N+T case, the upper side is similar with the other two cases. 
On the side below the rotor center, the tip vortex breaks down very quickly due to the interaction with tower vortex. 
Because of the effect of tower vortex, the flow behind the nacelle turns into turbulence earlier.

\begin{figure}[!h]
\centering
\includegraphics[width=0.9\linewidth]{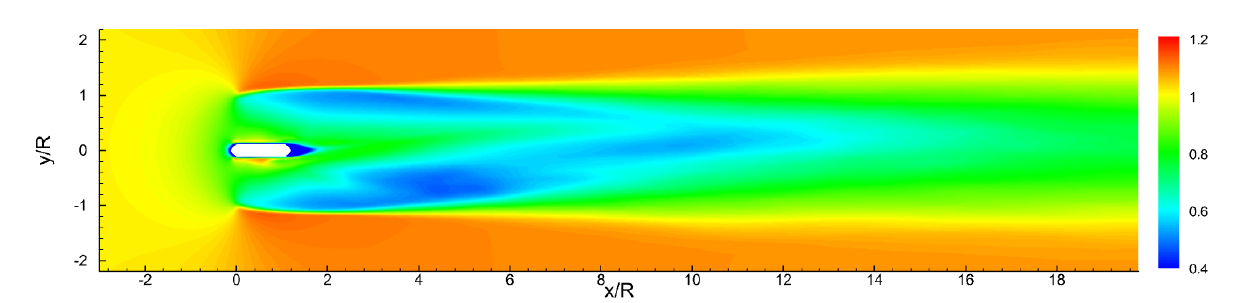}
\includegraphics[width=0.9\linewidth]{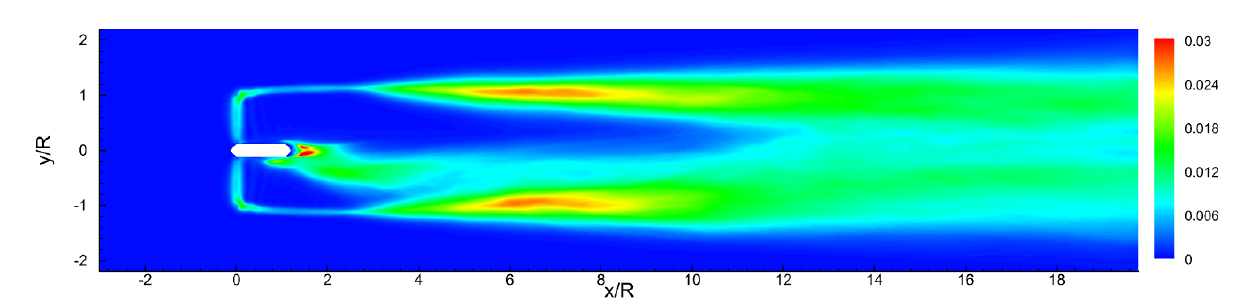}
\caption{\label{NTNU-xz-plane} NTNU rotor with nacelle and tower: the time-averaged streamwise velocity (top) and TKE (bottom) at the $z=0.902D$ (hub height) plane.}
\end{figure}
\begin{figure}[!h]
\centering
\includegraphics[width=0.45\linewidth]{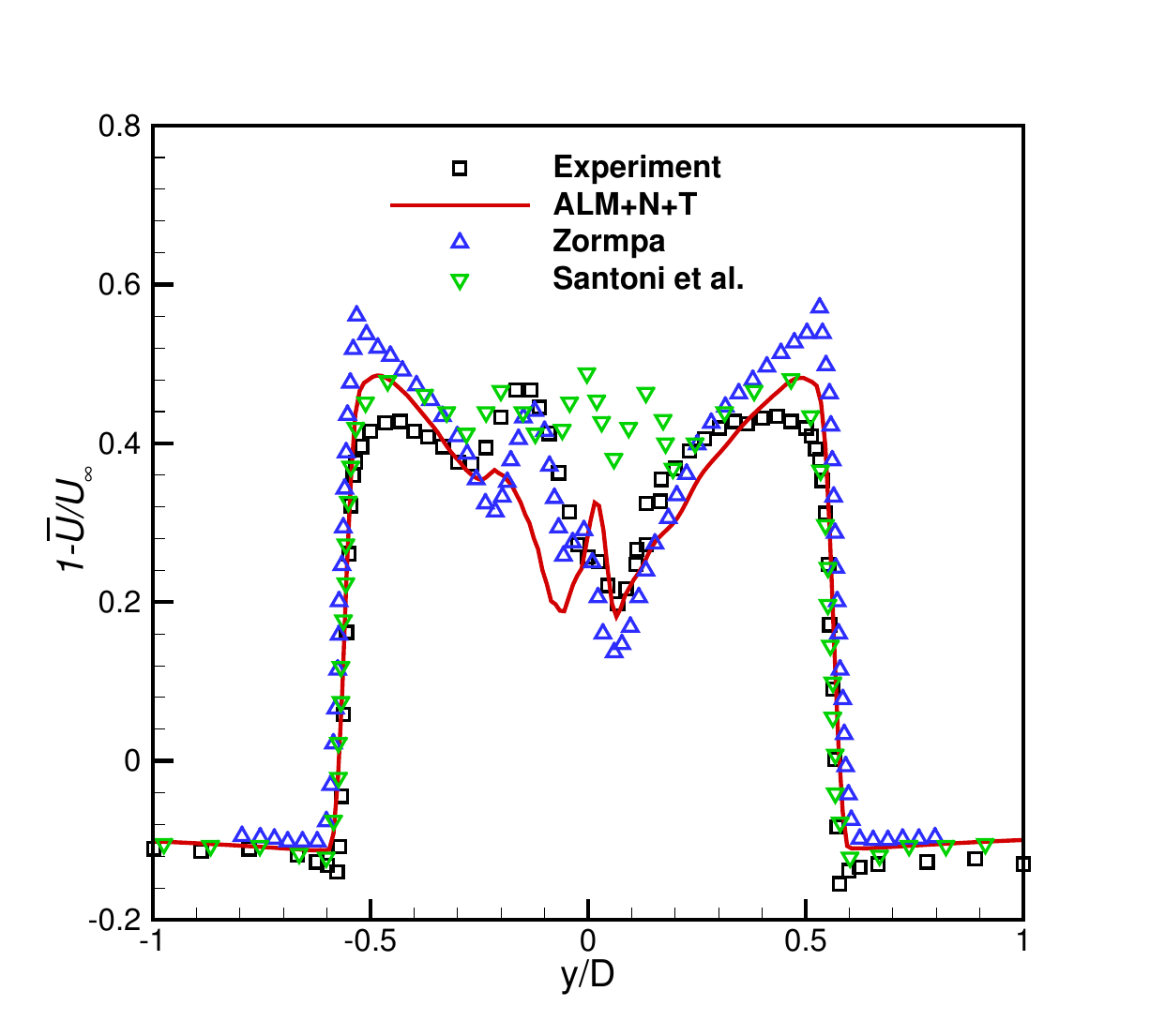}\includegraphics[width=0.45\linewidth]{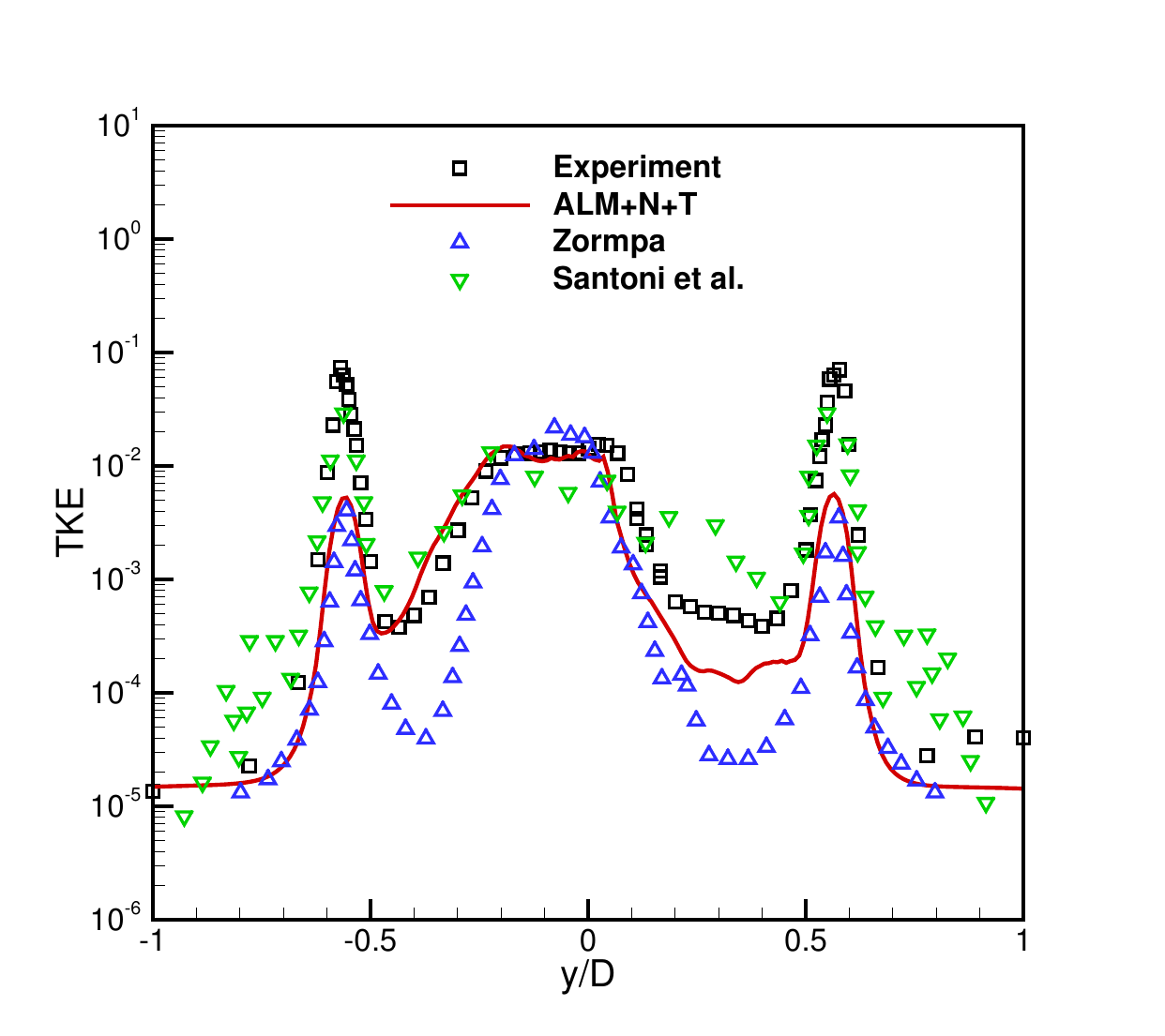}
\includegraphics[width=0.45\linewidth]{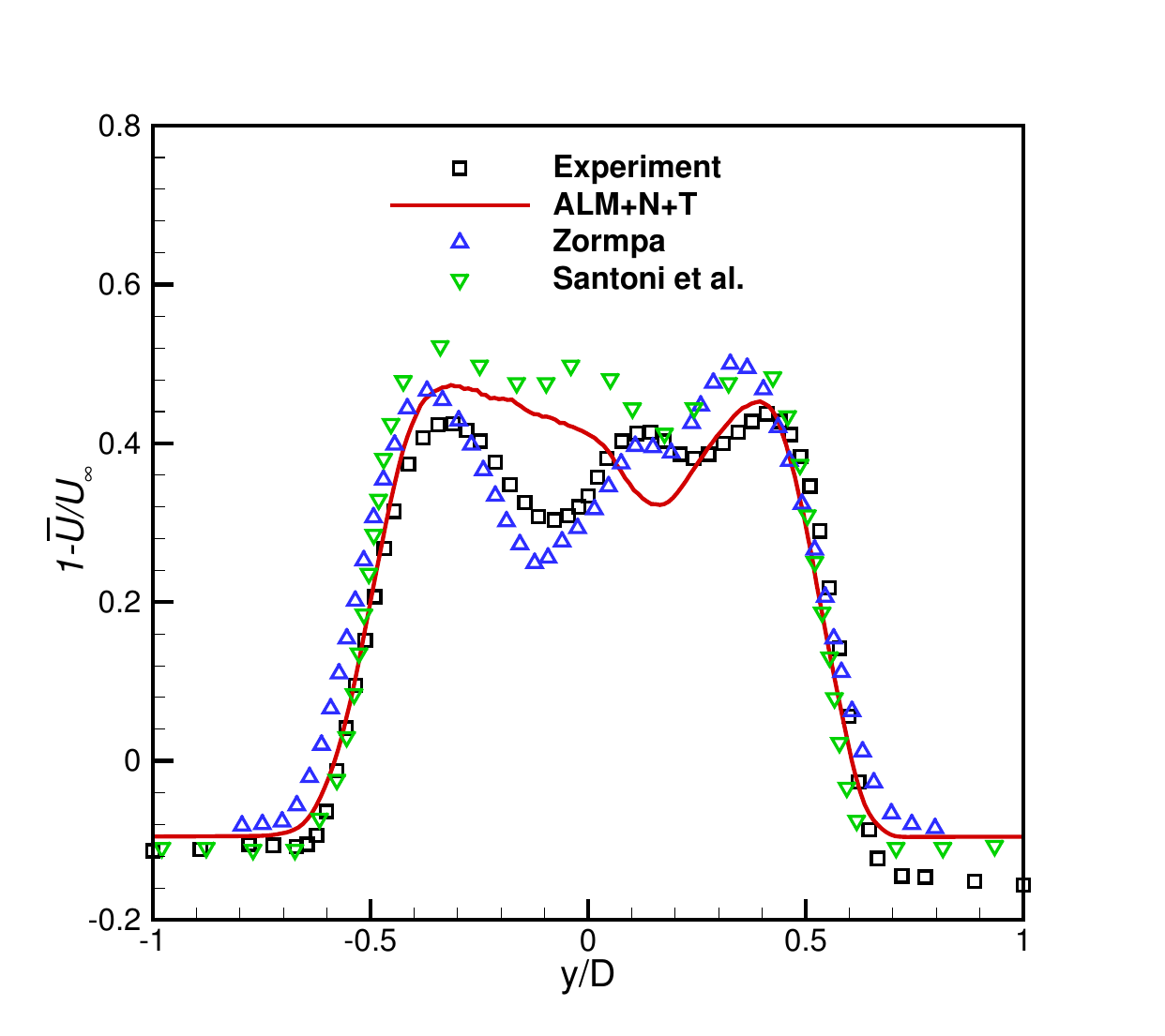}\includegraphics[width=0.45\linewidth]{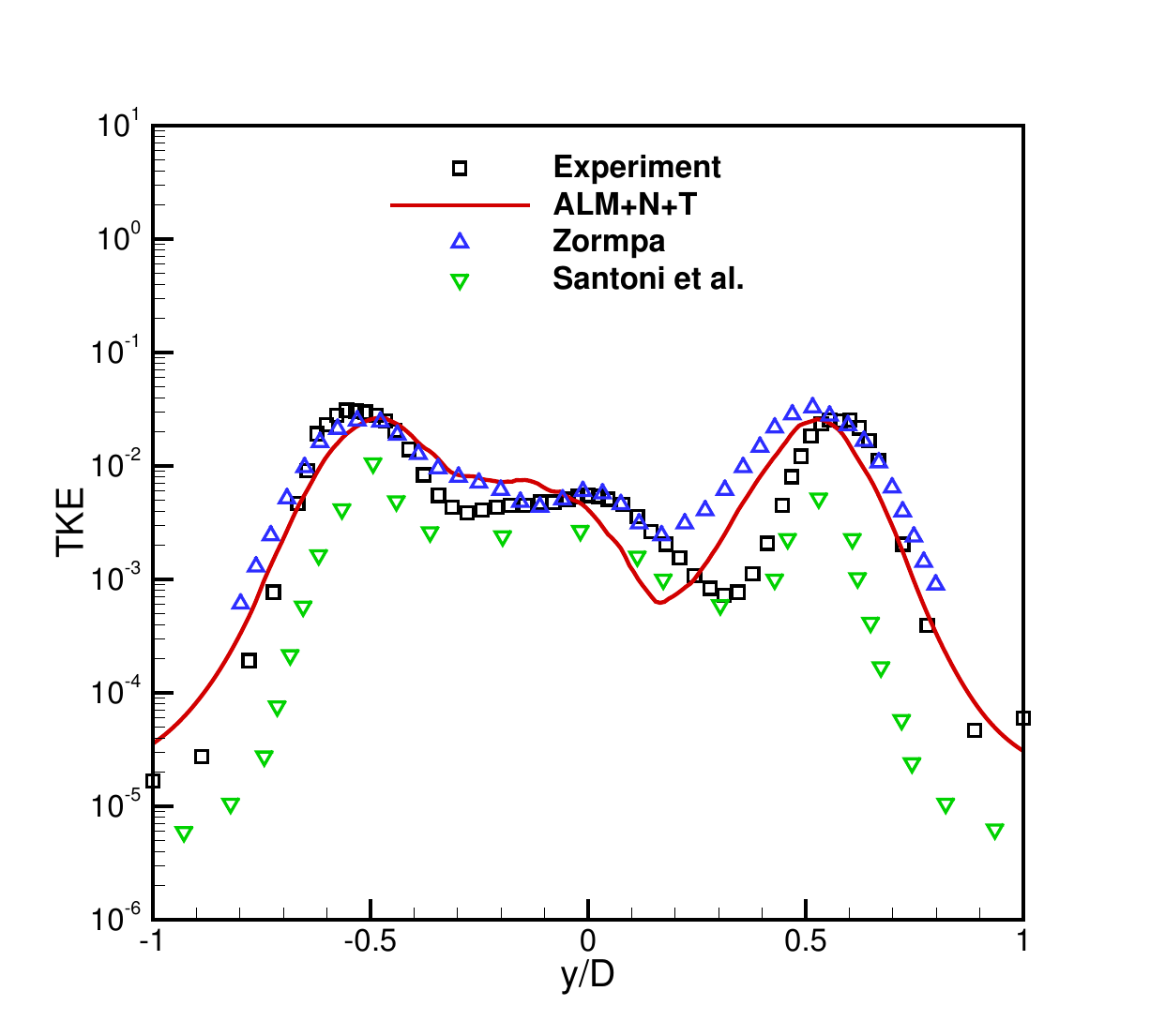}
\includegraphics[width=0.45\linewidth]{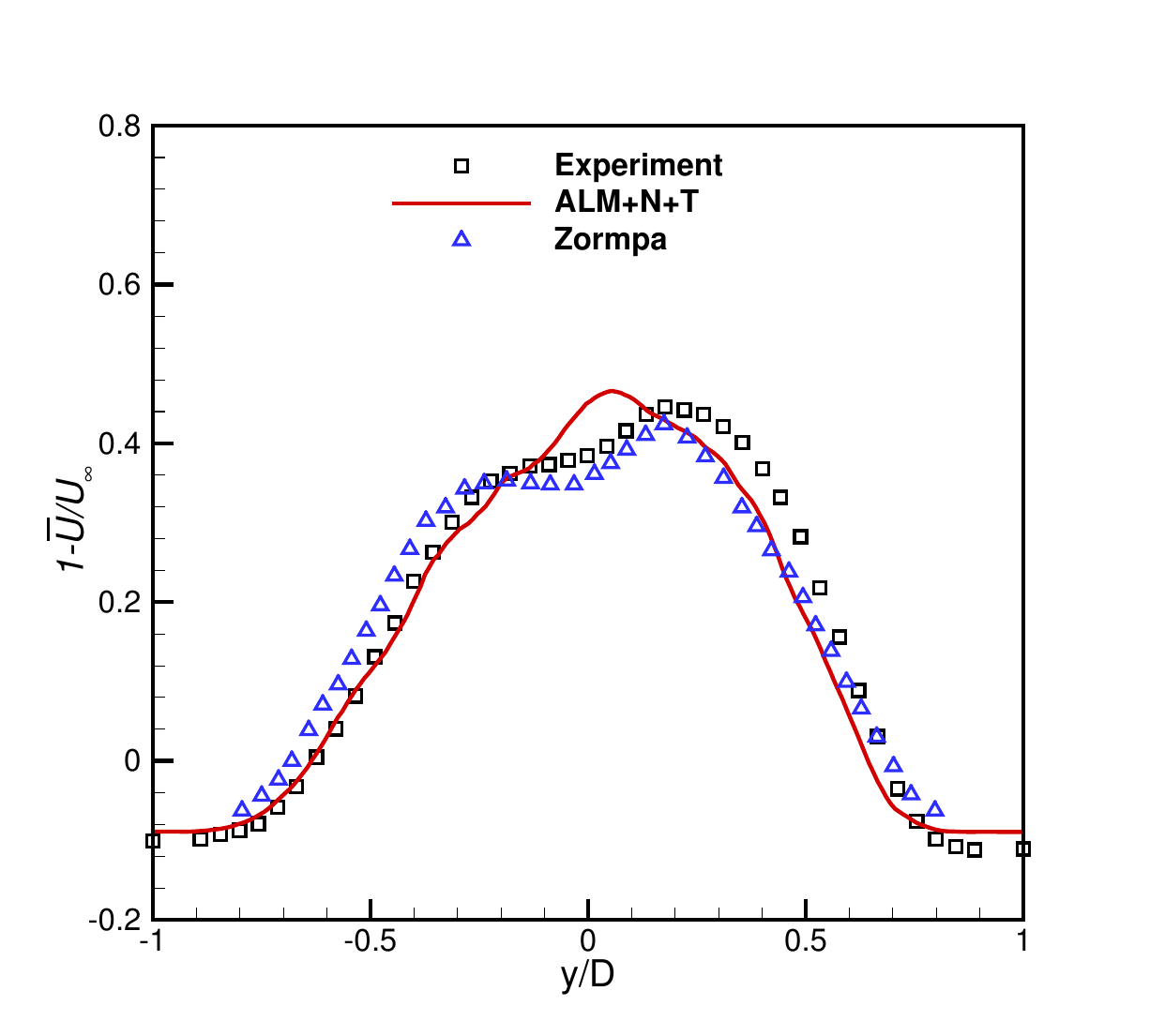}\includegraphics[width=0.45\linewidth]{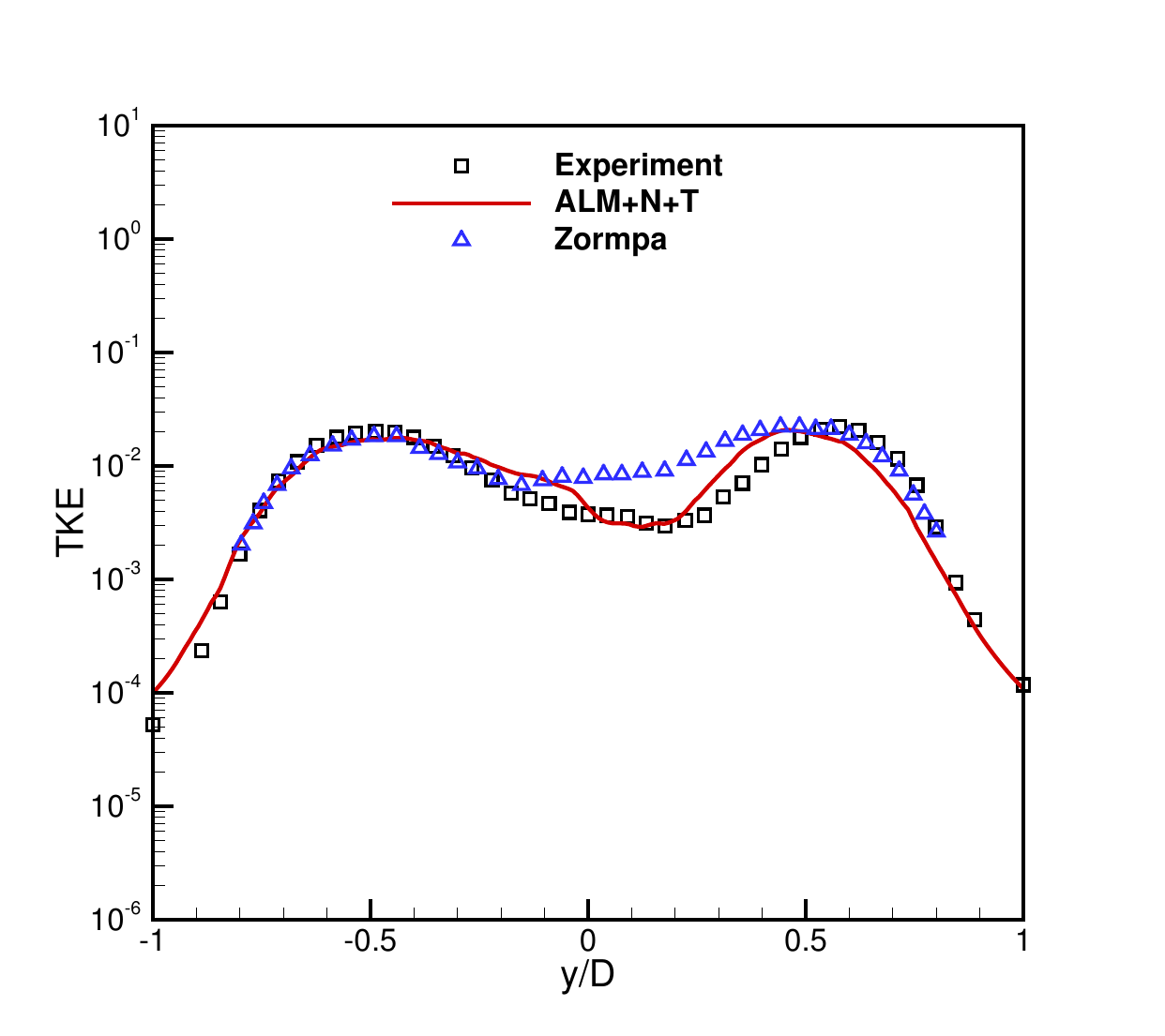}
\caption{\label{NTNU-profile} NTNU rotor with nacelle and tower: the time-averaged streamwise velocity (left) and TKE (right) profiles at $x= 1D, 3D$ and $5D$. The data at $x=5D$ is not provided in \cite{NTNU-IBM-Santoni}.}
\end{figure}

\subsubsection{Time-averaged turbulent turbine wakes}
For the ALM+N+T case, the time-averaged streamwise velocity and TKE contours are presented in Fig.\ref{NTNU-xz-plane} within $3$-$5$ flow-through time, 
in which the two-dimensional contours are given at the hub height plane. 
At $x=0D$ where the rotor plane is placed, the streamwise velocity is reduced and the TKE is triggered due to rotor rotation. 
On the lateral side of nacelle, the streamwise velocity increases. 
However, behind the nacelle, the formation of separated vortices creates a low-speed region and generates turbulent kinetic energy.
In the wake region, the wake velocity recovers and the TKE is fully developed. 
An asymmetry phenomenon is observed  both in velocity and TKE. At the side with $y<0$, the higher 
velocity deficit and higher TKE are observed. The asymmetrical distribution is consistent with the conclusion above.

The velocity deficit and TKE profiles at $x=1D, 3D$ and $5D$ behind the turbine are presented in Fig.\ref{NTNU-profile}, 
where the experimental data is given from \cite{NTNU-turbine}. The results agree well with the reference experimental results. 
The data in \cite{Zormpa-phd} and \cite{NTNU-IBM-Santoni} are also given as comparisons. 
The blade is modeled using ALM and the nacelle and tower are modeled by body-fitted mesh in \cite{Zormpa-phd}. 
The ALM is used for blade and the IBM is used for nacelle and tower in \cite{NTNU-IBM-Santoni}. 

For the velocity profiles, at $x=1D$, a high velocity deficit is observed in the region behind the blade tip due to the drag force of the blade. 
The numerical simulation gives slightly larger results than measured data. 
In the experiment, a velocity deficit peak can be observed at around $y=-0.2D$, while in the numerical simulation a velocity deficit peak is located 
in the $y=0$ region because of the nacelle drag force. 
At $x=3D$, the numerical simulation differs from the measured data in $y\in(-0.4D,0.2D)$. 
And the experiment predicts lower result in $y>0.7D$ region. 
At $x=5D$, the wake begins to recover and the numerical result is consistent with measured data. 
The width of the wake is also similar with the measured data at three locations. 

For the TKE profiles, the experiment predicts higher TKE in the region behind the blade tip at $x=1D$, 
where the turbulence is triggered due to the blade tip vortex. 
The TKE in $y<0$ side is higher than the other side, this asymmetrical distribution is caused 
by the blade rotation and is correctly predicted by the numerical simulation. 
At $x=3D$, there is slightly phase difference in $y>0$ region. 
At $x=5D$, the TKE distribution becomes smoother because of the wake mixing 
and the numerical result is consistent with experiment. 
For the AL-LES case, the TKE in the region behind the rotor center differs from other cases. 
It predicts lower TKE and there are two unphysical peaks triggered by the blade root vortex.

\section{Conclusion}
In this paper, the high-order gas-kinetic scheme (GKS) has been integrated with the actuator line model (ALM) 
and the immersed boundary method (IBM) to simulate the wind turbine wakes with nacelle and tower. 
We first extend the high-order GKS to simulate the three-dimensional weakly compressible flows.
The ALM is adopted to model the rotating blades and the IBM is developed to model the nacelle and tower.
Both the ALM and IBM are integrated into high-order GKS framework as an external body force term in the momentum equations. 
The low-speed turbulent channel flow demonstrates the ability of current extended high-order GKS to resolve the weakly compressible multi-scale turbulent flows accurately.
The turbulent flow past a circular cylinder confirms that the IBM captures the wake metrics accurately over the blunt body.

In the simulation of NREL $5$~MW reference wind turbine, the effect of the smearing kernel width $\varepsilon$ is investigated.
For point-based ALM sampling, the smearing kernel width affects the sampled angle of attack in the outboard region.
The typical $\varepsilon=3\Delta x$ is adopted, 
whose solutions agree well with the reference ALM and blade-resolved solutions.

The wind turbine wake is classified into near wake, transitional wake and far wake. 
For the aerodynamic loadings and the near wake region, 
there is limited impact varying the numerical scheme and grid size. 
Compared to the second-order GKS, the high-order GKS gives a higher resolution in the transitional wake and far wake region with the same grid size. 
The time-averaged velocity and turbulent kinetic energy from the second-order GKS with a locally-refined grid converge
to those from the high-order GKS with the uniform grid.  
In summary, the high-order GKS with ALM improves the
resolution of the vortex-dominated wake turbulence.

We simulate the benchmarking NTNU "Blind Test 1" wind turbine  using the IBM to model the nacelle and tower.
The periodic power and thrust coefficients of the rotor blade are captured due to the blade-tower interactions, while the steady coefficients are obtained without the tower.
We find the near-wake vortex triggered by the tower is similar to the turbulent circular cylinder flow. 
Compared to the turbine wakes without tower, the effect of the tower wake causes an asymmetrical wake distribution and the interaction between tower vortex and tip vortex leads to an earlier transition into turbulence. 
The time-averaged streamwise velocity and turbulent kinetic energy profiles at measured locations downstream the turbine from the high-order GKS with IBM agree well with the experimental data.

In conclusion, the high-order GKS with ALM and IBM offers a highly accurate and efficient approach for simulating the wind turbine with nacelle and tower, where the vortex-dominated high-Reynolds number wake turbulence can be well resolved. 
In order to study the aerodynamic loadings and turbulent wakes of wind turbines under realistic atmospheric conditions, the turbulent inflows and shear velocity profiles will be integrated into the current in-house solver in future work.

\section*{Acknowledgements.}
The current research of L. Pan is supported by National Natural
Science Foundation of China (12494543, 12471364) and Beijing Natural
Science Foundation (1232012). 
G.Y Cao thank the UKRI Future Leaders Fellowship MR/V02504X/1.


\section*{Data availability.}
The data that support the findings of this study are available from
the corresponding author upon reasonable request.

\section*{Appendix}
This section presents the determination of microscopic coefficients in Eq.\eqref{flux}.
The following notation for the moments of equilibrium state $g_{0}$ is introduced for the three-dimensional weakly compressible flows
\begin{align*}
\langle \dots\rangle=\frac{1}{\rho}\int\left(\dots\right)g_{0}\text{d}u\text{d}v\text{d}w.
\end{align*}
The moments $\langle u^{n} \rangle$ with integer $n \geq 0$ can be calculated as \begin{align*}
\langle u^{0} \rangle &=1,\\
\langle u^{1} \rangle &=U,\\
&\dots\\
\langle u^{n+2} \rangle= \frac{n+1}{2\lambda}&\langle u^{n} \rangle+U\langle u^{n+1} \rangle.
\end{align*}
The moments $\langle v^{n} \rangle$  and $\langle w^{n} \rangle$ can be calculated similarly, and the general moment formula reads
\begin{align*}
\langle u^{m}v^{n}w^{l}\rangle =\langle u^{m}\rangle \langle v^{n}\rangle  \langle w^{l}\rangle.
\end{align*}

The relation between the derivatives of macroscopic flow variables and the derivatives of $g_0$ can be given by
\begin{align*}
\int \boldsymbol{\psi}\frac{\partial g_{0}}{\partial x}\text{d}u\text{d}v\text{d}w=\frac{\partial \textbf{Q}_0}{\partial x}.
\end{align*}
The spatial derivatives of $g_0$ reads
\begin{align*}
\frac{\partial g_{0}}{\partial x}=a_1g_{0},
\end{align*}
with
\begin{align*}
a_1&=a_1^1+a_1^2u+a_1^3v+a_1^4w.
\end{align*}
By using the formula for moment evaluation, we have
\begin{align*}
\left(
\begin{matrix}
1 & U &V &W \\
U & U^2+\frac{1}{2\lambda} & UV & UW \\
V & UV & V^2+\frac{1}{2\lambda} & VW \\
W & UW & VW & W^2+\frac{1}{2\lambda}
\end{matrix}
\right)
\left(
\begin{array}{c}
a_1^1 \\
a_1^2 \\
a_1^3 \\
a_1^4
\end{array}
\right)=
\left(
\begin{array}{c}
d_1 \\
d_2 \\
d_3 \\
d_4
\end{array}
\right),
\end{align*}
with $\displaystyle\frac{\partial \textbf{Q}_0}{\partial x}=(d_1,d_2,d_3,d_4)^T$.
The solution to the above equation gives the microscopic coefficients as
\begin{align*}
a_1^4&=2\lambda r_4,\\
a_1^3&=2\lambda r_3,\\
a_1^2&=2\lambda r_2,\\
a_1^1&=d_1-Ua_2-Va_3-Wa_4.
\end{align*}
with
\begin{align*}
r_4&=d_4-Wd_1,\\
r_3&=d_3-Vd_1, \\
r_2&=d_2-Ud_1.
\end{align*}

Similarly, the coefficients $a_2$ and $a_3$ corresponding to $\displaystyle\frac{\partial \textbf{Q}_0}{\partial y}$
and $\displaystyle\frac{\partial \textbf{Q}_0}{\partial z}$ can be also determined.
After determining $a_1$, $a_2$ and $a_3$, the term $A$ can be obtained from the compatibility condition
\begin{align*}
\int  \boldsymbol{\psi} Ag_0\text{d}u\text{d}v\text{d}w=-\int\boldsymbol{\psi}(a_1u+a_2v+a_3w)g_0\text{d}u\text{d}v\text{d}w.
\end{align*}

\end{document}